\documentclass[a4paper,fleqn,usenatbib]{mnras}

\usepackage{newtxtext,newtxmath}

\usepackage[T1]{fontenc}
\usepackage{ae,aecompl}

\usepackage{graphicx}	
\usepackage{amsmath}	
\usepackage{amssymb}	

\title[Evidence of atomic diffusion in M67?]{The Gaia-ESO Survey: Evidence of atomic diffusion in M67?}

\author[Bertelli Motta et al.]{
	C. Bertelli Motta,$^{1}$\thanks{E-mail: cbertelli@ari.uni-heidelberg.de}
    A. Pasquali,$^{1}$
   J. Richer,$^{2}$
    G. Michaud, $^{2}$
	M. Salaris,$^{3}$
	A. Bragaglia,$^{4}$
	\newauthor
	L. Magrini,$^{5}$
	S. Randich,$^{5}$
	E. K. Grebel,$^{1}$
	V. Adibekyan,$^{6}$
	S. Blanco-Cuaresma,$^{7}$
	\newauthor
	A. Drazdauskas,$^{8}$
	X. Fu,$^{4,9}$
	S. Martell,$^{10}$
	G. Tautvai\v{s}ien\.{e},$^{8}$
G. Gilmore,$^{11}$
E.~J. Alfaro,$^{12}$
\newauthor
T. Bensby,$^{13}$
E. Flaccomio,$^{14}$
S.~E. Koposov,$^{11,15}$
A.~J. Korn,$^{16}$
A.~C. Lanzafame,$^{17}$
\newauthor
R. Smiljanic,$^{18}$
A. Bayo,$^{19,20}$
G. Carraro,$^{21}$
A.~R. Casey,$^{22,23}$
M.~T. Costado,$^{24}$
F. Damiani,$^{14}$
\newauthor
E. Franciosini,$^{5}$
U. Heiter,$^{25}$
A. Hourihane,$^{11}$
P. Jofr\'e,$^{11,26}$
C. Lardo,$^{27}$
J. Lewis,$^{11}$
\newauthor
L. Monaco,$^{28}$
L. Morbidelli,$^{5}$
G.~G. Sacco,$^{5}$
S.~G. Sousa,$^{6}$
C.~C. Worley,$^{11}$
S. Zaggia$^{21}$\\
}

\date{Accepted 2018 April 17. Received 2018 April 12; in original form 2017 November 24}

\pubyear{2015}

\begin{document}
\label{firstpage}
\pagerange{\pageref{firstpage}--\pageref{lastpage}}
\maketitle

\begin{abstract}
Investigating the chemical homogeneity of stars born from the same molecular cloud at virtually the same time is very important for our understanding of the  chemical enrichment of the interstellar medium and with it the chemical evolution of the Galaxy. One major cause of inhomogeneities in the abundances of open clusters is stellar evolution of the cluster members.
In this work, we investigate variations in the surface chemical composition of member stars of the old open cluster M67 as a possible consequence of atomic diffusion effects taking place during the main-sequence phase. The abundances used are obtained from high-resolution UVES/FLAMES spectra within the framework of the Gaia-ESO Survey. We find that the surface abundances of stars on the main sequence decrease with increasing mass reaching a minimum at the turn-off. After deepening of the convective envelope in sub-giant branch stars, the initial surface abundances are restored. We found the measured abundances to be consistent with the predictions of stellar evolutionary models for a cluster with the age and metallicity of M67. Our findings indicate that atomic diffusion poses a non-negligible constraint on the achievable precision of chemical tagging methods.
\end{abstract}




\section{Introduction}
\label{sec:intro}

Based on the commonly accepted knowledge that most stars form in associations or gravitationally bound clusters that dissolve with time and release their members into the field population \citep{lada2003}, the suggestion that stars can be traced back to their parent cluster based on their chemical composition, the so-called chemical tagging (see, e.g., \citealt{freeman2002}), has become very popular over the past years. 

While chemical tagging is a potentially powerful tool for the understanding of the Galactic evolutionary history, there are limitations to this method that need to be accounted for. In fact, the common and reasonable assumption that stars born as a single stellar population from the same molecular cloud share the same chemical composition only holds for the \textit{initial} surface abundances, which may vary as stars start following their evolutionary track. During the life of a star, several physical processes are at play in modifying its surface chemical composition. Although these changes are not dramatic, they do put constraints on the resolution achievable in studies of chemical tagging (see, e.g., \citealt{blanco2015} for chemical tagging applied to open clusters). In addition, two stars formed from two different molecular clouds might happen to present the same surface abundances. This could be because the clouds of origin shared the same chemical composition or because the two objects have different mass and find themselves in different evolutionary stages, which might have led by chance to the same surface abundances, even if the initial ones were different (see, e.g., \citealt{ness2018,dotter2017}). In this scenario it would be impossible to tag the two stars back to a particular cluster of origin based on their present-day chemical composition alone. As \citet{dotter2017} suggested, one would need to use stellar evolutionary models to infer the initial abundances of stars from the observed ones. 

Stellar evolutionary effects are best studied with a sample of stars belonging to the same stellar population and most likely sharing the same initial chemical composition. While globular clusters are known to host two or more populations of stars with different light-element abundances among their members (see, e.g., \citealt{gratton2004,kayser2008,carretta2009,gratton2012}), this is not the case for open clusters (see, e.g., \citealt{bragaglia2012,carraro2014,bragaglia2014}), which are therefore ideal test benches for our study.  Unfortunately, for most open clusters only the brighter stars in the red clump and upper red giant branch are usually considered for spectroscopic studies, mainly because the main sequence and the turn-off regions are usually relatively faint and require very long exposure times. This makes it difficult to find in the literature suitable data for the investigation of evolutionary effects, which require stars at different evolutionary stages. Furthermore, we also need to test these effects in a range of metallicity and age (i.e., stellar mass).

In this framework M67 plays a special role. With an age of $3.75-4$ Gyr (see, e.g., \citealt{sarajedini2009, bellini2010,kharchenko2013}), a distance of $800-900$ pc from the Sun (see, e.g., \citealt{sarajedini2009}) and approximately solar metallicity \citep[see, e.g.,][]{tautv2000,shetrone2000,yong2005,randich2006,pace2008,onehag2014,bertelli2017} this old open cluster is considered a proxy for solar studies and is used as a calibrator in many spectroscopic surveys. M67 hosts a solar twin with one of the most similar chemical compositions to the Sun discovered so far (M67-1194, \citealt{onehag2011}) and is subject to a search for extrasolar planets around its member stars (\citealt{pasquini2012,brucalassi2014,brucalassi2016,brucalassi2017}). It has been speculated that this old open cluster could be very similar to (or even be) the environment in which the Sun formed. For all these reasons M67 has been studied in great detail in the past and, unlike most open clusters, high-resolution spectra of a large number of cluster members in all evolutionary stages are available for analysis, making M67 an excellent test-case for evolutionary studies.

Atomic diffusion (see \citealt{michaud2015}) is expected to alter the surface abundance of stars during the main-sequence phase due to the combined effect of gravitational settling, causing different elements to sink towards the interior of the star, and radiative acceleration working against it. In low-mass stars the overall trend is similar for all elemental species: surface abundances decrease along the main sequence with increasing stellar mass, reaching a minimum at the turn-off. The amplitude of the abundance variation between the early main sequence and the turn-off depends on the efficiency of radiative acceleration, i.e. on the degree of absorption of the outgoing photon flux for the different species. When the outer convective zone becomes deeper (after the turn-off), material from the stellar interior is brought to the surface. For most elements this means a recovery of the initial surface abundances, with the exception of species that undergo nuclear processing, such as $^{3}\mathrm{He}$, Li, Be, and B \citep[the surface abundances of C and N also vary due to the dredge-up of the material involved in the CNO cycle once the convective envelope reaches the innermost parts of the star, see][and references therein]{bertelli2017}. While metals sink towards the stellar interior, hydrogen is pushed into the outer layers, depriving the nucleus of fuel and thus shortening the life of the star on the main sequence (see, e.g., \citealt{jofre2011} and \citealt{salaris2001}). This poses a further constraint on chemical tagging, since taking into account diffusion when estimating the age of a cluster through isochrone fitting can reduce ages by up to $\sim10\%$ for the oldest globular clusters. The reduction in age is proportionally smaller for younger systems such as open clusters, but still present.

Thus, in order to study atomic diffusion, stars along the main sequence (MS), turn-off (TO), subgiant (SGB), and red giant branch (RGB) are needed.  In \citet{bertelli2017}, we investigated the effects of the first dredge-up (FDU) affecting stars on the sub-giant branch and lower red giant branch of  M67 with data from the Apache Point Observatory Galactic Evolution Experiment --APOGEE--  \citep[see][]{majewski2017}. M67 was furthermore included in the study of \citet{smiljanic2016} where, using data from the Gaia-ESO Survey \citep[GES,][]{gilmore2012,randich2013}, the authors investigated variations in Na and Al abundances as a consequence of mixing processes in the stellar interiors. In this work, we extend this study to the variations in surface abundances due to atomic diffusion by investigating the MS, TO, SGB, and RGB phase of M67 with GES data.

In Section~\ref{sec:data} and ~\ref{sec:models} we present the data set and the models used for comparison in this study. In Section~\ref{sec:res} we present the results and in Section~\ref{sec:dis} we discuss them together with possible caveats. Finally, in Section~\ref{sec:end} we draw our conclusions.

\section{Data}
\label{sec:data}

One of the main goals of GES is the observation of a large number of open clusters covering a wide range in age, mass, metallicity, and Galactocentric distance. This is achieved by combining observations with FLAMES--GIRAFFE and FLAMES--UVES at the Very Large Telescope (VLT, see, e.g., \citealt{dekker2000,pasquini2002}). Main-sequence and turn-off stars in old open clusters were normally observed with the GIRAFFE set-ups HR9B ($514.3-535.6$ nm, $R\sim32000$) and HR15N ($647-679$ nm, $R\sim20000$). They cover a wavelength range that allows for precise measurements of the radial velocity, in addition to the abundances of Fe, Cr, and Ti for HR9B, and of Ca, Ti, Si, Al, and Li for HR15N, although with low precision. High-precision abundances of a large number of chemical species were calculated from FLAMES--UVES spectra generally obtained for stars on the MS, RGB, and on the red clump (RC) with the set-up U580 (200 nm around the central wavelength 580 nm, $R\sim47000$).

M67 is included in the list of open clusters analysed within GES. The observations were collected from the ESO archive. In particular, they were part of the program 082.D--0726 (PI Gustafsson), which observed with FLAMES--UVES 25 stars in the field of M67 in all evolutionary stages between the main sequence and the lower red giant branch (see Table~\ref{tab:all}). Thus, M67 is one of the few clusters in GES  for which detailed chemical abundances from high-resolution spectroscopy are available for stars on the MS, TO, and RGB. 

Fourteen of the stars distributed on the MS, TO, and early SGB branch were analysed in \citet{onehag2014} who searched the data for possible diffusion effects. They found differences in abundance between stars at the upper MS and TO and stars on the SGB of the order of $\sim0.02$ dex (we note that they divide the sample that we define as TO in this work into TO and early SGB and compare the abundances of the first two with those of the third group). Although their results were not conclusive, also due to the small variation in abundances predicted by the models between these groups of stars, they seemed to support the presence of atomic diffusion. 

We used the recommended results of the GES fifth internal data release (GES iDR5) for all the 25  stars and performed a membership analysis based on their radial velocities. We first excluded all stars known as binaries from \citet{geller2015}. We then computed the mean radial velocity of the remaining 17 stars ($RV_{\mathrm{mean}}=34.54\pm0.83 \mathrm{\,km\, s}^{-1}$ ) and found that all stars lie within $3\sigma$ from this value. The mean error on the radial velocity ($0.36\mathrm{\,km\, s}^{-1}$) is smaller than the internal dispersion. This result is consistent with other mean radial velocity determinations of M67 known from the literature (see, e.g., \citealt{geller2015, yadav2008}) within few $\sigma$. In addition, we cross-matched the sample with the HSOY catalogue \citep{altmann2017} in order to check also the mean proper motion (PM) of the cluster ($\bar{\mu}_{RA}=-9.67\pm0.64\mathrm{\,mas\, yr}^{-1},  \bar{\mu}_{Dec}=-3.35\pm0.88\mathrm{\,mas\, yr}^{-1}$). We excluded one star that does not have an entry in the HSOY catalogue, and one because its PM is inconsistent with that of the cluster. Moreover, we verified that the 2MASS photometry \citep{skrutskie2006} of the remaining 15 stars is consistent with a 3.75 Gyr PARSEC isochrone (\citealt{bressan2012}) calculated with $(m-M)_0=9.64$ mag, $E(B-V)=0.023$ mag, and $\mathrm{[Fe/H]}=0.06$ dex (\citealt{bellini2010}; a list of the selected members and their parameters can be found in Table~\ref{tab:memb}). This leaves 11 stars in common between our sample and the sample from \citet{onehag2014}. We excluded 3 of their stars during our membership analysis (one as a binary and two because of PM criteria) and gained 4 additional ones along the SGB and lower RGB (see Table~\ref{tab:all}). In particular, the inclusion of three RGB stars into our analysis is of great importance for the study of diffusion effects. In fact, in absence of stars in their early main-sequence phase and thus still unaffected by diffusion, RGB stars are the only objects in our sample that show a surface chemical abundance similar to the initial one (except for C, N, and Li) and that can thus be compared to stars on the upper MS and TO, where  diffusion effects reach their peak. 

\begin{table}
	\caption{All stars in the field of M67 included in GES iDR5 from the observing program 082.D--0726 (PI Gustafsson). In the last column we include the results of our membership analysis: single members (SM), binary members (BM), and non-members (NM).}
	\begin{tabular}{|l|l|l|l|}
		\hline
		\multicolumn{1}{|c|}{ID} &
		\multicolumn{1}{c|}{RA (J2000)} &
		\multicolumn{1}{c|}{Dec (J2000)} &
	    \multicolumn{1}{c|}{Memb.} \\
		\hline
		08505182+1156559$^b$ & 08:50:51.82 & +11:56:55.9 & BM\\
		08505600+1153519$^a$ & 08:50:56.00 & +11:53:51.9 & SM\\
		08505891+1148192$^a$ & 08:50:58.91 & +11:48:19.2 & SM\\
		08510017+1154321 & 08:51:00.17 & +11:54:32.1 & SM\\
		08510080+1148527$^{a,d}$ & 08:51:00.80 & +11:48:52.7 & SM\\
		08510325+1145473$^a$ & 08:51:03.25 & +11:45:47.3 & SM\\
		08510524+1149340$^a$ & 08:51:05.24 & +11:49:34.0 & SM\\
		08510838+1147121 & 08:51:08.38 & +11:47:12.1 & SM\\
		08510969+1159096 & 08:51:09.69 & +11:59:09.6 & NM\\
		08511267+1150345$^c$ & 08:51:12.67 & +11:50:34.5 & NM\\
		08511799+1145541 & 08:51:17.99 & +11:45:54.1 & NM\\
		08511854+1149214$^a$ & 08:51:18.54 & +11:49:21.4 & SM\\
		08511868+1147026 & 08:51:18.68 & +11:47:02.6 & NM\\
		08511901+1150056 & 08:51:19.01 & +11:50:05.6 & NM\\
		08512012+1146417$^a$ & 08:51:20.12 & +11:46:41.7 & SM\\
		08512291+1148493 & 08:51:22.91 & +11:48:49.3 & NM\\
		08512940+1154139 & 08:51:29.40 & +11:54:13.9 & NM\\
		08513045+1148582 & 08:51:30.45 & +11:48:58.2 & NM\\
		08513322+1148513$^c$ & 08:51:33.22 & +11:48:51.3 & NM\\
		08513577+1153347 & 08:51:35.77 & +11:53:34.7 & SM\\
		08513740+1150052$^a$ & 08:51:37.40 & +11:50:05.2 & SM\\
		08514081+1149055$^a$ & 08:51:40.81 & +11:49:05.5 & SM\\
		08514122+1154290$^a$ & 08:51:41.22 & +11:54:29.0 & SM\\
		08514507+1147459 & 08:51:45.07 & +11:47:45.9 & SM\\
		08514995+1149311$^a$ & 08:51:49.95 & +11:49:31.1 & SM\\
		\hline\end{tabular}\\
$^a$: stars in common with the selection of  \citet{onehag2014} and ours.\\
$^b$: stars from the selection of \citet{onehag2014} excluded from our analysis since they are binaries.\\
$^c$: stars from the selection of \citet{onehag2014} excluded from our analysis because of PM criteria.\\
$^d$: the solar twin M67-1194 used to normalise the abundances of our sample stars.
	\label{tab:all}
\end{table}

For all the 15 stars in our sample the chemical analysis was performed by GES working group (WG) 11, responsible for the analysis of UVES F, G, K stars (see \citealt{smiljanic2014} for  a description of the abundances derivation and \citealt{sacco2014} for details about the reduction of UVES data).

In the GES iDR5 archive abundances are given in the form:
\begin{equation}
\mathrm{A(X)}=\log{\mathrm{(X/H)}}+12,
\end{equation}
and we transformed them for our purposes into 
\begin{equation}
\mathrm{[X/H]}=\log{\mathrm{(X/H)}}-\log{(\mathrm{X}_{1194}/\mathrm{H}_{1194})}
\end{equation}
subtracting the GES abundances of the solar twin M67-1194 (ID 08510080+1148527, $\mathrm{A(X)}_{1194}$) from $\mathrm{A(X)}$.

\section{Models}
\label{sec:models}

\begin{figure}
	\includegraphics[width=\columnwidth]{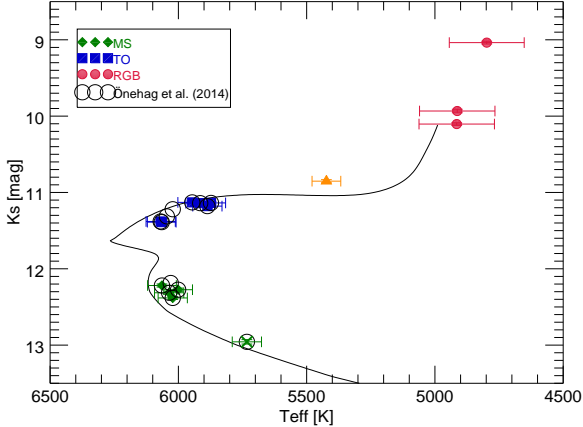}
	\caption{K$_\text{s}$ magnitude of the M67 members selected from the GES archive plotted using 2MASS photometry as a function of their $T_{eff}$ derived within GES. Main-sequence stars are plotted as green diamonds, turn-off stars as blue squares, and red giant stars as red circles. The star S806 is plotted as an orange triangle. The solar twin M67 1194 is plotted as a green cross. The stars selected in \citet{onehag2014} are plotted as black circles. The solid line represents the isochrone from \citet{michaud2004} for an age of 3.7 Gyr. }
	\label{fig:fig_iso}
\end{figure}

We compared the results with the stellar evolutionary models calculated in \citet{michaud2004} (with an extension of the calculation to a slightly older age for the 1.35 solar mass model) for stars of solar metallicity, ages of 3.7-4 Gyr, and masses ranging from 0.5 to $1.4 M_{\odot}$ . The calculations follow the description of \citet{turcotte1998} and \citet{richard2001}, including gravitational settling and radiative acceleration. Whereas all chemical species are affected by gravitational settling during the  MS, different elements experience a varying amount of radiative acceleration, depending on the fraction of the total photon flux they absorb. After the TO the convective envelope deepens into the stellar interior and the effects of atomic diffusion are subverted, thus restoring in most cases the original surface abundances. Fig.~\ref{fig:fig_iso} compares the K$_\text{s}$ magnitude (2MASS), corrected for reddening, of the stars in our sample as a function of the effective temperatures derived by GES with the theoretical isochrones from \citet{michaud2004} corresponding to an age of 3.7 Gyr. From the plot it is evident that the calculated effective temperatures are systematically slightly colder but nevertheless agree within the errors with the predicted ones.

The models from \citet{michaud2004} are divided into two sets of calculations. While the first one does not include turbulent transport, the second one is calculated including a density-dependent turbulent diffusion coefficient that is 400 times the He diffusion coefficient at $\log T = 6.09$, and that decreases inwards from there as $\rho^{-3}$ (hereafter we will refer to this calculation as T6.09). This parametrisation of turbulent transport was chosen by \citet{michaud2004} since it showed a minimisation of  the Li depletion in Pop II stars (see \citealt{richard2002}). The presence of turbulence reduces the effects of gravitational settling and as a consequence also the under-abundances at the TO, as can be seen in Fig.~\ref{fig:fig_fe}, where the models are plotted as isochrones, each data point representing a different mass (see Table~\ref{tab:mod} for a summary of the $T_{\mathrm{eff}}$ and $\log g$ corresponding to each mass). The turbulent mixing efficiency necessary to match the observed variations in surface abundances due to atomic diffusion has been shown to vary for different globular clusters, more metal-rich clusters needing more efficient mixing (see, e.g., \citealt{gruyters2016,gruyters2013,korn2007}). In the models, the outer atmosphere pressure is obtained using an 
approximate T$-\tau$ relationship   \citep[where T is the temperature and $\tau$ is the mean optical depth; for details see][] {krishna1966,michaud2004}

In interpreting abundance isochrones, for instance of M67, it is worth 
	remembering that for T$_{\text{eff}}\lesssim 5000$ ( or $\log g < 3.6$) the dredge up is essentially complete for those species that have not been influenced by nuclear reactions.  This implies, for instance, that the lowest $\log g$ points in Fig.~\ref{fig:fig_fe} ($\log g \sim 3.3$, dashed lines)  would have been at very nearly the same abundance  when $\log g = 3.6$ for that star.  The shape of the isochrones is affected by the finite (and small) number of models calculated and so by the small number of straight line segments available for the 	isochrones.  The convergence of abundances for the models with and without turbulence is shown on Fig.~\ref{fig:fig_end}.

\begin{table}
	\caption{$T_{\mathrm{eff}}$ and $\log g$ for the models presented in this work in the age range 3.7--4.0 Gyr.}
\begin{tabular}{|l|l|l|}
	\hline
	\multicolumn{1}{|c|}{M*/M$_{\odot\mathrm{NoTurb}}$} &
	\multicolumn{1}{c|}{$T_{\mathrm{effNoTurb}}$[K]} &
	\multicolumn{1}{c|}{$\log g_{\mathrm{NoTurb}}$} \\
	\hline
	0.5 & 3907.8 & 4.843\\
	0.6 & 4078.7 & 4.741\\
	0.7 & 4421.7 & 4.660\\
	0.9 & 5404.7 & 4.557\\
	1.0 & 5780.2 & 4.467\\
	1.06 & 5956.0 & 4.394\\
	1.07 & 5981.2 & 4.378\\
	1.075 & 5994.9 & 4.371\\
	1.08 & 6004.8 & 4.365\\
	1.081 & 6006.9 & 4.363\\
	1.082 & 6008.1 & 4.365\\
	1.083 & 6010.8 & 4.360\\
	1.085 & 6014.5 & 4.360\\
	1.09 & 6023.2 & 4.350\\
	1.15 & 6083.1 & 4.267\\
	1.2 & 6083.3 & 4.201\\
	1.25 & 6221.3 & 4.065\\
	1.3 & 5977.2 & 3.886\\
	1.35 & 4933.1 & 3.240\\
	\hline\end{tabular}

\begin{tabular}{|l|l|l|}
	\hline
	\multicolumn{1}{|c|}{M*/M$_{\odot\mathrm{Turb}}$} &
	\multicolumn{1}{c|}{$T_{\mathrm{effTurb}}$[K]} &
	\multicolumn{1}{c|}{$\log g_{\mathrm{Turb}}$} \\
	\hline
	0.5 & 3907.6 & 4.844\\
	0.6 & 4078.0 & 4.742\\
	0.7 & 4420.6 & 4.661\\
	0.8 & 4928.1 & 4.618\\
	0.9 & 5399.5 & 4.560\\
	1.0 & 5781.3 & 4.469\\
	1.05 & 5934.5 & 4.414\\
	1.1 & 6055.1 & 4.350\\
	1.15 & 6116.0 & 4.283\\
	1.2 & 6114.5 & 4.212\\
	1.25 & 6095.3 & 4.129\\
	1.3 & 6111.6 & 3.938\\
	1.35 & 5059.9 & 3.525\\
	\hline\end{tabular}
	\label{tab:mod}
\end{table}

\section{Results}
\label{sec:res} 

\begin{figure*}
	\includegraphics[width=0.69\columnwidth]{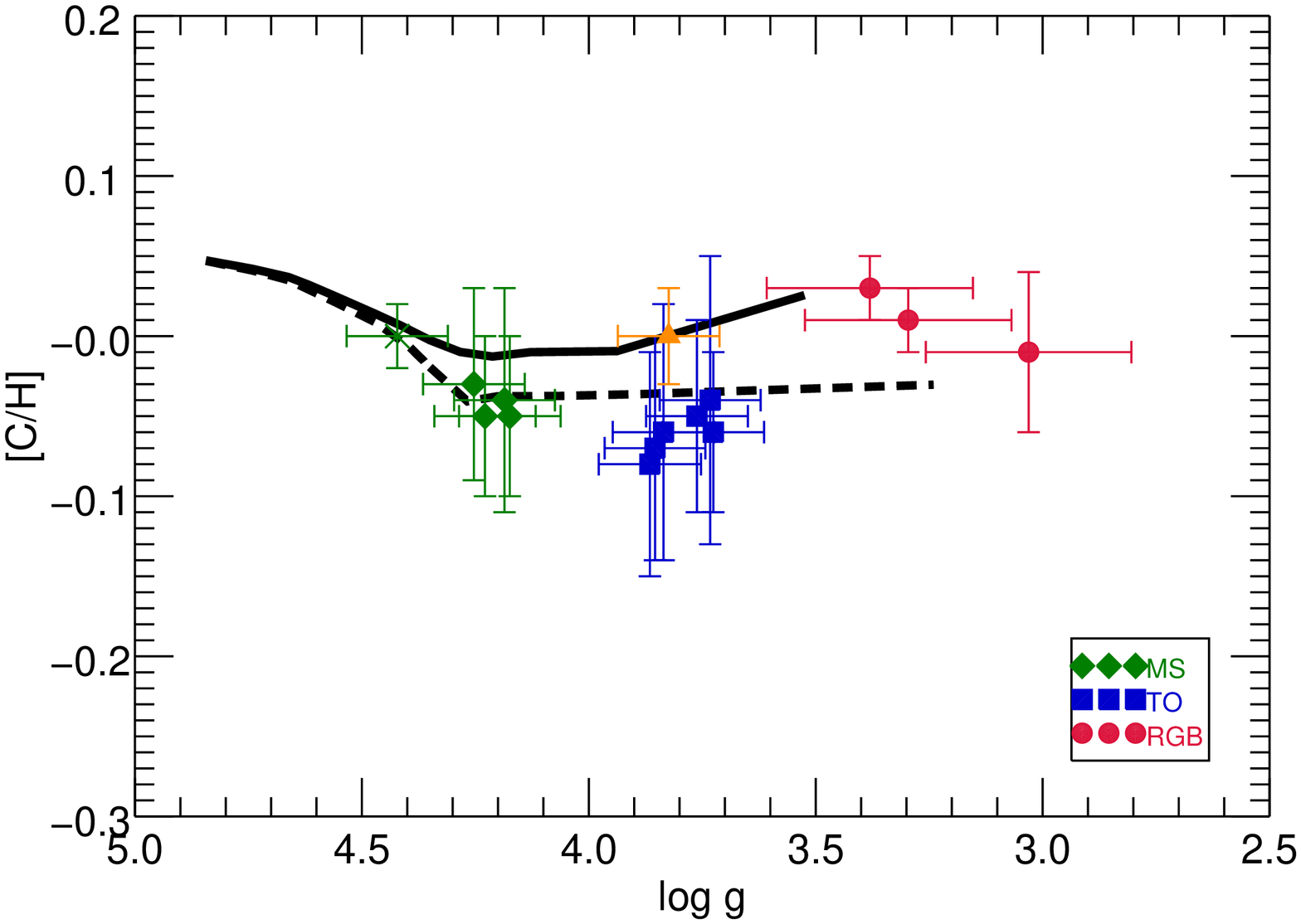}
	\includegraphics[width=0.69\columnwidth]{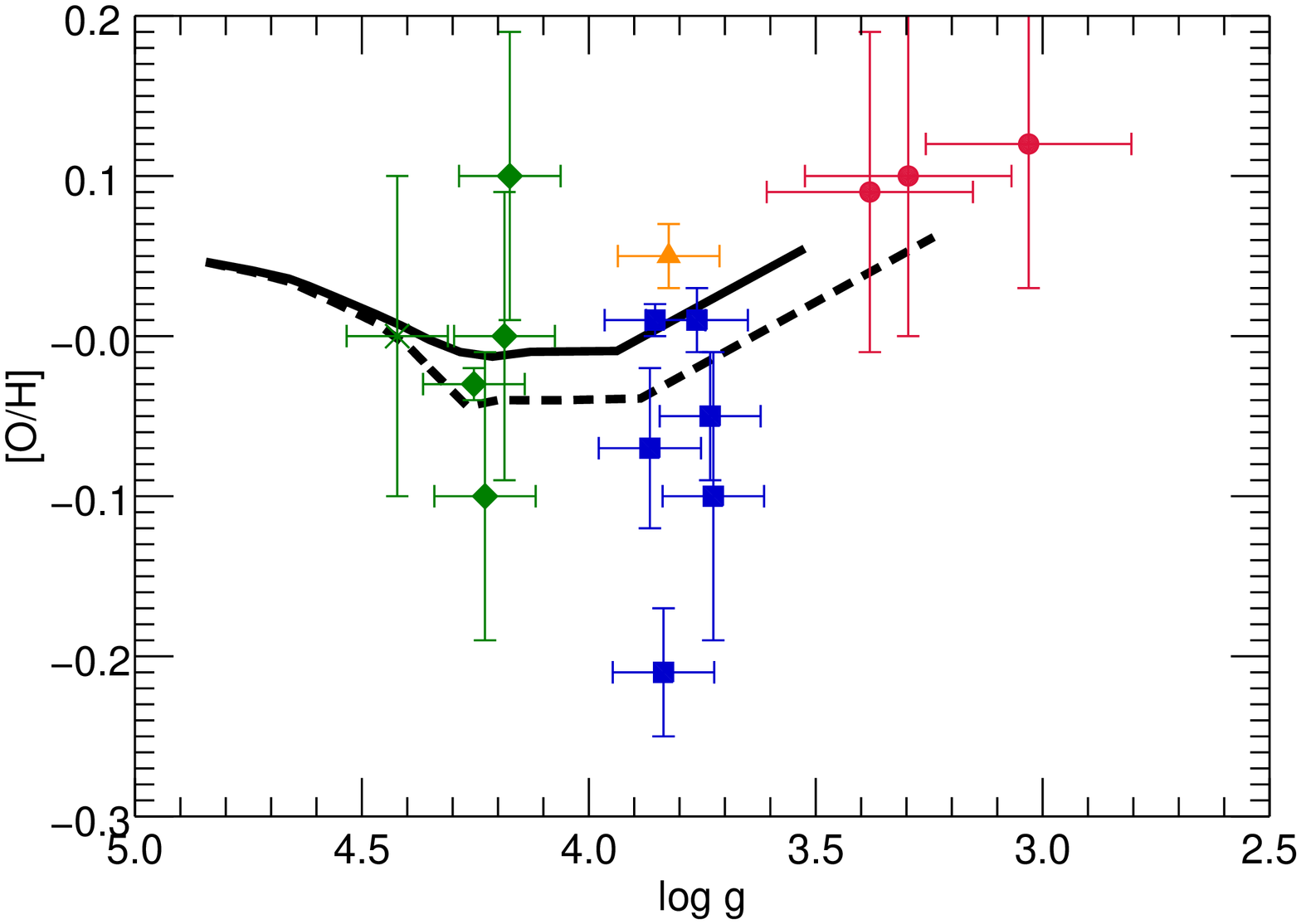}
	\includegraphics[width=0.69\columnwidth]{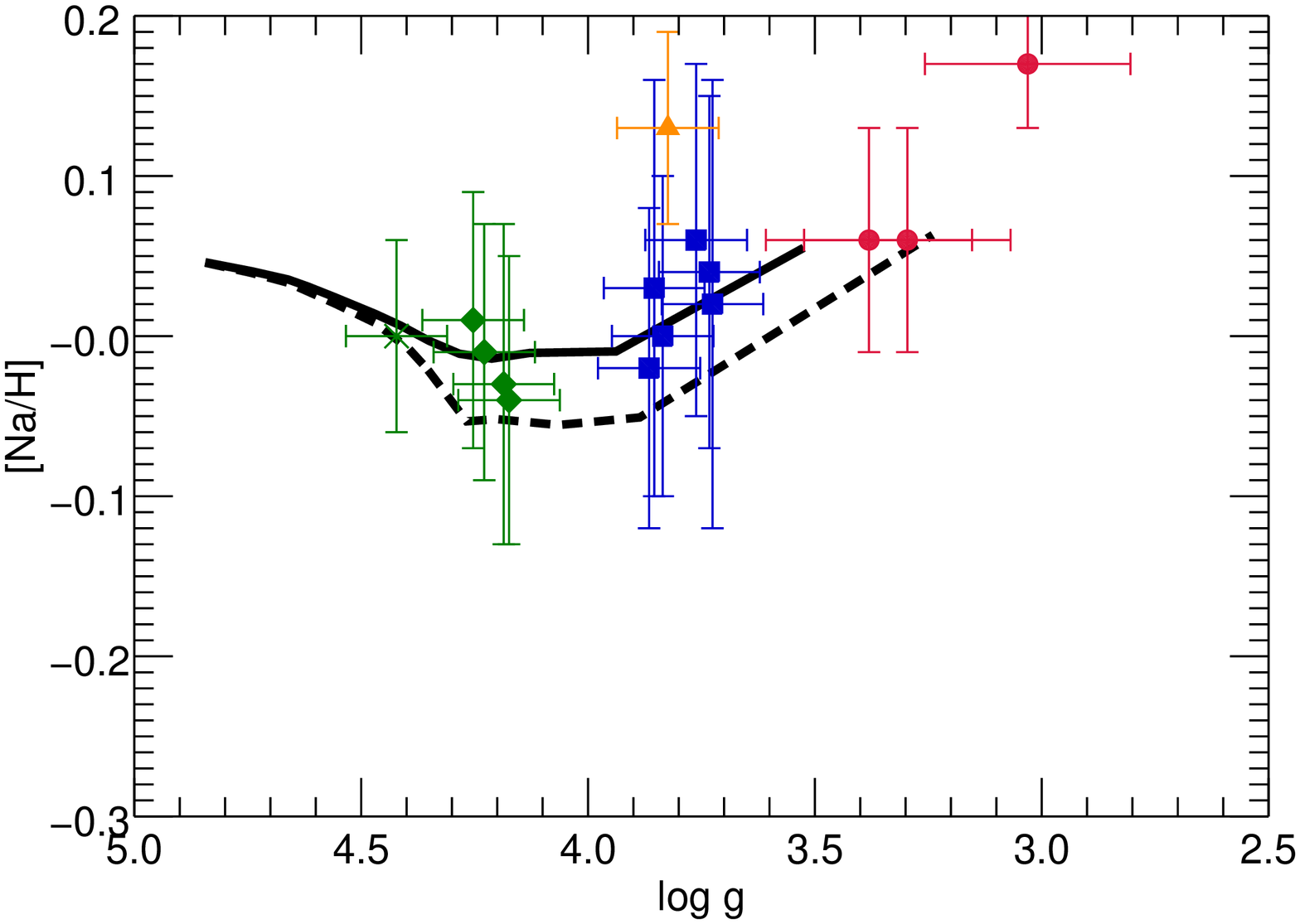}
	\includegraphics[width=0.69\columnwidth]{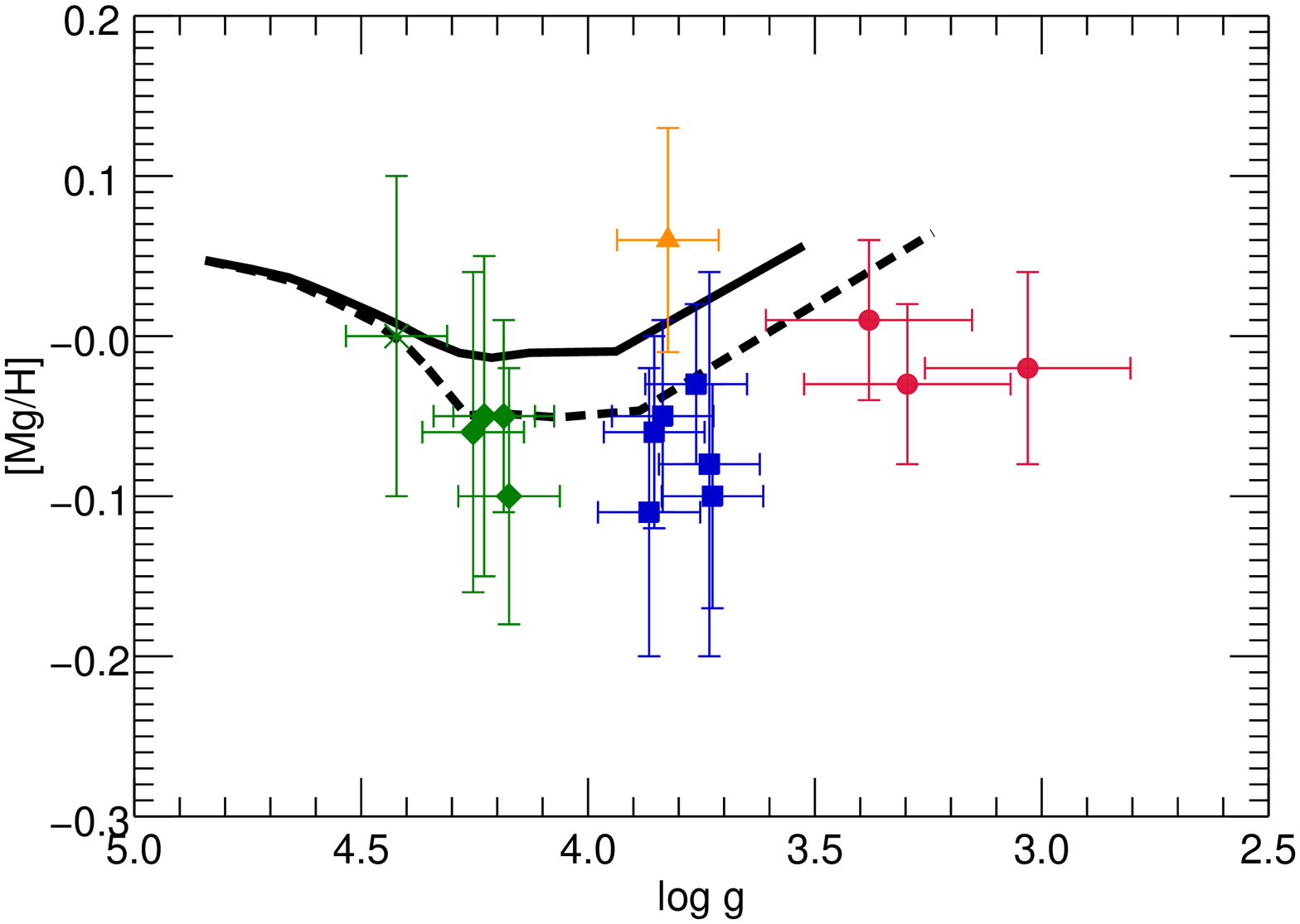}
	\includegraphics[width=0.69\columnwidth]{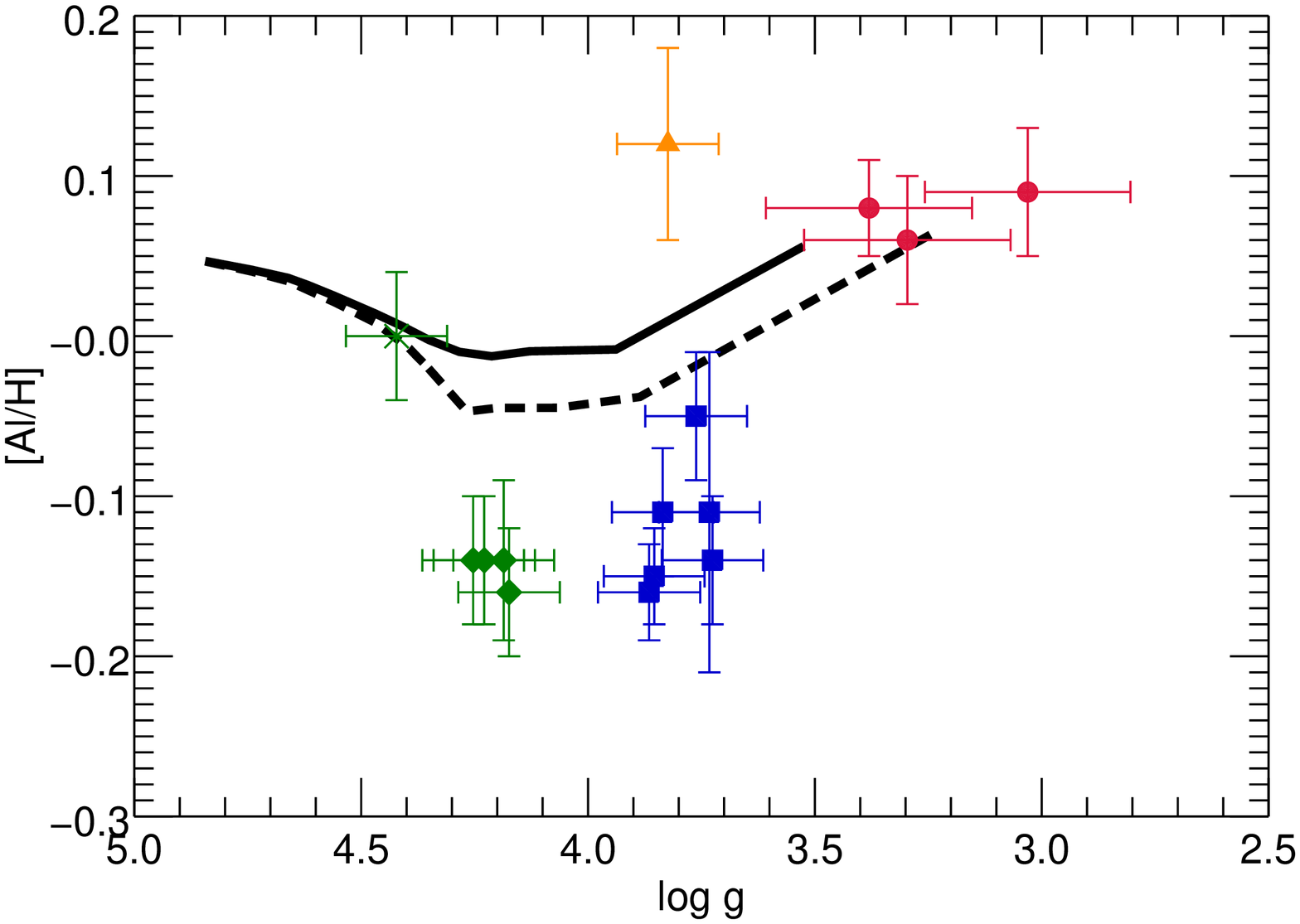}
	\includegraphics[width=0.69\columnwidth]{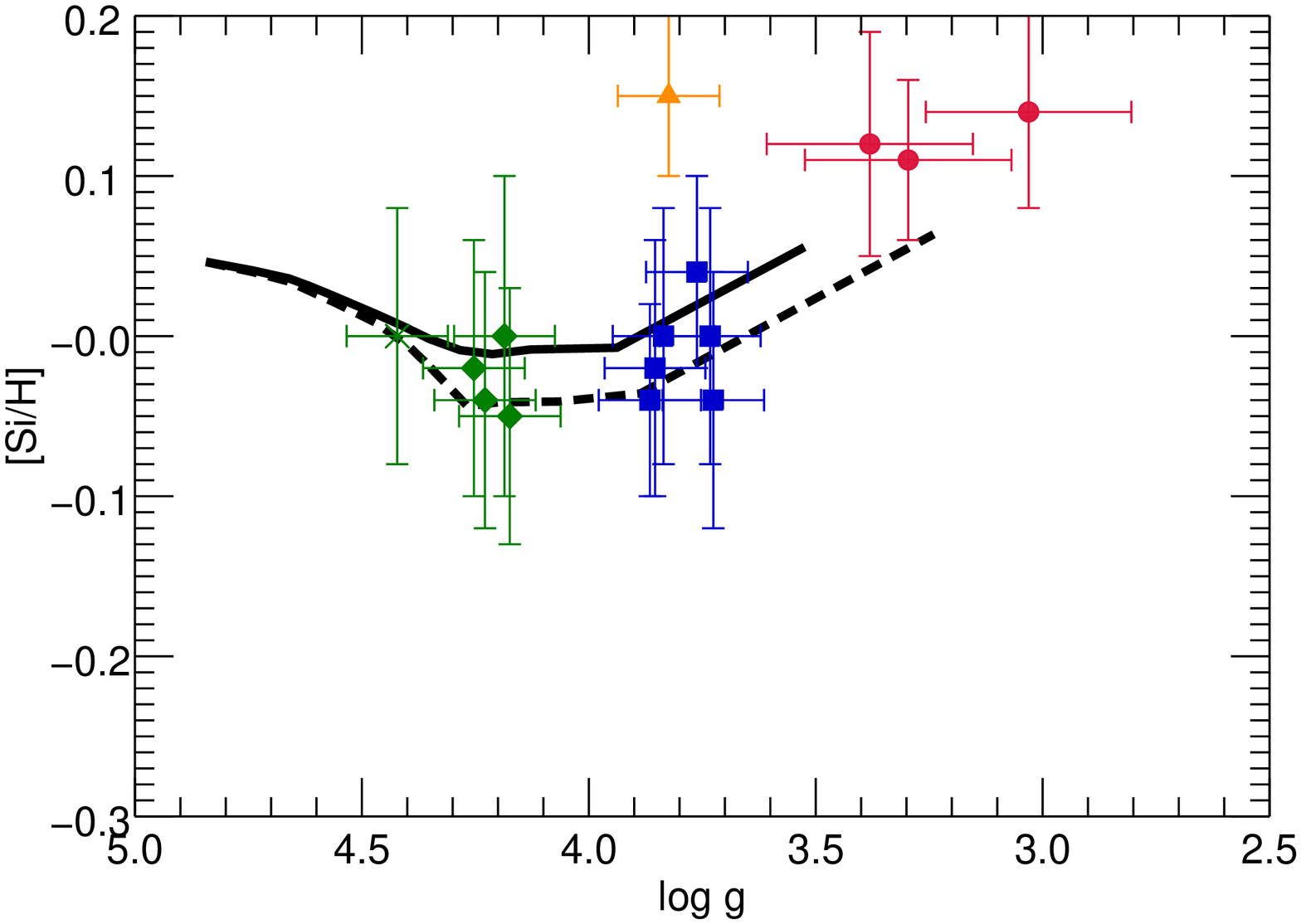}
	\includegraphics[width=0.69\columnwidth]{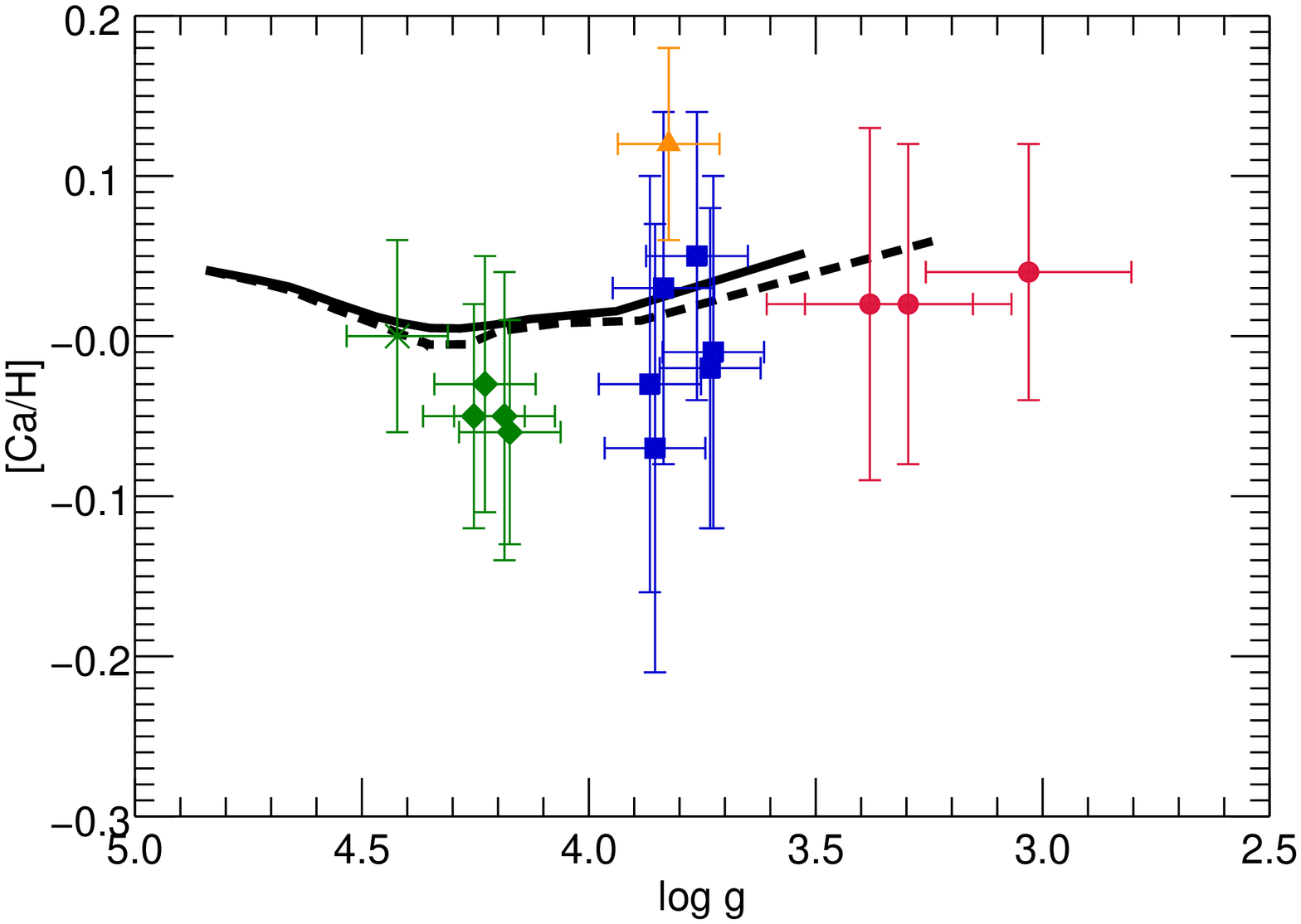}
	\includegraphics[width=0.69\columnwidth]{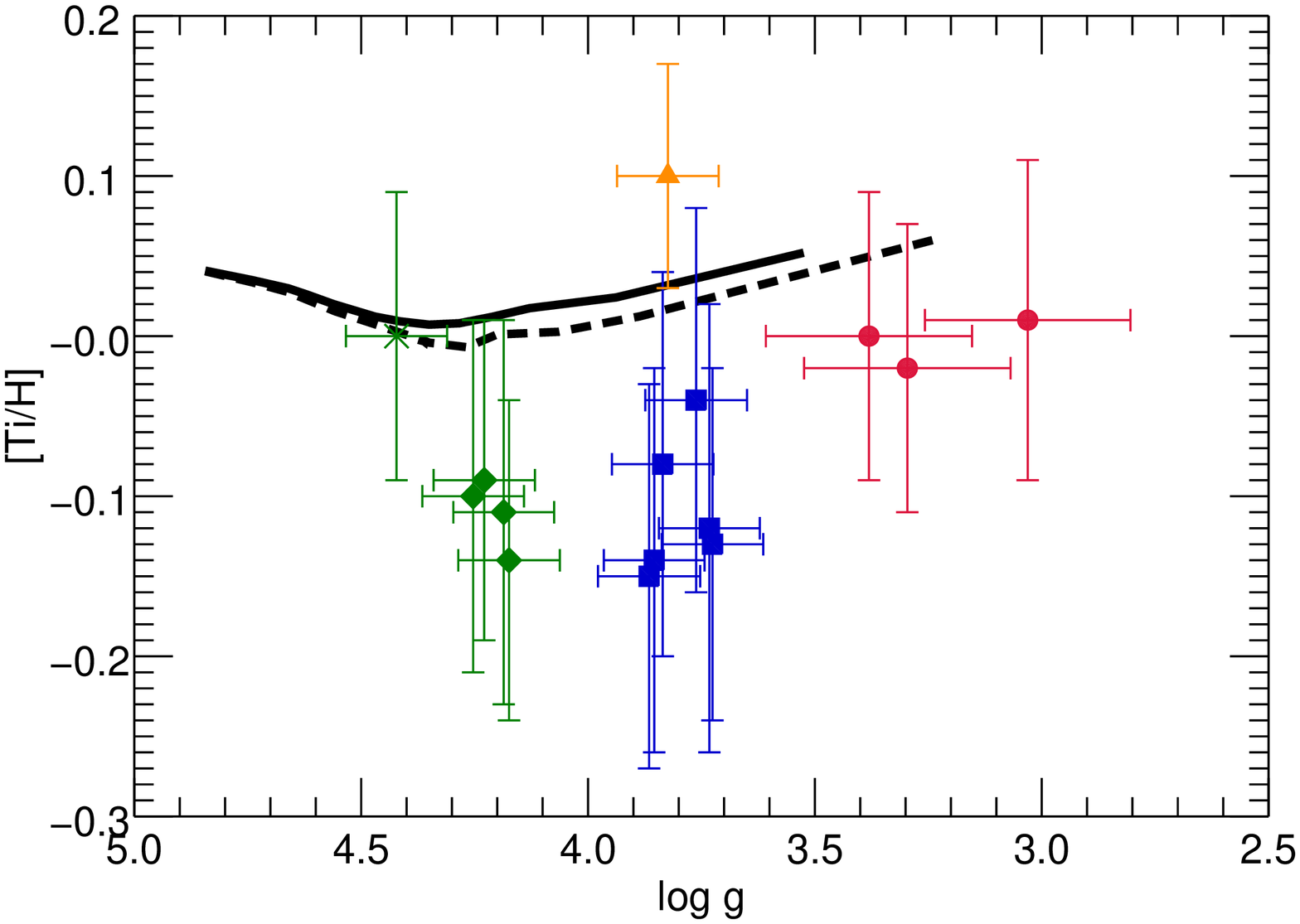}
	\includegraphics[width=0.69\columnwidth]{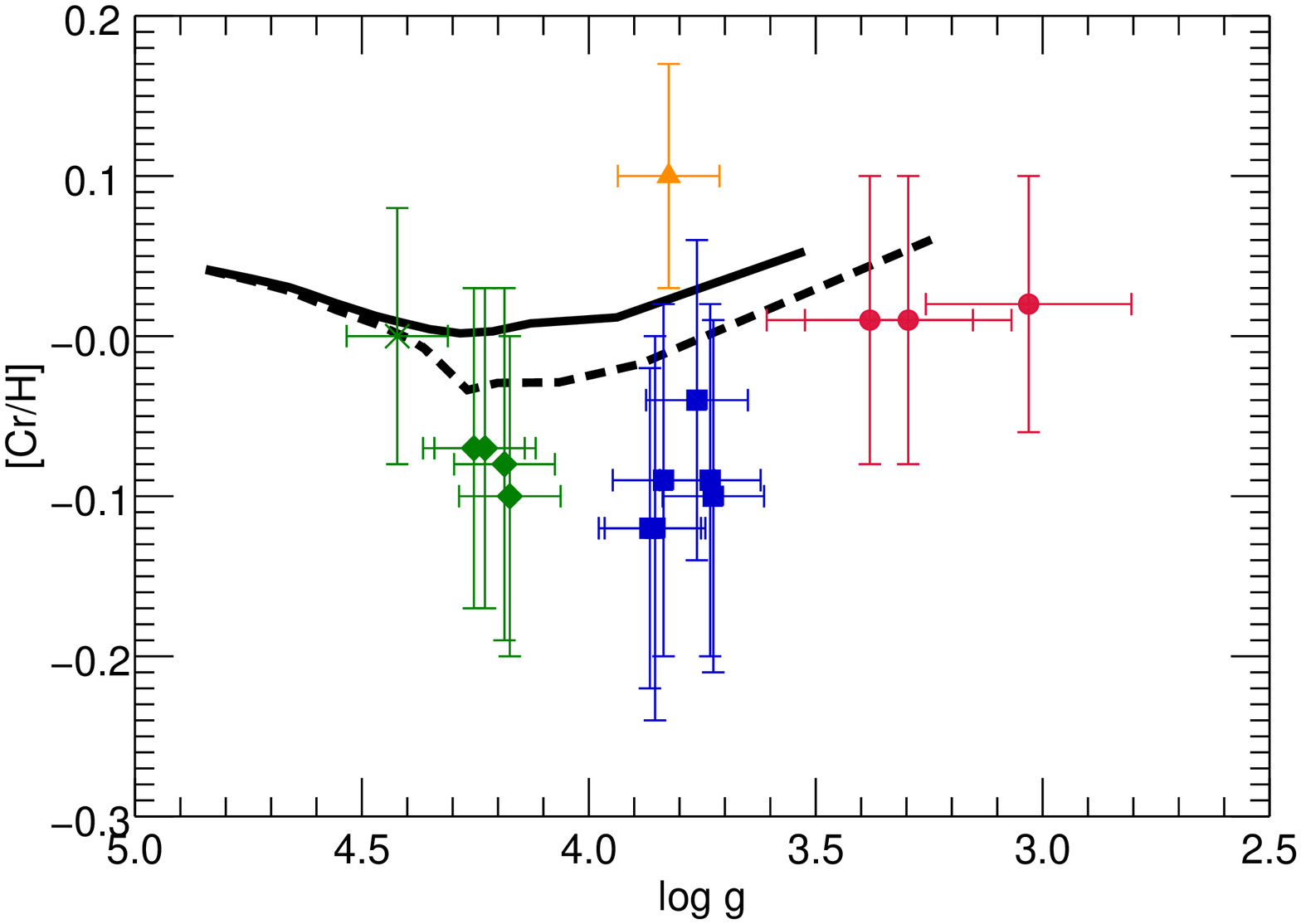}
	\includegraphics[width=0.69\columnwidth]{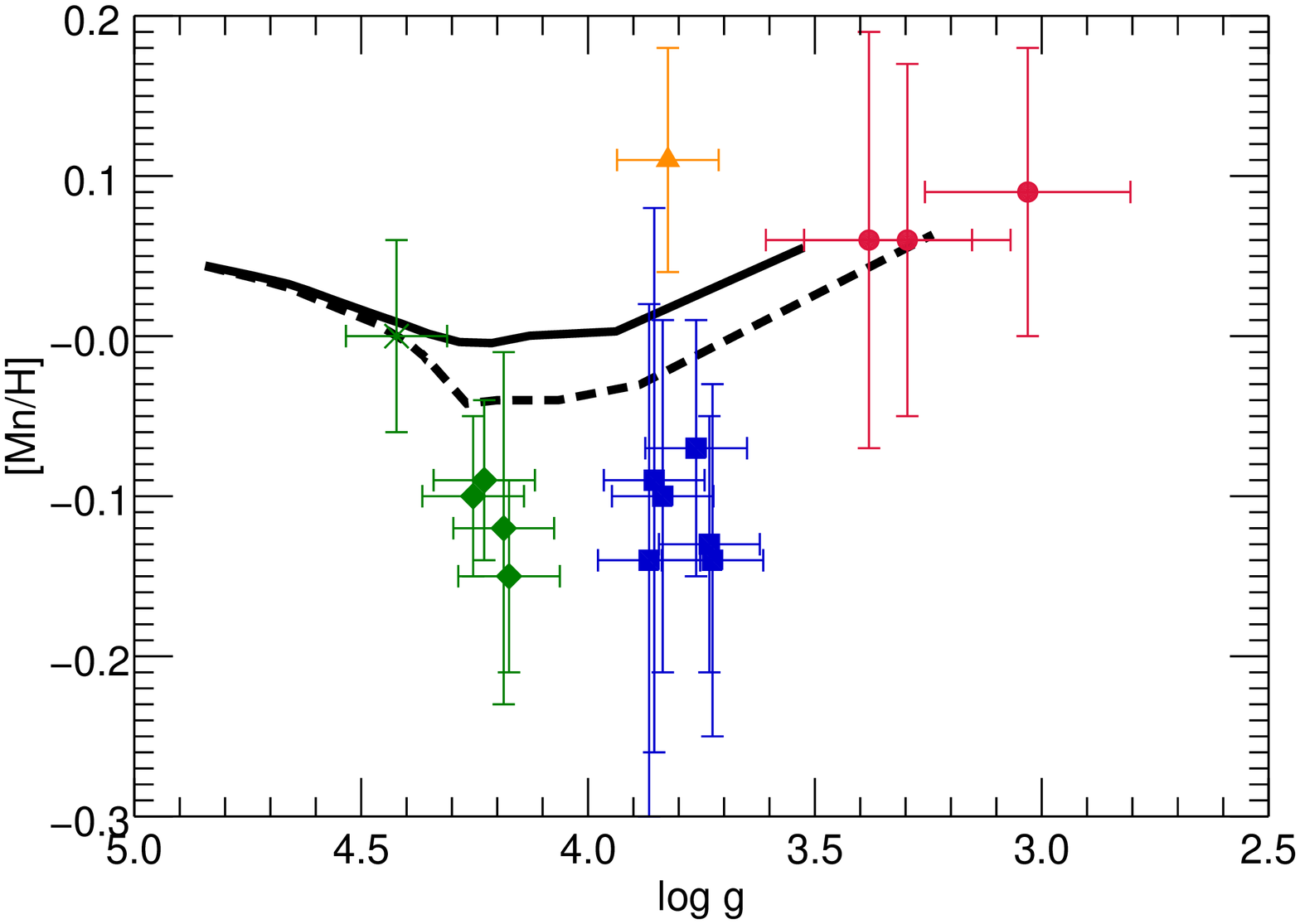}
	\includegraphics[width=0.69\columnwidth]{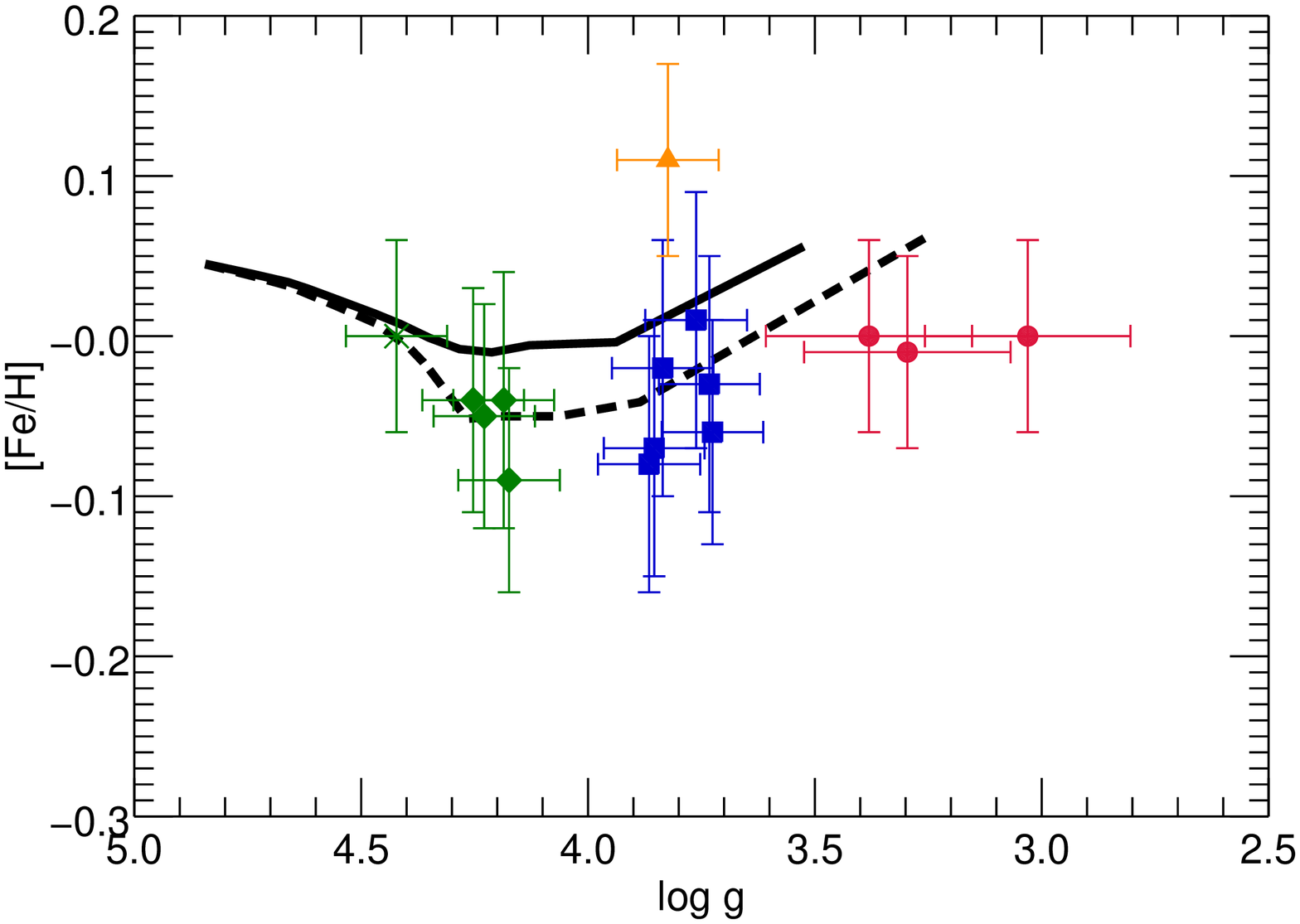}
	\includegraphics[width=0.69\columnwidth]{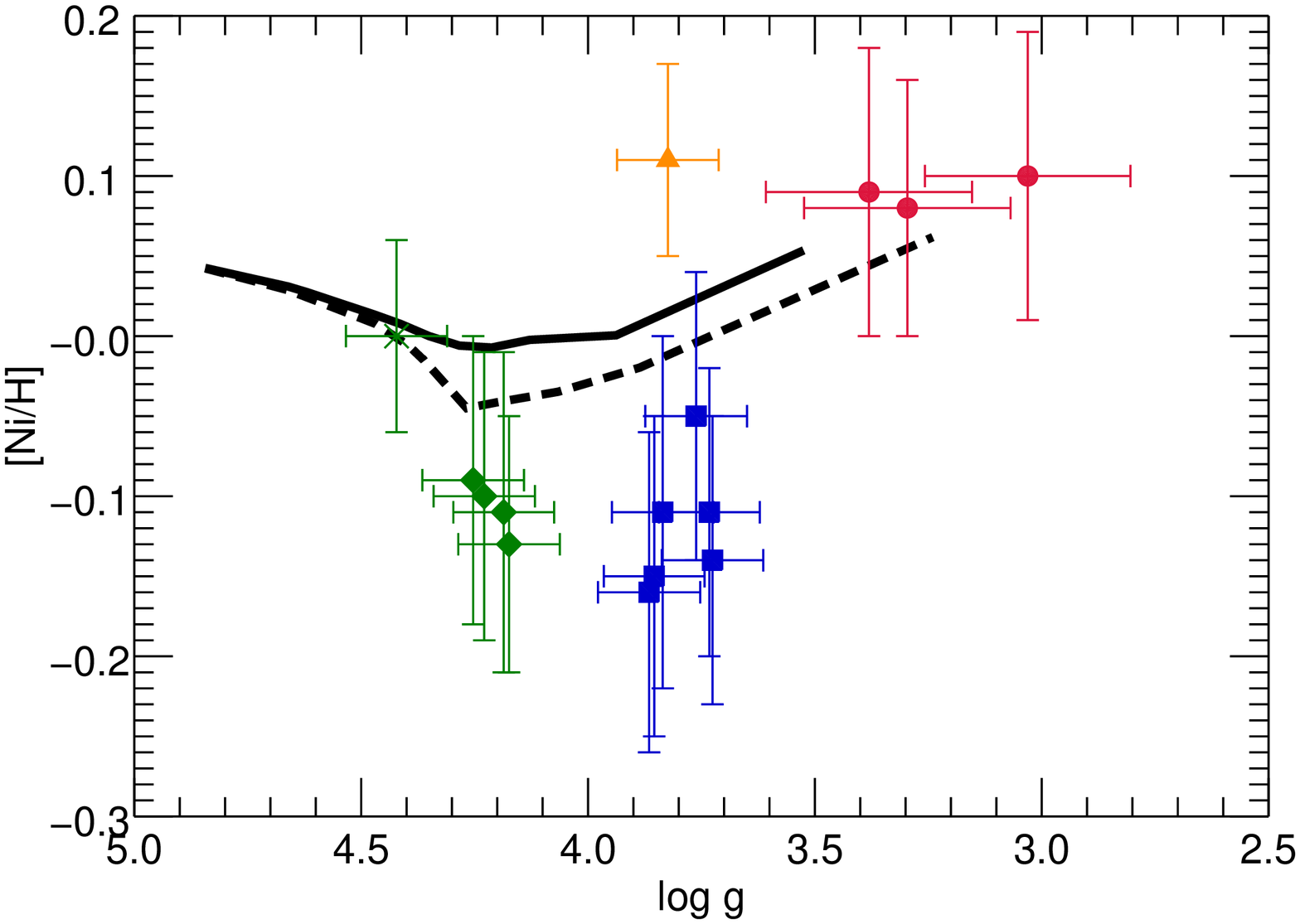}
	\caption{The different panels show from top to bottom and from left to right the abundances of the species analysed in this work as a function of $\log g$. The symbols are the same as in Fig.~\ref{fig:fig_iso}. We also overplotted models of stellar evolution with (black solid line) and without (dashed black line) turbulence in the stellar interior. Note that the lowest $\log g$ point of the dashed lines could be horizontally shifted to the left to the same $\log g$ as the solid lines.  The large shift is due to the rapid variation of $\log g$ with age for the 1.35 solar mass model at that evolutionary stage and the slight age difference between the models with and without turbulence.  }
	\label{fig:fig_fe}
\end{figure*}

In the following, we present the results of our study and compare the measured abundances from the GES data with the models of \citet{michaud2004}. In Fig.~\ref{fig:fig_fe} the elemental abundances of stars from the MS, SGB, and RGB are plotted as a function of $\log g$. The abundances shown and discussed in this work refer to the elements in their neutral state. We present and discuss in this work only elements for which we have both the measured abundances from GES iDR5 and the predictions from the models by \citet{michaud2004}. The star with GES-ID 08510017+1154321 is highlighted as an orange triangle for the reasons explained in subsection~\ref{sec:s806}. The solar twin M67-1194 is plotted as a green cross and is not included in the calculations when computing the average abundances of the MS stars. In fact, M67-1194 lies on the 'lower' MS with respect to the rest of the sample and presents for most elements an abundance more similar to the RGB than to the MS and TO samples, as predicted by the models of \citet{michaud2004} for stars of the same mass. Models of stellar evolution including turbulent diffusion are shown as a black, solid line, while models without turbulence are displayed as a dashed, black line. 

Since the effects of diffusion on the stellar surface chemical composition is very small (typically less than $\sim 0.1$ dex for stars in the mass range of M67 members) precise measurements are required. Indeed, if the scatter of the observationally derived abundances is larger than or even comparable to the expected variation, no sensible conclusion can be drawn. For this reason, we performed a statistical analysis of the abundances aimed at determining which of the elements derived within GES clearly show trends consistent with the effects of atomic diffusion predicted by the models of \citet{michaud2004}.

Table~\ref{tab:diff} summarizes the mean elemental abundances and relative standard deviations obtained within GES for two groups of stars in M67: the upper MS and the RGB stars. Table~\ref{tab:diff_mod} shows the difference in surface chemical abundances between these two groups for the species studied in this work, with errors calculated as 
\begin{equation}
\mathrm{err\_}\Delta[X/\mathrm{H}]=\sqrt{\sigma_{\mathrm{ms}}^2+\sigma_{\mathrm{g}}^2}
\end{equation}
where $\sigma_{\mathrm{ms}}$ and $\sigma_{\mathrm{g}}$ are the standard deviations of the abundance distributions of the upper MS and RGB stars, respectively; the last two columns show the maximum difference in surface abundance predicted by the models. We calculated this abundance difference between the models with mass $1.2M_{\odot}$ 
	along the upper MS, respectively with and without turbulence,  
	and the model with $1.35M_{\odot}$ on the RGB without turbulence. As
	already pointed out in Section~\ref{sec:models}, the models with and without turbulence converge to the same 
	abundances on the RGB (see Fig.~\ref{fig:fig_end}). Since the models with turbulence are evolved to a slightly younger age than those without turbulence, we compared the $1.2 M_{\odot}$ model with turbulence with the more evolved $1.35 M_{\odot}$ model without turbulence, assuming that the corresponding model with turbulence would show the same abundance at the same age. TO stars were excluded from this discussion because the maximum abundance variation predicted by the models is reached at the $\log g$ corresponding to  the upper MS ($\sim1.2M_{\odot}$ in the models).

For Al, Si, Ca, Ti, Cr, Mn, and Ni,  $\Delta[\mathrm{X}/\mathrm{H}]$ is larger than $3\times\mathrm{err\_}\Delta[X/\mathrm{H}]$, while for Fe $\Delta[\mathrm{X}/\mathrm{H}]$ is larger than $2\times\mathrm{err\_}\Delta[X/\mathrm{H}]$. We conclude that for these elements the presence of a trend in $\log g$ is clear and the offset in surface abundance between upper main sequence and red giant branch can be considered statistically significant. $\Delta[\mathrm{X}/\mathrm{H}]$ is larger than $1\times\mathrm{err\_}\Delta[\mathrm{X}/\mathrm{H}]$ for C, O, Na, and Mg making the offset between the two samples less clear, but still visible.  [S/H] is available only for stars on the upper MS and on the TO. We will thus exclude S from the following discussion.

For the elements of the oxygen group (O, Na, Mg, Al, Si, and S), the models predict the largest difference between initial surface abundance and values at the turn-off. This is due to the small radiative acceleration that these elements experience below the surface convection zone (for details, see \citealt{michaud2004}, Fig. 3). 
As shown in Fig.~\ref{fig:fig_fe}, the models are consistent with GES [Na/H], [Mg/H], and [Si/H] abundances of the M67 stars within $1-3\times\mathrm{err\_}\Delta[X/\mathrm{H}]$. For Al the difference between $\Delta[\mathrm{X}/\mathrm{H}]$ and the offset predicted by the models is larger than $3\times\mathrm{err\_}\Delta[X/\mathrm{H}]$. As explained above, the measured [O/H] abundances present a large scatter, even if  $\Delta[\mathrm{X}/\mathrm{H}]$ is consistent with the predictions of the models. This is due to the weakness of the forbidden [OI] line at 630.03 nm used to derive the O abundance and to the fact that the telluric lines in that region are not corrected for by the pipeline.

Species between P and Ti are expected to experience a larger radiative acceleration as compared to lighter elements (see Fig. 3 in \citealt{michaud2004}). This reduces the effect of gravitational settling, resulting in a much smaller difference between the surface abundances of TO and RGB stars as compared to Al or O. 
Fig.~\ref{fig:fig_fe} shows how the [Ca/H] and [Ti/H] both suffer from an offset with respect to the model abundances, which is more prominent for [Ti/H] than for [Ca/H]. Nevertheless, looking at Table~\ref{tab:diff} and ~\ref{tab:diff_mod} one can see that a trend in $\log g$ is present for Ca and Ti and is slightly more pronounced than predicted by the models, but still consistent within $3\times\mathrm{err\_}\Delta[X/\mathrm{H}]$.

Elements of the iron group (Cr, Mn, Fe, Ni) are predicted to experience less radiative acceleration as compared to Ca or Ti, but more than the elements in the oxygen group, resulting in very similar but slightly smaller under-abundances at the TO. Cr is systematically under-abundant compared to the theoretical predictions. The trend between main sequence and giant branch is consistent with the models in the case of Cr and slightly smaller for Fe. A similar trend between the TO and the SGB/low-RGB was also found for [Fe/H] by \citet{casey2016}, analysing lower resolution spectra. For Mn and Ni $\Delta[\mathrm{X}/\mathrm{H}]$ differs by $3\times\mathrm{err\_}\Delta[X/\mathrm{H}]$ or more from the abundance differences predicted by the models. 

 We note that, since we are interested in a relative comparison between two groups of stars analysed within GES iDR5 rather than in the absolute abundances of the stars, $\mathrm{err\_}\Delta[X/\mathrm{H}]$ is calculated only from the internal scatter of the measured abundances and does not take into account the error of the measurements.

\begin{table}
	\caption{Mean surface chemical abundances and standard deviations for stars on the main sequence ($[\mathrm{X}/\mathrm{H}]_{\mathrm{MS}}$) and giant stars ($[\mathrm{X}/\mathrm{H}]_{\mathrm{RGB}}$) in M67.}
	\begin{tabular}{|l|r|r|r|r|}
		\hline
		\multicolumn{1}{c|}{X} &
		\multicolumn{1}{c|}{$[\mathrm{X}/\mathrm{H}]_{\mathrm{MS}}$} &
		\multicolumn{1}{c|}{$\sigma_{\mathrm{MS}}$} &
		\multicolumn{1}{c|}{$[\mathrm{X}/\mathrm{H}]_{\mathrm{RGB}}$} &
		\multicolumn{1}{c|}{$\sigma_{\mathrm{RGB}}$} \\
		\hline
		C & -0.043& 0.010 & 0.010 & 0.020\\
		O & -0.008 & 0.083 & 0.103 & 0.015 \\
		Na & -0.018 & 0.022 & 0.097 & 0.064 \\
		Mg &  -0.065 & 0.024 & -0.013 & 0.021 \\
		Al & -0.145 & 0.010 & 0.077 & 0.015 \\
		Si & -0.028 & 0.022 & 0.123 & 0.015\\
		Ca & -0.048 & 0.013 & 0.027 & 0.012 \\
		Ti & -0.110 & 0.022 & -0.003 & 0.015 \\
		Cr & -0.080 & 0.014 & 0.013 & 0.006\\
		Mn & -0.115 & 0.027 & 0.070 & 0.017 \\
		Fe & -0.055 & 0.024 & -0.003 & 0.006 \\
		Ni & -0.108 & 0.017 & 0.090 & 0.010 \\
		\hline
		
	\end{tabular}
	\label{tab:diff}
\end{table}

\begin{table}
	\caption{Difference between the mean abundances (and errors) of selected elements on the upper-main
			sequence and on the red giant branch (column two and three), compared to the theoretical counterpart 
			calculated employing for the upper main-sequence a 
			$1.2M_{\odot}$ model, both with (column 5) and without (column 4) turbulence,
			and the $1.35M_{\odot}$ model without turbulence for the red giants (see text for details).}
	\begin{tabular}{|l|r|r|r|r|}
		\hline
		\multicolumn{1}{c|}{X} &
		\multicolumn{1}{c|}{$\Delta[\mathrm{X}/\mathrm{H}]$} &
		\multicolumn{1}{c|}{$\mathrm{err\_}\Delta[\mathrm{X}/\mathrm{H}]$} &
		\multicolumn{1}{c|}{$\Delta[\mathrm{X}/\mathrm{H}]_\mathrm{NoTurb}$} &
		\multicolumn{1}{c|}{$\Delta[\mathrm{X}/\mathrm{H}]_\mathrm{Turb}$} \\
		\hline
		C &  0.053 & 0.022 & 0.007 & -0.018\\
		O &  0.111 & 0.084 & 0.102 & 0.075\\
		Na &  0.114 & 0.067 & 0.115 & 0.077\\
		Mg & 0.052 & 0.032 & 0.113 & 0.078\\
		Al & 0.222 & 0.018 & 0.109 & 0.077\\
		Si &  0.151 & 0.027 & 0.105 & 0.075\\
		Ca & 0.074 & 0.017 & 0.057 & 0.053\\
		Ti & 0.107 & 0.027 & 0.059 & 0.048\\
		Cr &  0.093 & 0.015 & 0.090 & 0.058\\
		Mn & 0.185 & 0.032 & 0.104 & 0.068\\
		Fe & 0.052 & 0.025 & 0.114 & 0.074\\
		Ni & 0.198 & 0.020 & 0.103 & 0.069\\
		\hline
		
	\end{tabular}
	\label{tab:diff_mod}
\end{table}

\section{Discussion}
\label{sec:dis} 
In the following we discuss the interpretation of the results presented so far as well as possible caveats, such as corrections for NLTE effects, abundance trends in a cluster for which we do not expect to see diffusion effects, and trends of the abundances of M67 with temperature derived from APOGEE infrared spectra.

\subsection{S806}
\label{sec:s806}

Figure~\ref{fig:fig_fe} shows that one star (GES-ID: 08510017+1154321, S806 in the notation from \citealt{sanders1977}) in M67 presents for almost all elements investigated in this work higher abundances with respect to its companions. The star matches all criteria (kinematic and photometric) considered in the membership analysis of the cluster, but its chemical composition is not consistent with that of other cluster members with the same gravity. 

Of course the possibility exists that S806 is not a member of M67, although this is not very  likely given the perfect match with the clusters kinematics. Nevertheless, the surface abundances measured in S806 cannot be explained by simple stellar evolutionary processes. We therefore need to explain its position on the sub-giant branch and its peculiar chemical composition.

A possible explanation for the peculiar surface abundances of S806 (with respect to its $\log g$) is that the star might be part of a binary system initially composed of one massive companion, and a secondary companion with a mass of approximately $1.3 M_{\odot}$ now lying on the SGB. If the massive companion evolved into an AGB star and expelled part of its envelope, the surface of the secondary companion would have been polluted by the ejected material (the s-element abundances for S806 are also higher than in the other cluster members). At an age of $\sim4$ Gyr, the massive companion would have evolved into a white dwarf and thus would not contribute significantly to the luminosity of the binary system. In this scenario the binary would have to be face-on, since S806 has been classified as a single star in \citet{geller2015}. Alternatively, the system could have been disrupted by encounters with other stars within the cluster. In this case we would only be observing the low-mass companion.

\subsection{NLTE effects}
\label{sec:nlte}

The abundances of GES iDR5 are calculated for 1D-atmospheres and LTE. It is therefore interesting to discuss how corrections for NLTE would affect the abundances of the elements under study and if these alone can account for the abundance trends observed. 

The upper main sequence stars of M67 have a mean $T_{\mathrm{eff}}=6031$ K and $\log g=4.21$ dex. For the giant stars we find a mean $T_{\mathrm{eff}}=4875$ K and $\log g=3.24$. In \citet{lind2011} this corresponds to corrections for the Na abundance of  $\sim-0.15$ dex for the 568.2 nm line and of $\sim-0.1$ dex for the 615.4 nm line for both dwarfs and giants (assuming [Fe/H]=0 dex and [Na/Fe]=0 dex), thus leaving the offset between the two groups unchanged. 

For Al, \citet{nordlander2017} predict a correction of $\sim0.0$ dex for the MS stars and of $\sim-0.1$ dex for the RGB stars for the line at $669.6$ nm and [Fe/H]=0 dex, [Al/Fe]=0 dex. This would indeed diminish the offset observed between MS and RGB in the GES Al abundances, but a clear trend in $\log g$ would still be visible.

\citet{zhang2017} predict for a star with  $T_{\mathrm{eff}}=6070$ K and $\log g=4.08$ a Mg correction in the optical of $+0.01$ dex. This corresponds to our MS group. In the sample of stars analysed by these authors we could not find any star with the temperature and gravity of the M67 giants. The most similar one has $T_{\mathrm{eff}}=4901$ K and $\log g=2.76$ and corresponds to a correction of $-0.02$ dex in the optical. Thus, in the case of magnesium the NLTE correction would not significantly change the difference in abundance between MS and RGB.

Similarly, for Si the corrections calculated by \citet{zhang2016} in the optical are $-0.01$ dex and $-0.04$ dex for the MS and the RGB, respectively (using the same objects as for Mg).

Going to heavier elements, NLTE corrections are available only for Fe from \citet{lind2012}. The authors calculate corrections of $\sim +0.02$ dex and  $\sim +0.01$ dex for the MS and RGB, respectively, at [Fe/H]=0 dex. These corrections would not change significantly the overall difference in abundance between the two groups.

\subsection{NGC 6633}
\label{sec:ngc6633}

In the previous section we have discussed that the abundance trends visible in the stars of M67 cannot be explained by NLTE effects, which remain unaccounted for in the GES iDR5 analysis. In order to be reasonably confident that the trends we observe are indeed due to diffusion effects, we need to rule out systematic offsets deriving from the analysis of stars in different evolutionary stages. We therefore choose a second cluster, NGC 6633, for which stars both on the main sequence and on the RGB/RC have been observed and analysed in GES iDR5. This cluster is much younger than M67, with an age of $\sim 540$ Myr \citep{randich2017} and has a slightly sub-solar metallicity of $\mathrm{[Fe/H]}= -0.05\pm0.06 $ dex \citep{jacobson2016}. 

\begin{figure}
	\includegraphics[width=\columnwidth]{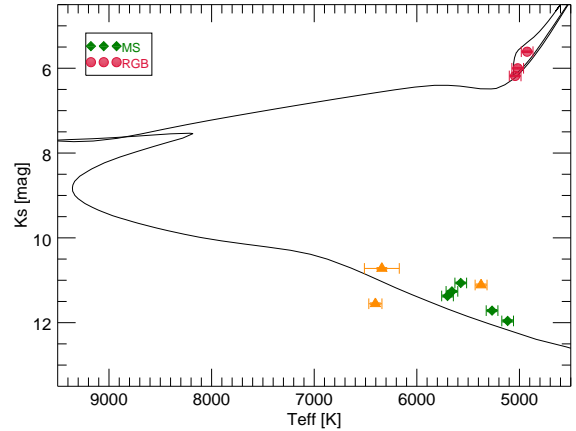}
	\caption{K$_\text{s}$ magnitude of the NGC6633 members selected from the GES archive plotted using 2MASS photometry as a function of their $T_{eff}$ derived within GES. Green circles are stars on the main sequence, red circles represent giant stars and orange triangle stars with flags that we do not take into account in our analysis. The solid line is a PARSEC isochrone for an age of 540 Myr.}
	\label{fig:fig_iso_ngc6633}
\end{figure}

After performing a membership analysis of the stars observed with UVES in the field of NGC 6633 based on radial velocities and HSOY proper motions, we found 5 main sequence stars and 3 stars that could be red giants or red clump stars based on their position in the CMD. Their Ks magnitude corrected for reddening is plotted as a function of effective temperature (green and red dots, respectively for MS and RGB stars) in Fig.~\ref{fig:fig_iso_ngc6633}, together with a Padova isochrone with age 540 Myr, $\mathrm{[Fe/H]}=-0.05$ dex and distance modulus $\mathrm{(m-M)}_0=8.07$ mag \citep{randich2017}. We find also three further stars on the main sequence that are probable members of the cluster based on their kinematic properties (orange triangles in Fig.~\ref{fig:fig_iso_ngc6633}), but that we do not take into consideration for our analysis due to the following reasons. One of them, 18265591+0635559, is flagged as a binary in GES iDR5. The $\log g$ of 18264026+0637500 is on the grid edge and the star is therefore flagged for "suspicious stellar parameters".  18272787+0620520 has abundances and errors in the abundances that are very different from the cluster distribution for most elements. Since it lies on the edge of the proper motion distribution of our sample, it is probably not a member of the cluster.

Due to its young age, we do not expect to observe diffusion effects in the observed stars of NGC 6633, since there was not enough time for diffusion to become efficient in those MS stars. Besides, we do not have any stars at the TO where we could in principle observe diffusion effects, if any. Stars on the main sequence and on the giant branch are expected to present the same abundances. \textit{If they do not it would mean that there are systematic offsets between the abundances of dwarf and giant stars that are due to the analysis alone.}

\begin{figure*}
	\includegraphics[width=0.69\columnwidth]{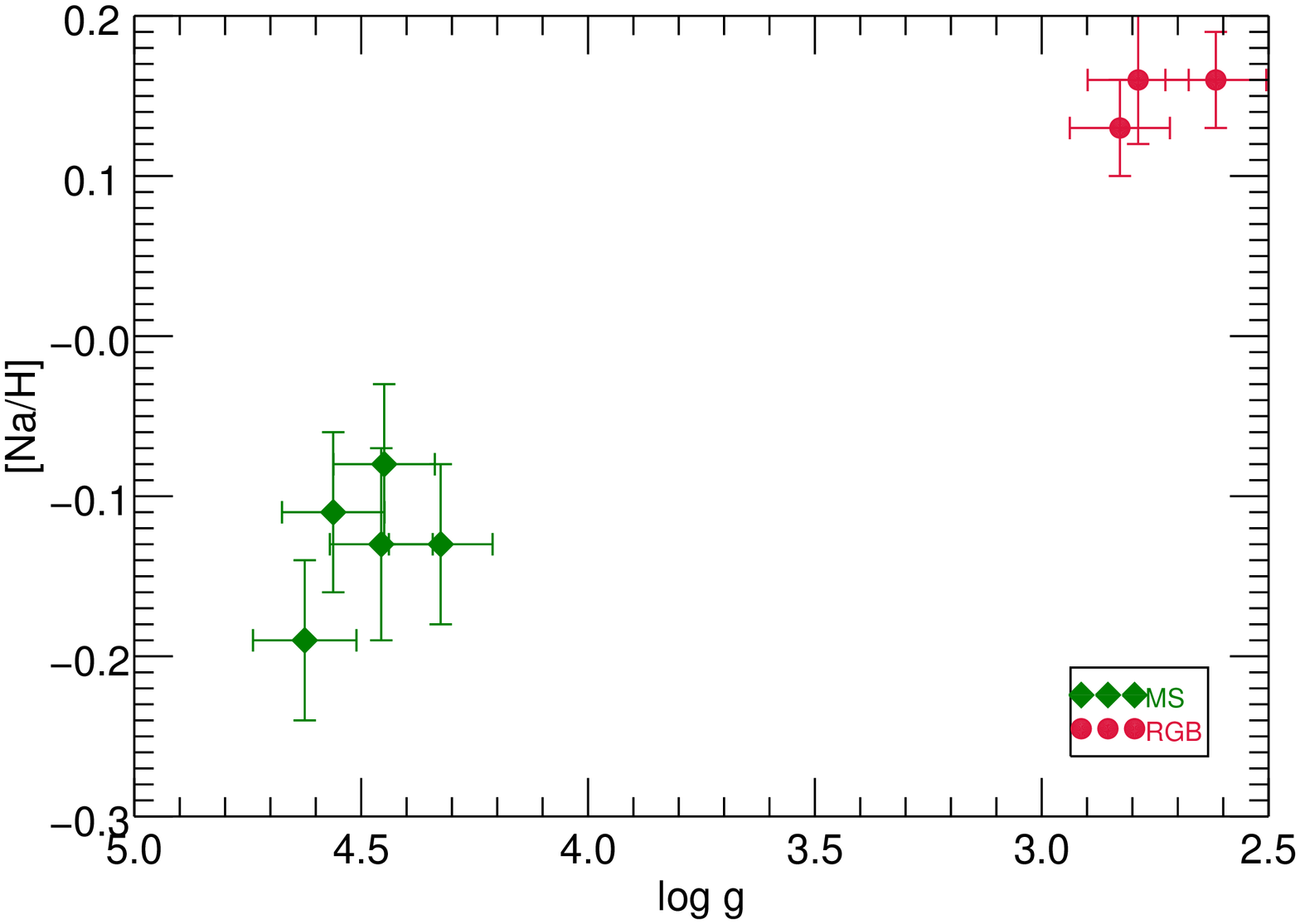}
	\includegraphics[width=0.69\columnwidth]{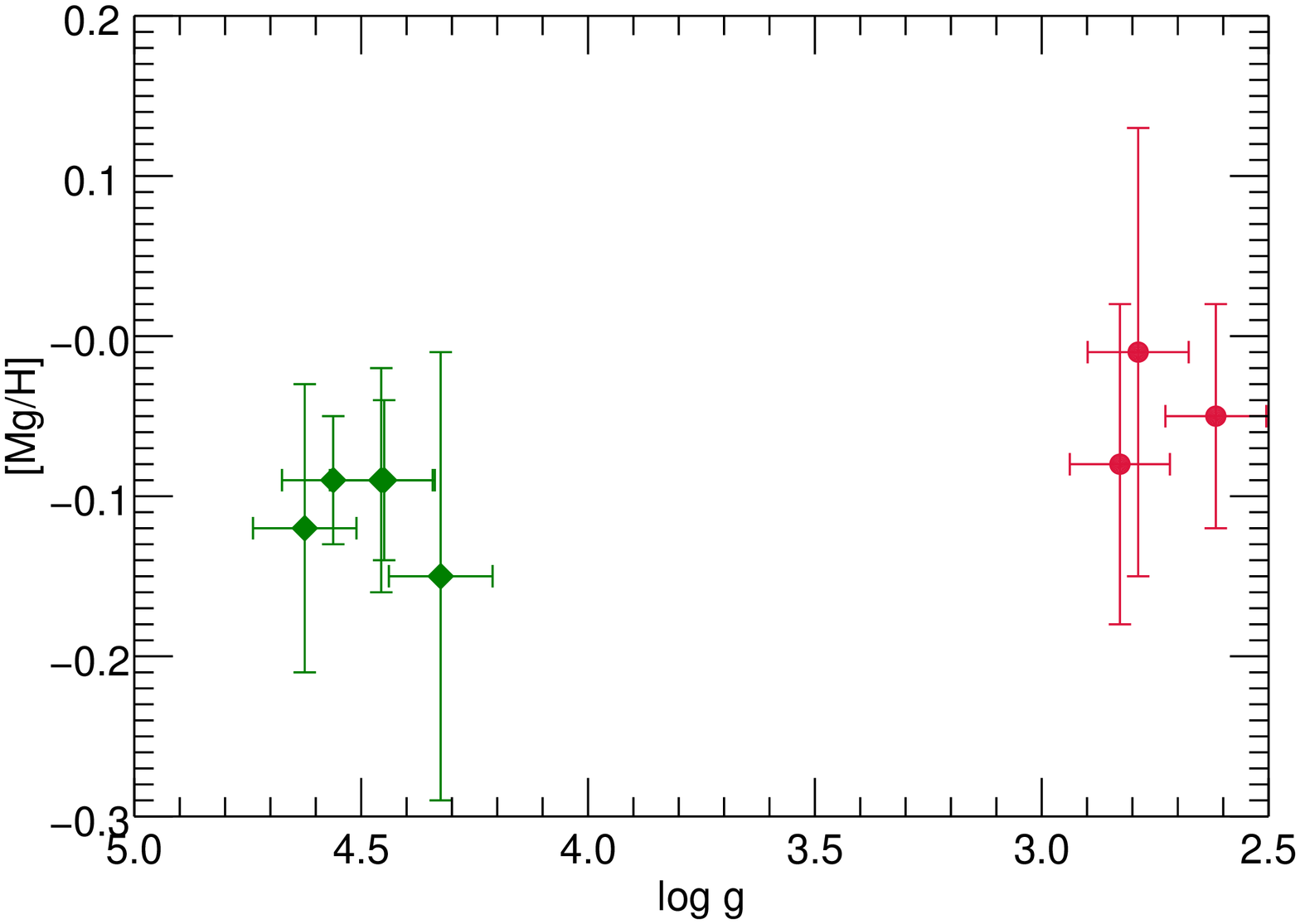}
	\includegraphics[width=0.69\columnwidth]{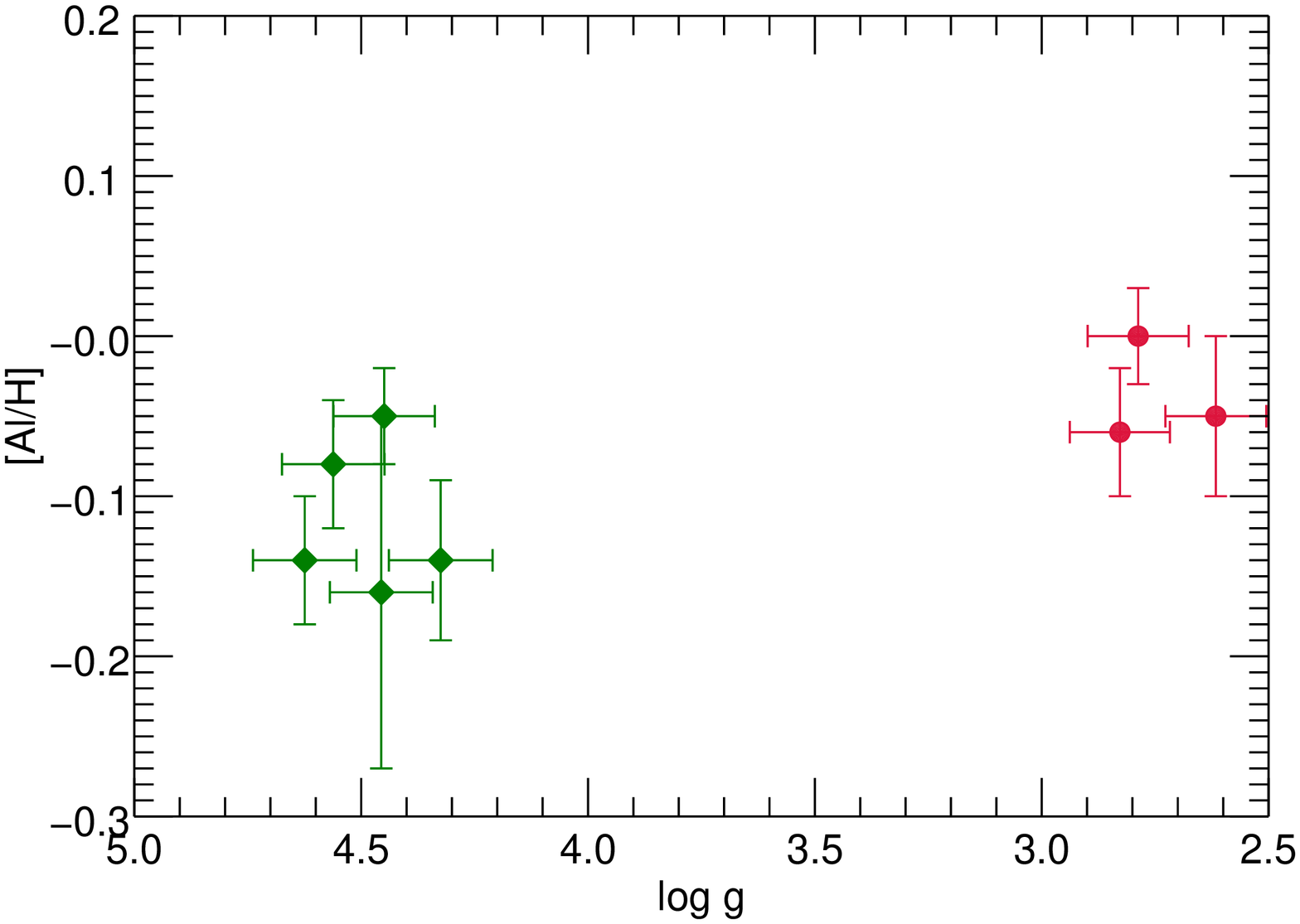}
	\includegraphics[width=0.69\columnwidth]{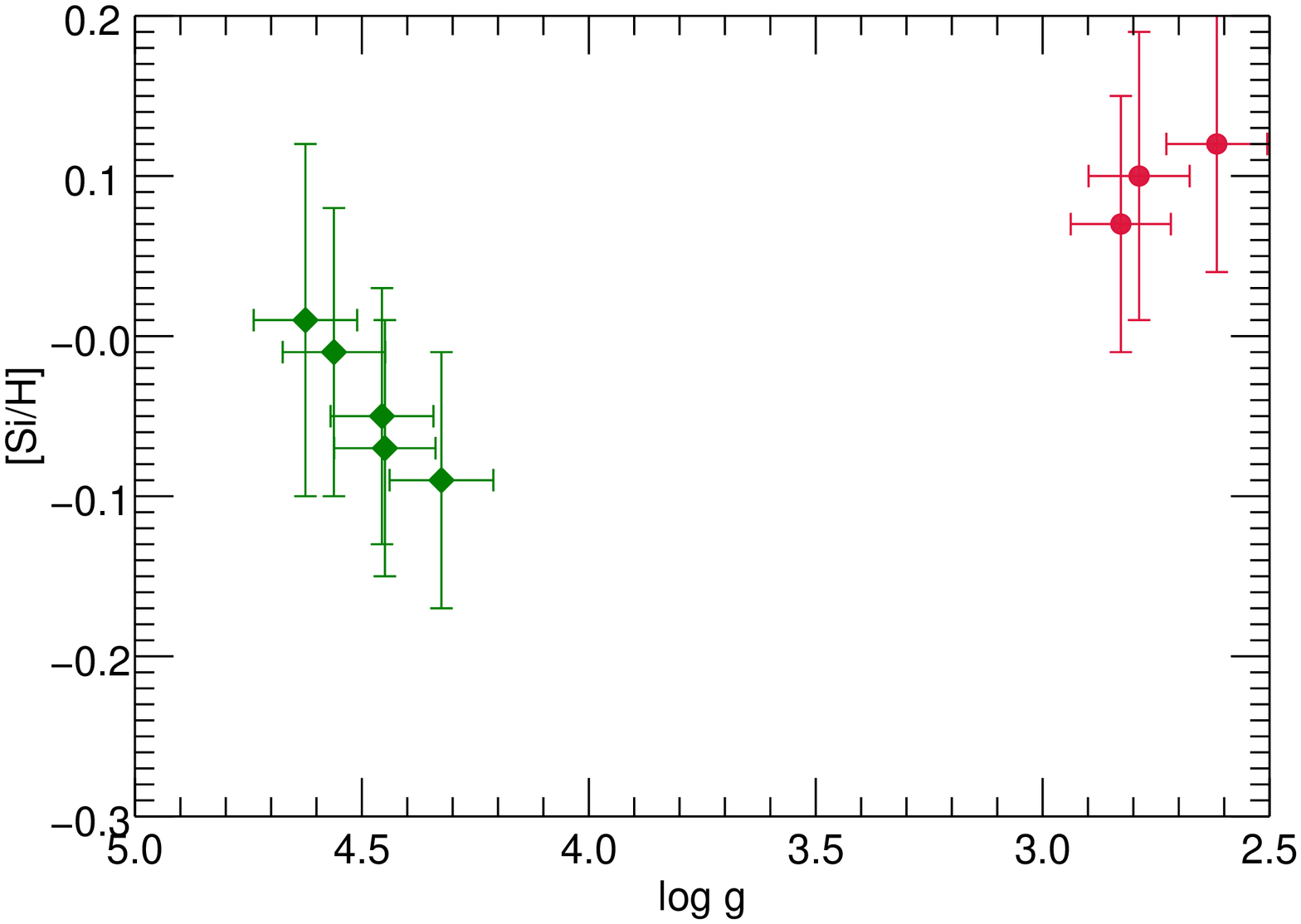}
	\includegraphics[width=0.69\columnwidth]{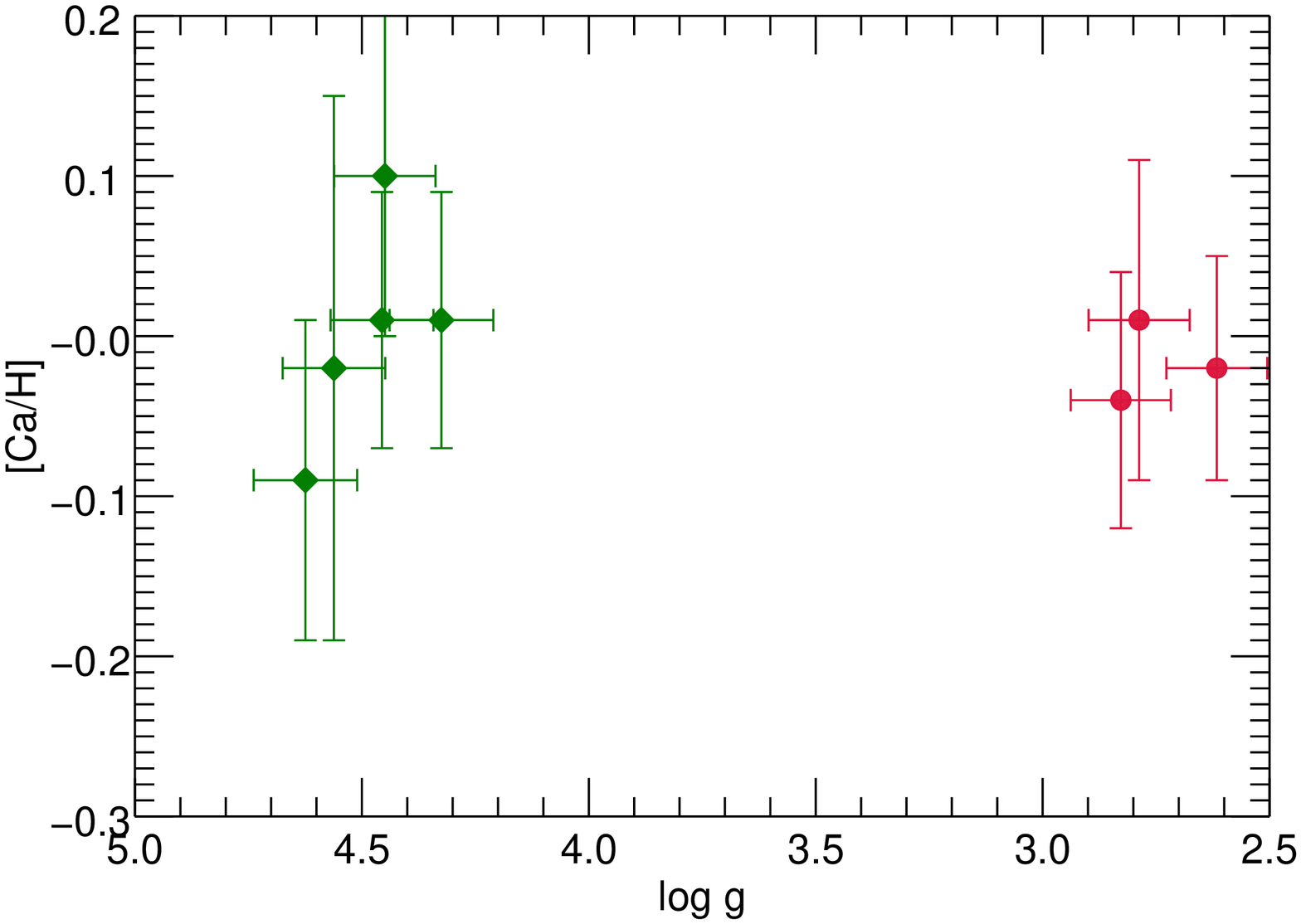}
	\includegraphics[width=0.69\columnwidth]{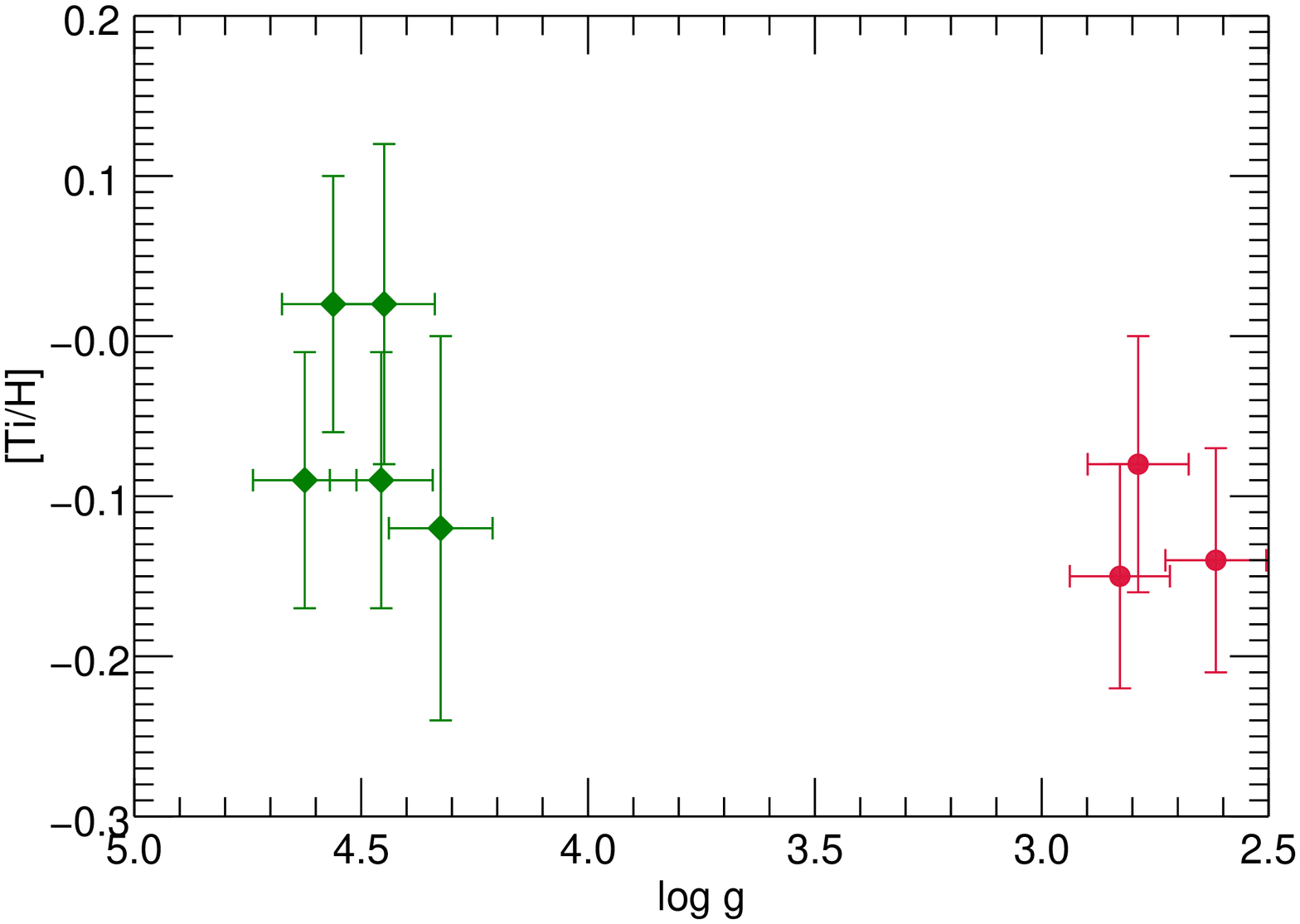}
	\includegraphics[width=0.69\columnwidth]{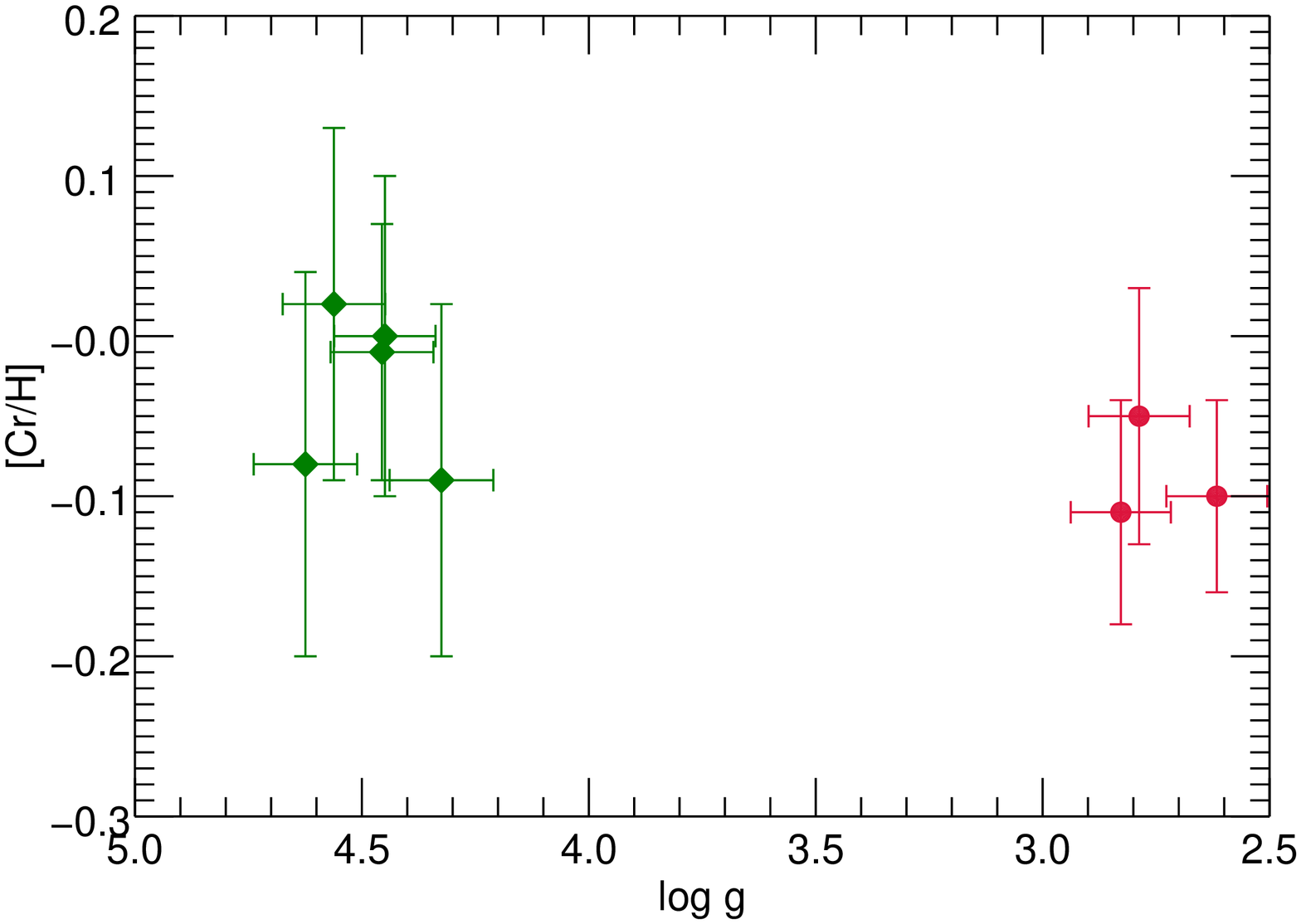}
	\includegraphics[width=0.69\columnwidth]{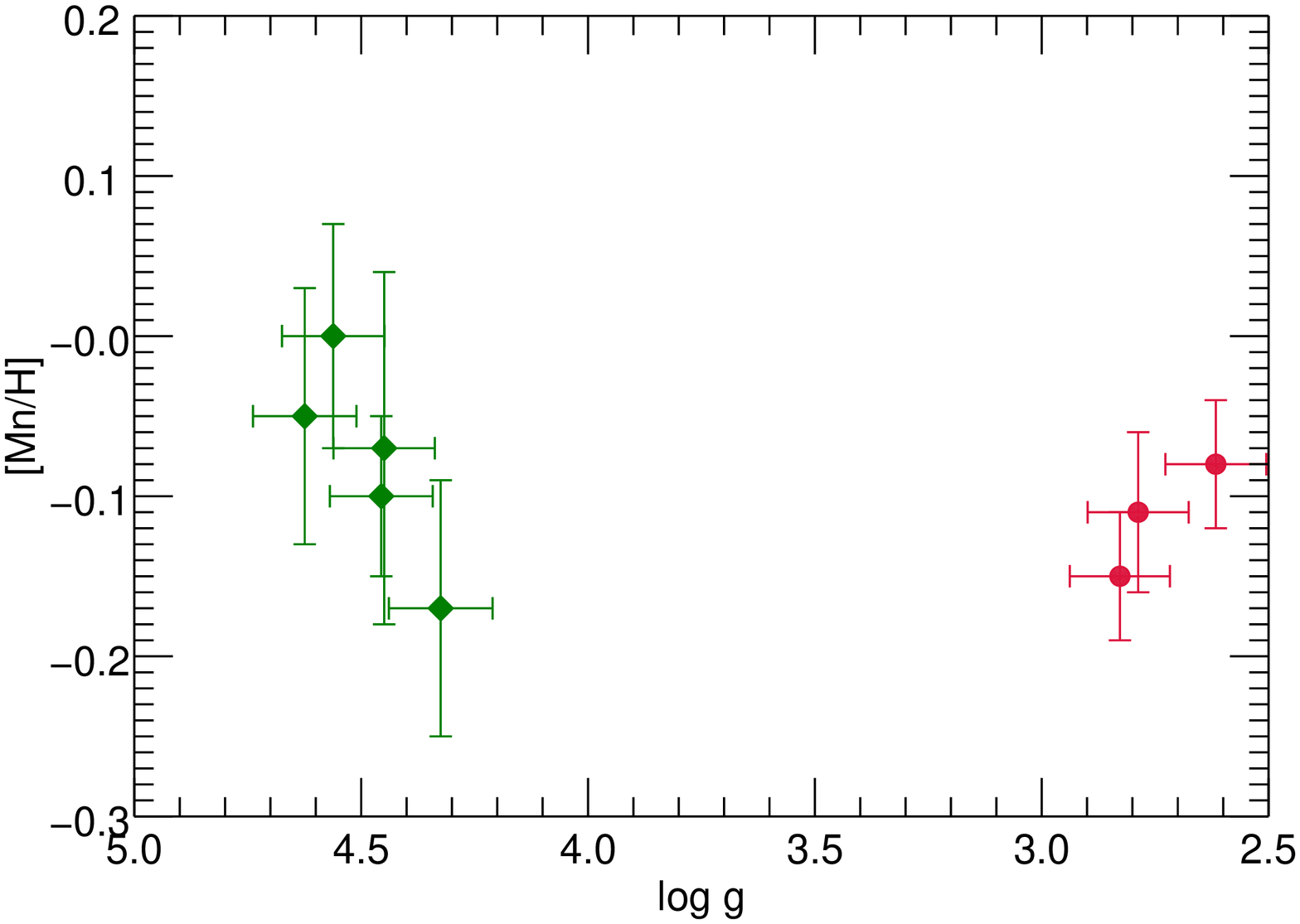}
	\includegraphics[width=0.69\columnwidth]{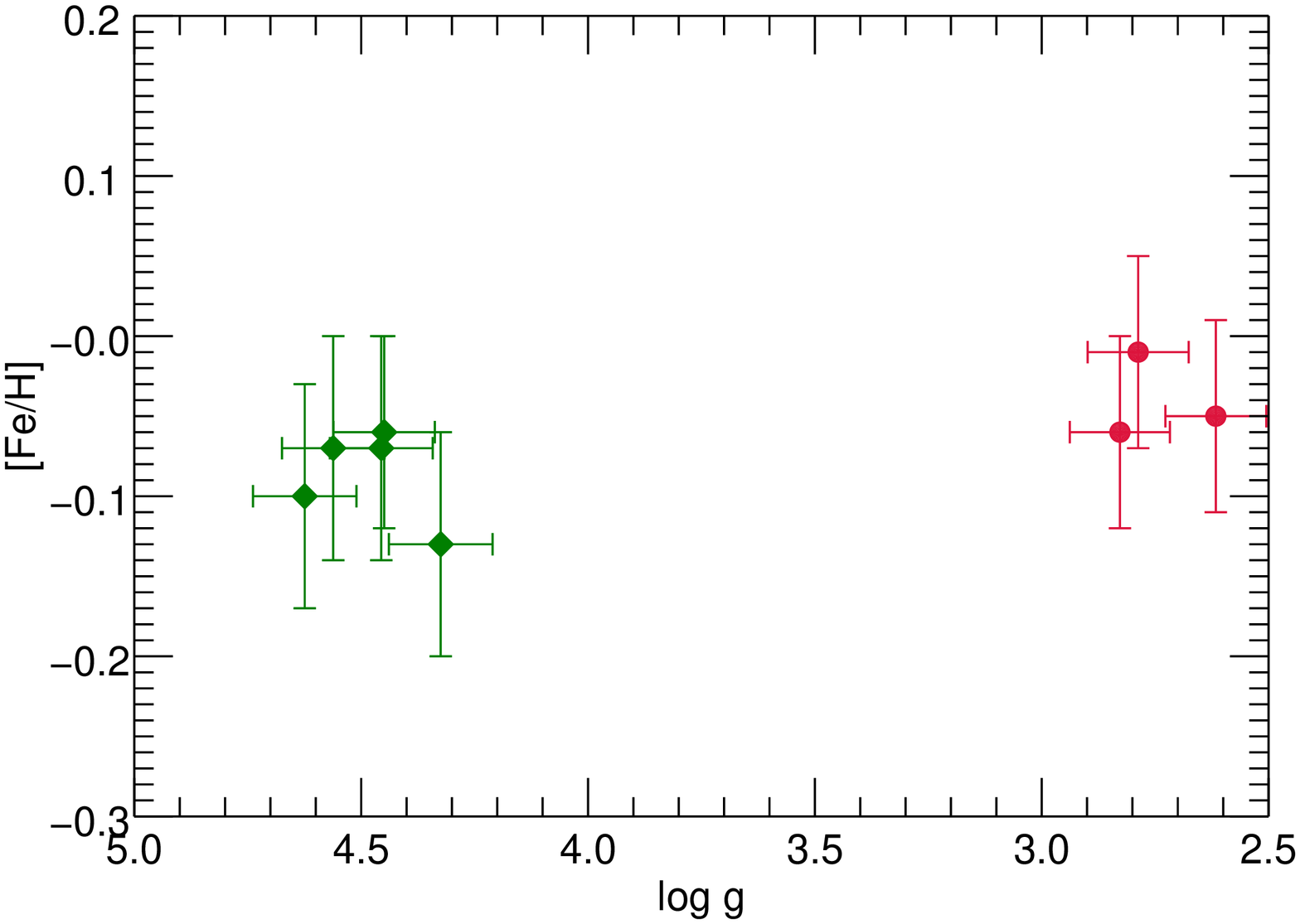}
	\includegraphics[width=0.69\columnwidth]{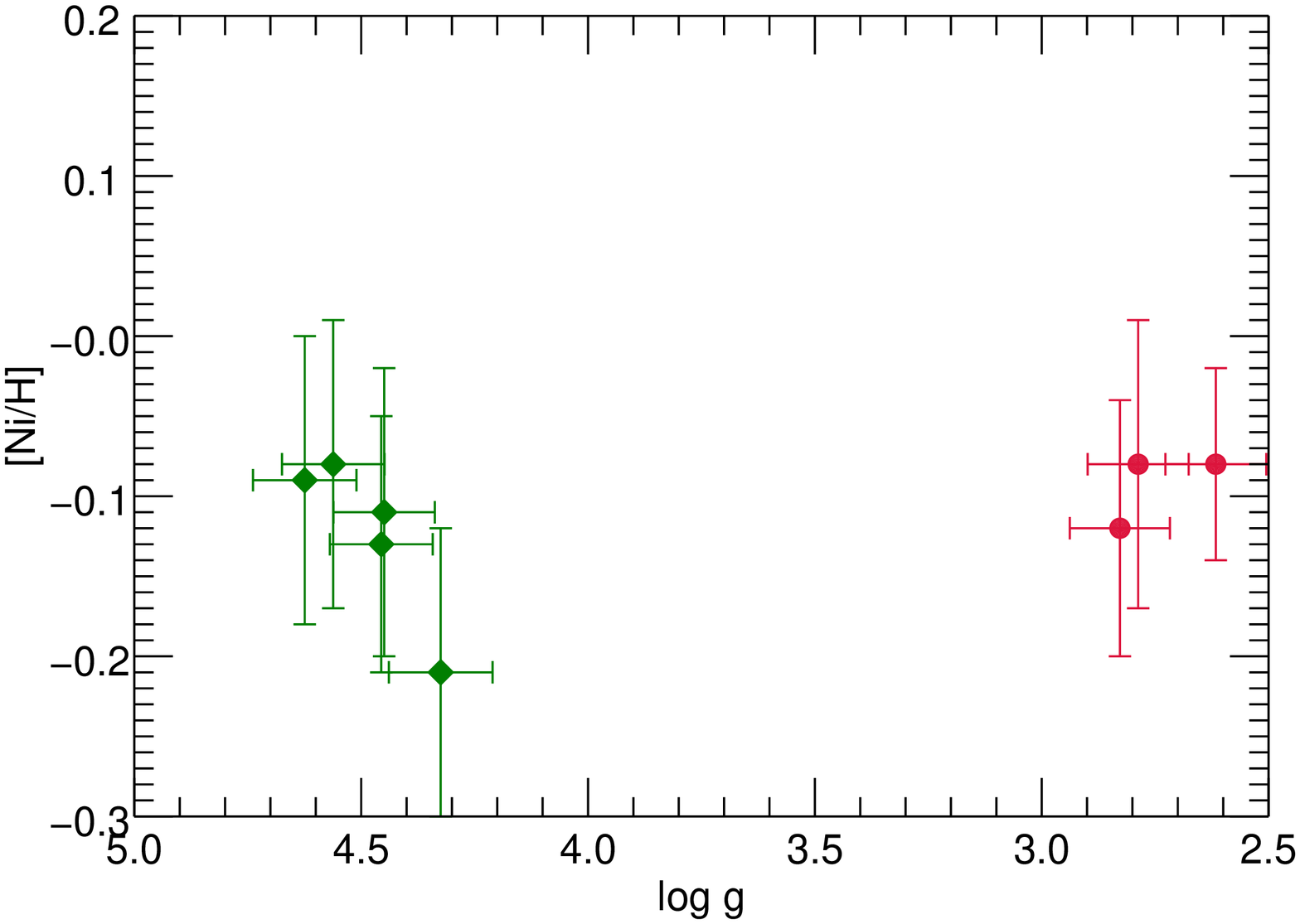}
	\caption{The different panels show from top to bottom and from left to right the abundances of the species discussed in this work as a function of $\log g$ in NGC 6633. The symbols are the same as in Fig.~\ref{fig:fig_iso_ngc6633}. }
	\label{fig:fig_ngc6633}
\end{figure*}

Figure~\ref{fig:fig_ngc6633} shows, similarly to Fig.~\ref{fig:fig_fe}, the abundances as a function of $\log g$. For elements of the Ca and Fe group, we do not find any significant offsets between the MS and the giant stars. As shown in Table~\ref{tab:diff_ngc6633} for all elements of the two groups $\Delta[\mathrm{X}/\mathrm{H}]$ is smaller than $(1-2)\times\mathrm{err\_}\Delta[\mathrm{X}/\mathrm{H}]$, i.e. the measured offsets are less significant than for the case of M67.

This is not the case for lighter elements of the oxygen group. Na in particular presents a large difference (0.278 dex) between the mean MS abundance and the average abundance of the giant stars. Nevertheless, this offset could be a physical effect: for stars in the mass range of the NGC 6633 giants ($\sim2.5 M_{\odot}$) we expect to see an enhancement in Na due to mixing effects after the dredge-up (see, e.g., \citealt{smiljanic2016} and references therein). Besides, also NLTE effects might play a role: \citet{lind2011} predict a correction for NLTE effects of the Na abundance between -0.1 and -0.15 dex for the NGC 6633 MS stars (with mean $T_{\mathrm{eff}}=5463$ K and $\log g=4.48$ ) and of $\sim-0.2$ dex for the giant stars (with mean $T_{\mathrm{eff}}=4995$ K and $\log g=2.74$ ) for the 568.2 nm line at solar metallicity. For the 615.4 nm line the correction would be $\sim-0.1$ dex for both groups. 

For Mg and Al an offset is also present, although not as pronounced as for Na ($\Delta[\mathrm{X}/\mathrm{H}]<2\times\mathrm{err\_}\Delta[\mathrm{X}/\mathrm{H}]$ ). The effects of mixing for Al should not be visible in stars less massive than $3 M_{\odot}$ \citep{smiljanic2016} and this offset would probably not disappear if NLTE effects were taken into account in the abundance analysis. \citet{nordlander2017} predict for Al a correction for dwarf and giant stars of $\sim-0.1$ dex at solar metallicity for the line at 669.6 nm. For Mg we do not expect NLTE effects to have a strong influence on the abundances. \citet{zhang2017} calculate corrections of the order of $\pm0.01$ dex for stellar parameters similar to our NGC 6633 sample. For the same stars, \citet{zhang2016} predict corrections of the order of -0.03 (RGB) and 0.00 dex (MS) for Si. The latter, however, has $\Delta[\mathrm{X}/\mathrm{H}]\sim3\times\mathrm{err\_}\Delta[\mathrm{X}/\mathrm{H}]$, i.e. the offset is relatively large and cannot be explained by NLTE effects. Since also in the case of M67, we found that the offset in Si between the upper MS and RGB is larger than predicted by the models, this could mean that indeed for Si there is a bias in the analysis of dwarf and giant stars. 

\begin{table*}
	\caption{Mean surface chemical abundances and standard deviations for stars on the main sequence ($[\mathrm{X}/\mathrm{H}]_{\mathrm{MS}}$) and giant stars ($[\mathrm{X}/\mathrm{H}]_{\mathrm{RGB}}$) in NGC 6633. The last two columns contain the difference in abundance between the two groups with the respective error calculated as in Sec.~\ref{sec:res}.}
	\begin{tabular}{|l|r|r|r|r|r|r}
		\hline
		\multicolumn{1}{c|}{X} &
		\multicolumn{1}{c|}{$[\mathrm{X}/\mathrm{H}]_{\mathrm{MS}}$} &
		\multicolumn{1}{c|}{$\sigma_{\mathrm{MS}}$} &
		\multicolumn{1}{c|}{$[\mathrm{X}/\mathrm{H}]_{\mathrm{RGB}}$} &
		\multicolumn{1}{c|}{$\sigma_{\mathrm{RGB}}$} &
	  	\multicolumn{1}{c|}{$\Delta[\mathrm{X}/\mathrm{H}]$} &
		\multicolumn{1}{c|}{$\mathrm{err\_}\Delta[\mathrm{X}/\mathrm{H}]$} \\
		\hline
		C & 0.072& 0.109 & -0.200 & 0.044 & -0.272 & 0.118\\
		Na & -0.128 & 0.040 & 0.150 & 0.017 & 0.278 & 0.044\\
		Mg &  -0.108 & 0.027 & -0.047 & 0.035 & 0.061 & 0.044\\
		Al & -0.114 & 0.047 & -0.037 & 0.032 & 0.077 & 0.057\\
		Si & -0.042 & 0.042 & 0.097 & 0.025 & 0.139 & 0.049\\
		Ca & 0.002 & 0.068 & -0.017 & 0.025 & -0.019 & 0.073\\
		Ti & -0.052 & 0.067 & -0.123 & 0.038 & -0.071 & 0.077\\
		Cr & -0.032 & 0.050 & -0.087 & 0.032 & -0.055 & 0.059\\
		Mn & -0.078 & 0.063 & -0.113 & 0.035 & -0.035 & 0.072\\
		Fe & -0.086 & 0.029 & -0.040 & 0.027 & 0.046 & 0.039\\
		Ni & -0.124 & 0.052 & -0.093 & 0.023 & 0.031 & 0.057\\
		\hline
		
	\end{tabular}
	\label{tab:diff_ngc6633}
\end{table*}

\subsection{APOGEE DR14}
\label{sec:apdr14}

The recent fourteenth data release of the Apache Point Observatory Galactic Evolution Experiment \citep[hereafter APOGEE DR14, see][]{majewski2017,sdss14} contains very important results concerning the abundances of M67 stars. In APOGEE DR14 stars in every evolutionary stage of M67 have been observed and analysed, from the main sequence to the red clump. We decided to be as conservative as possible in the choice of the stars to take into consideration and we selected only those with STARFLAG, ASPCAPFLAG, and the ELEMFLAG of the element under study equal to zero. By doing so we aim at excluding as many possible biases due to unreliable results as possible.

APOGEE DR14 contains the abundances determined by the APOGEE Stellar Parameters and Chemical Abundances Pipeline (ASPCAP) \citep[see][]{garciaperez2015,holtzman2015} as well as calibrated abundances. These are calculated under the assumption that abundances of open cluster members should be homogeneous and that therefore any trend with temperature must be corrected for. Since for our purposes it is important to take into consideration any trend of the chemical abundances as a function of temperature, we decided to use the uncalibrated abundances. We present the results in the form [X/H]. For the elements that were calculated as [X/M] we add the uncalibrated [M/H] of the star. The error bars only represent the errors on the fit of the uncalibrated abundances presented in the APOGEE DR14 catalogue (for the uncalibrated [M/H] no errors are given). These are very small and are not to be considered realistic errors.

\begin{figure}
	\includegraphics[width=\columnwidth]{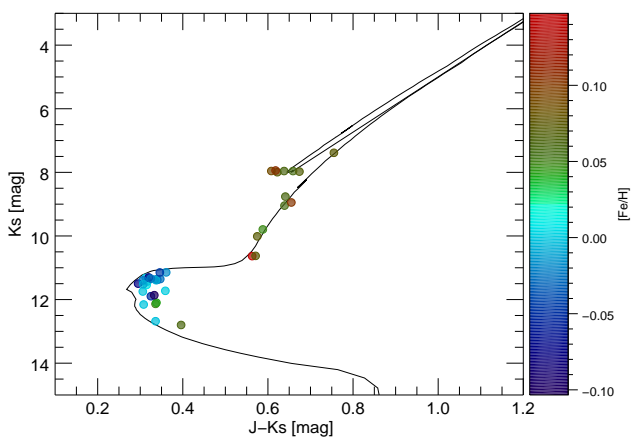}
	\caption{CMD of the stars selected as members of M67 from APOGEE DR14. Stars with STARFLAG, ASPCAPFLAG or FE\_H\_FLAG not equal to zero were excluded from the plot. The stars are colour coded by their iron abundance.The solid line represents a PARSEC isochrone with an age of 3.75 Gyr.}
	\label{fig:fig_ap_iso}
\end{figure}

\begin{figure*}
	\includegraphics[width=0.69\columnwidth]{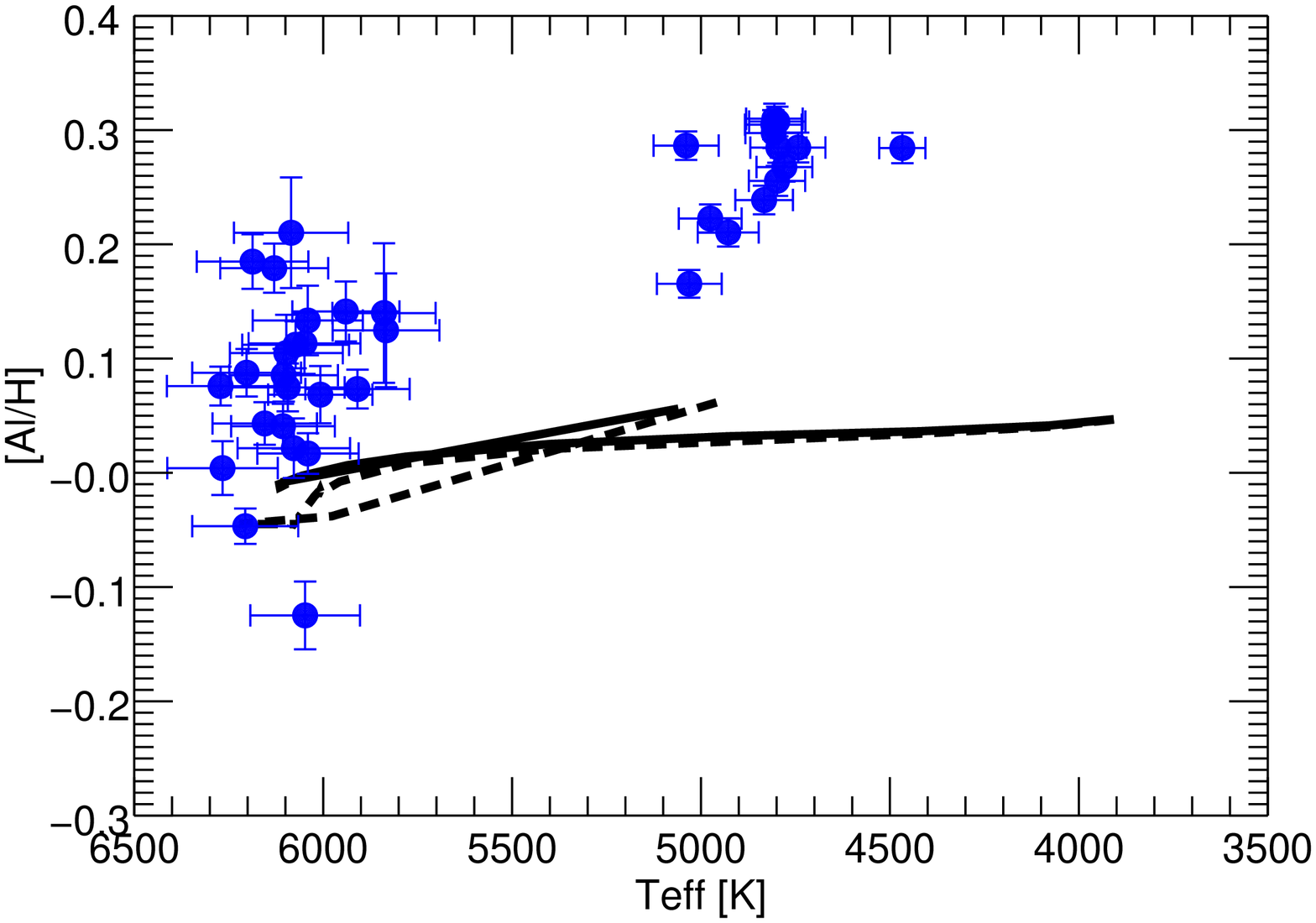}
	\includegraphics[width=0.69\columnwidth]{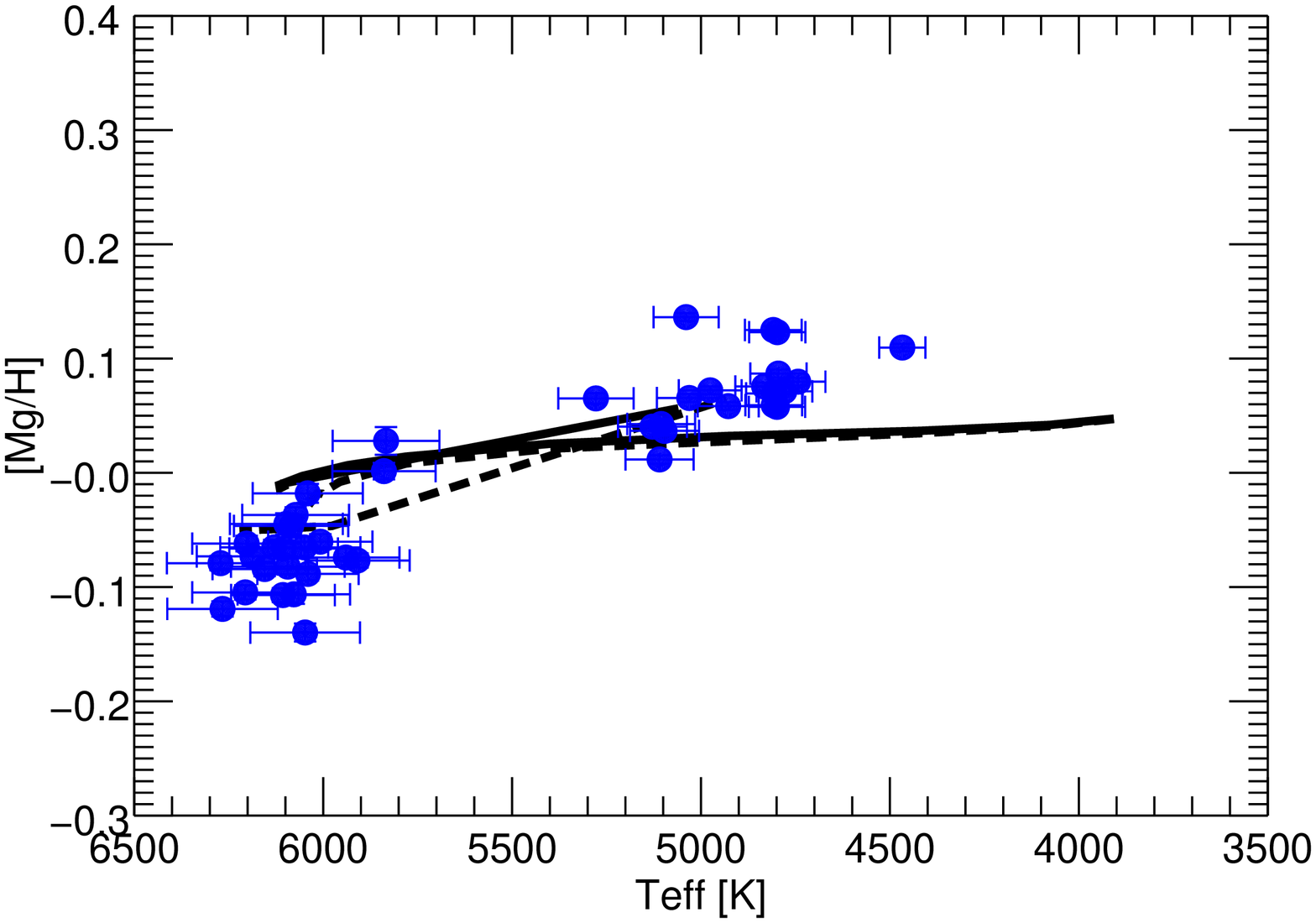}
	\includegraphics[width=0.69\columnwidth]{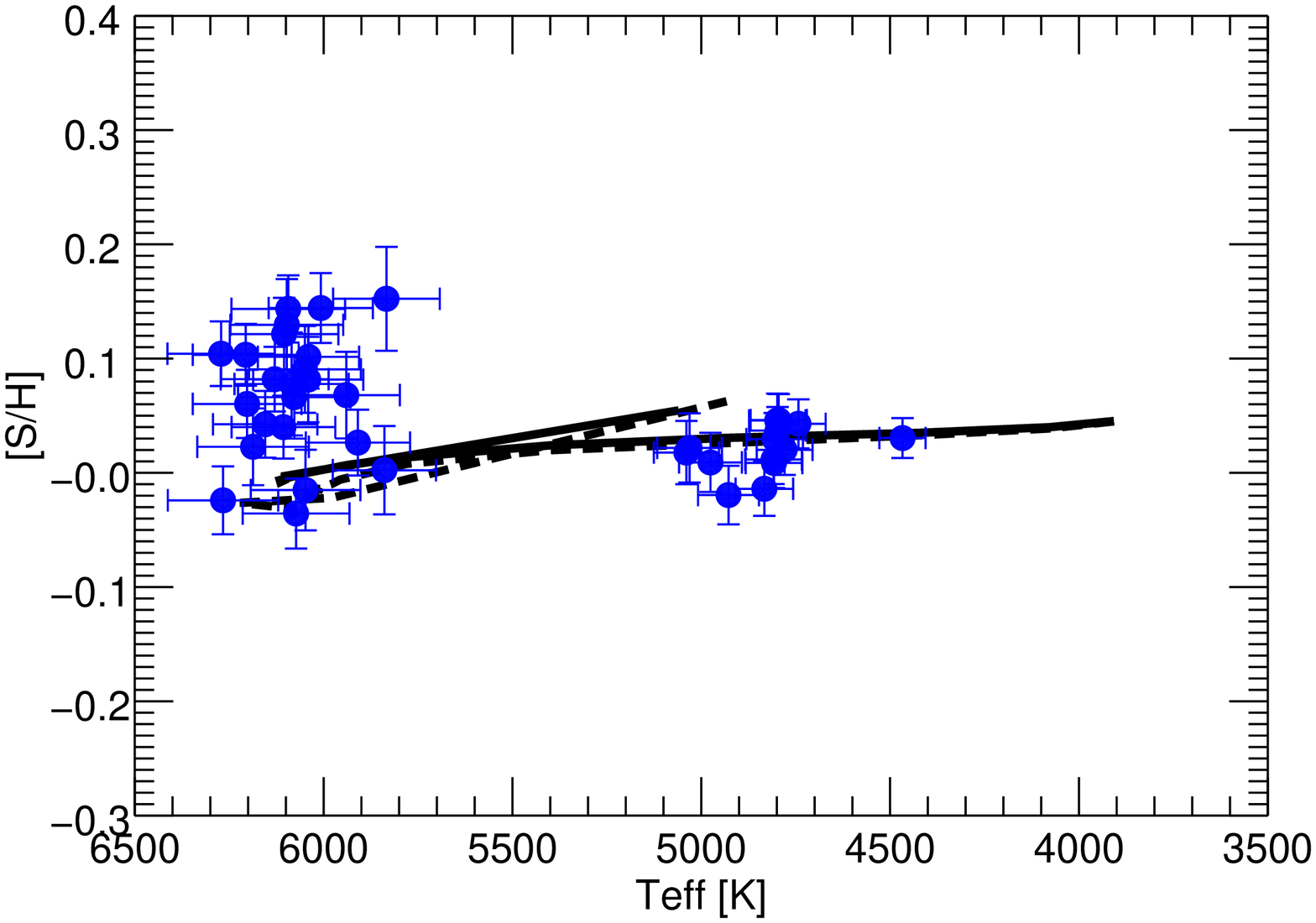}
	\includegraphics[width=0.69\columnwidth]{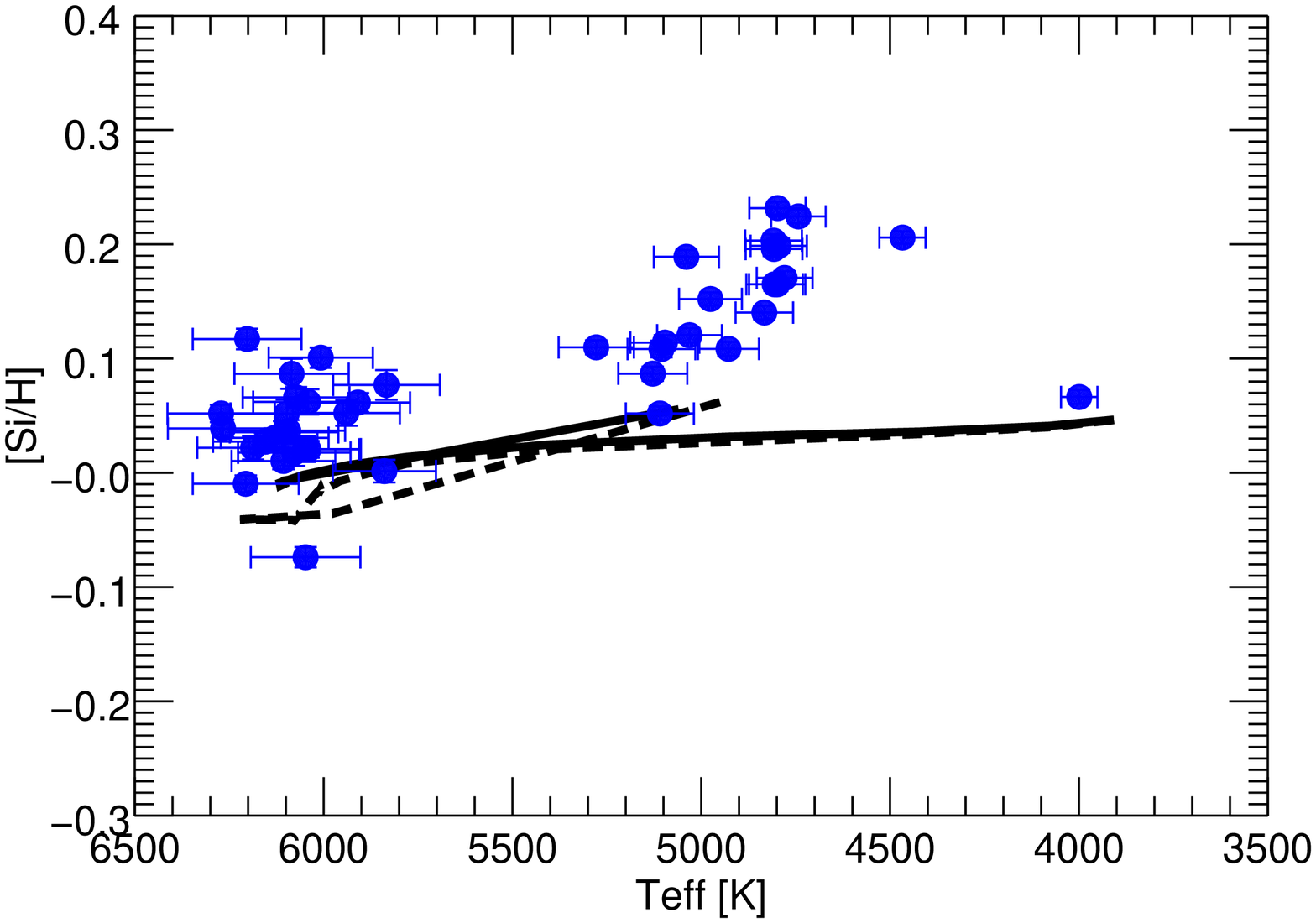}
	\includegraphics[width=0.69\columnwidth]{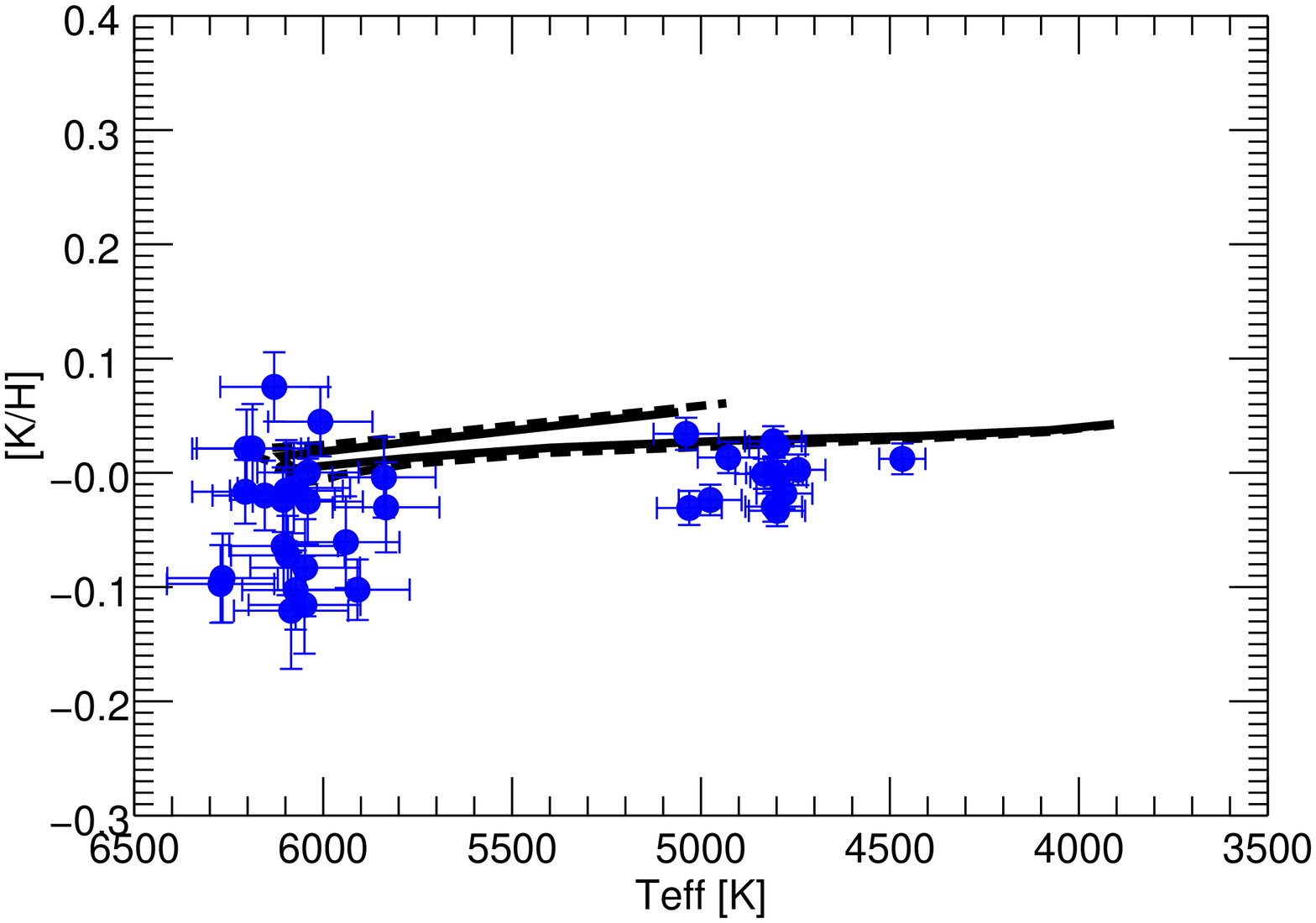}
	\includegraphics[width=0.69\columnwidth]{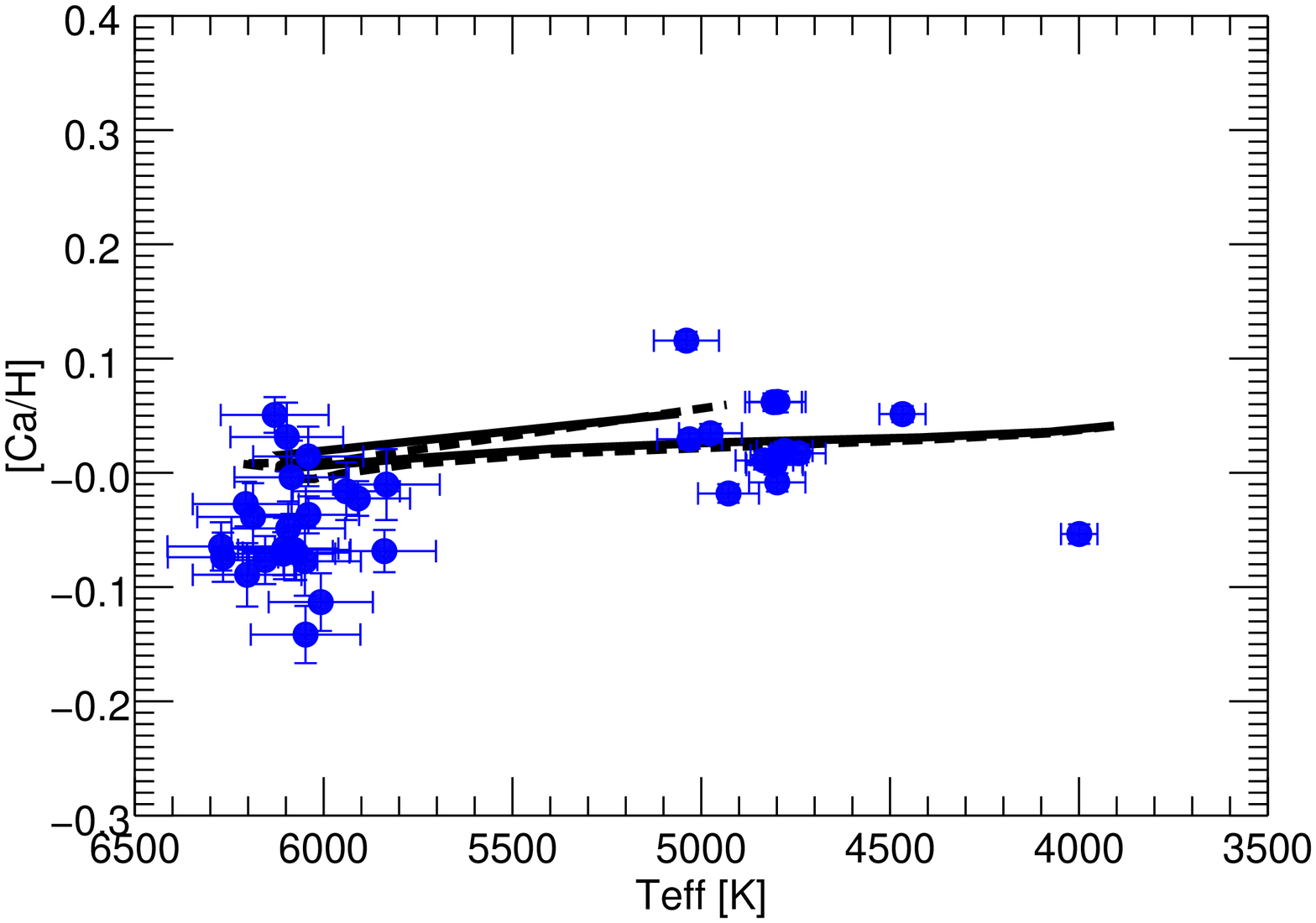}
	\includegraphics[width=0.69\columnwidth]{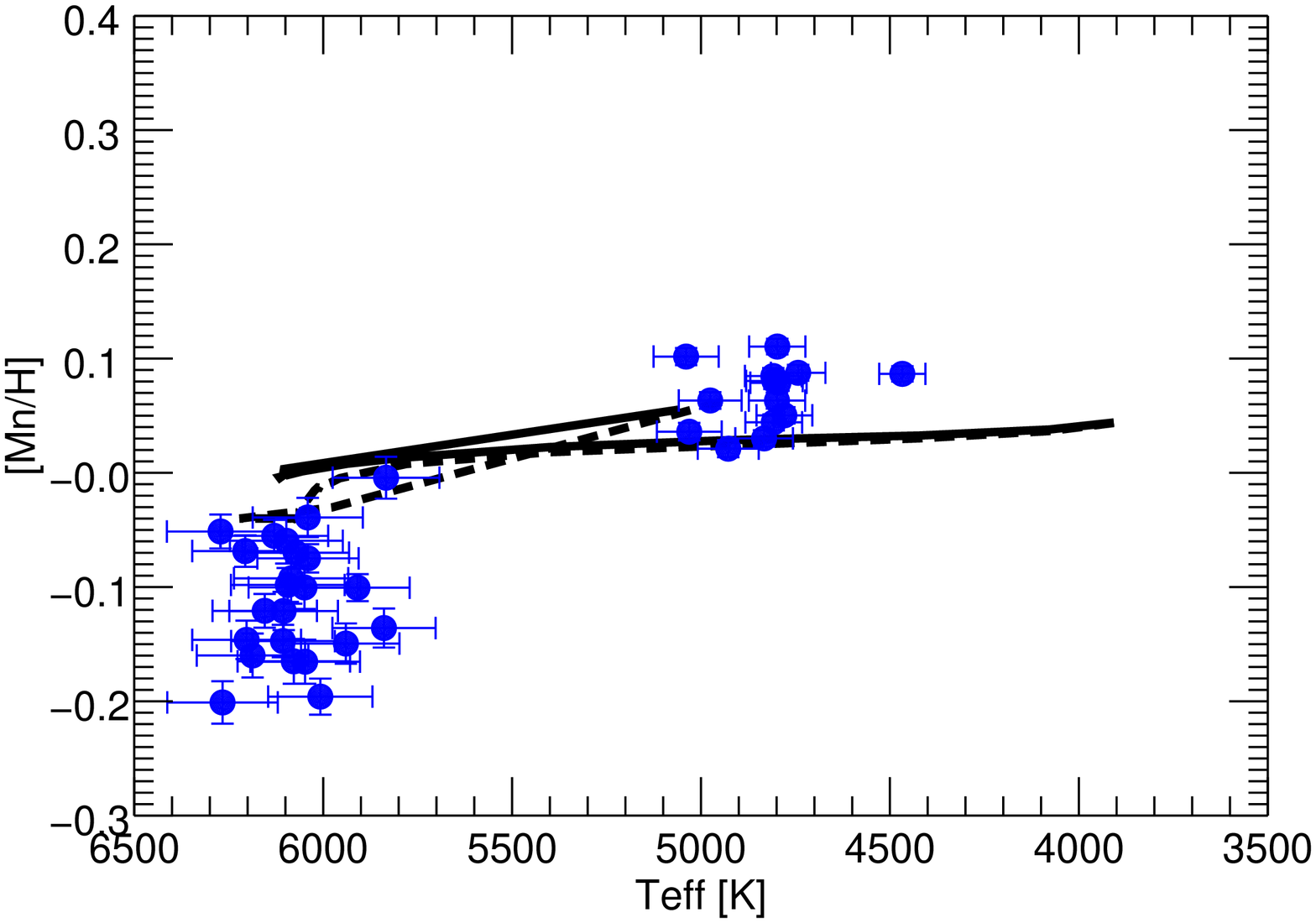}
	\includegraphics[width=0.69\columnwidth]{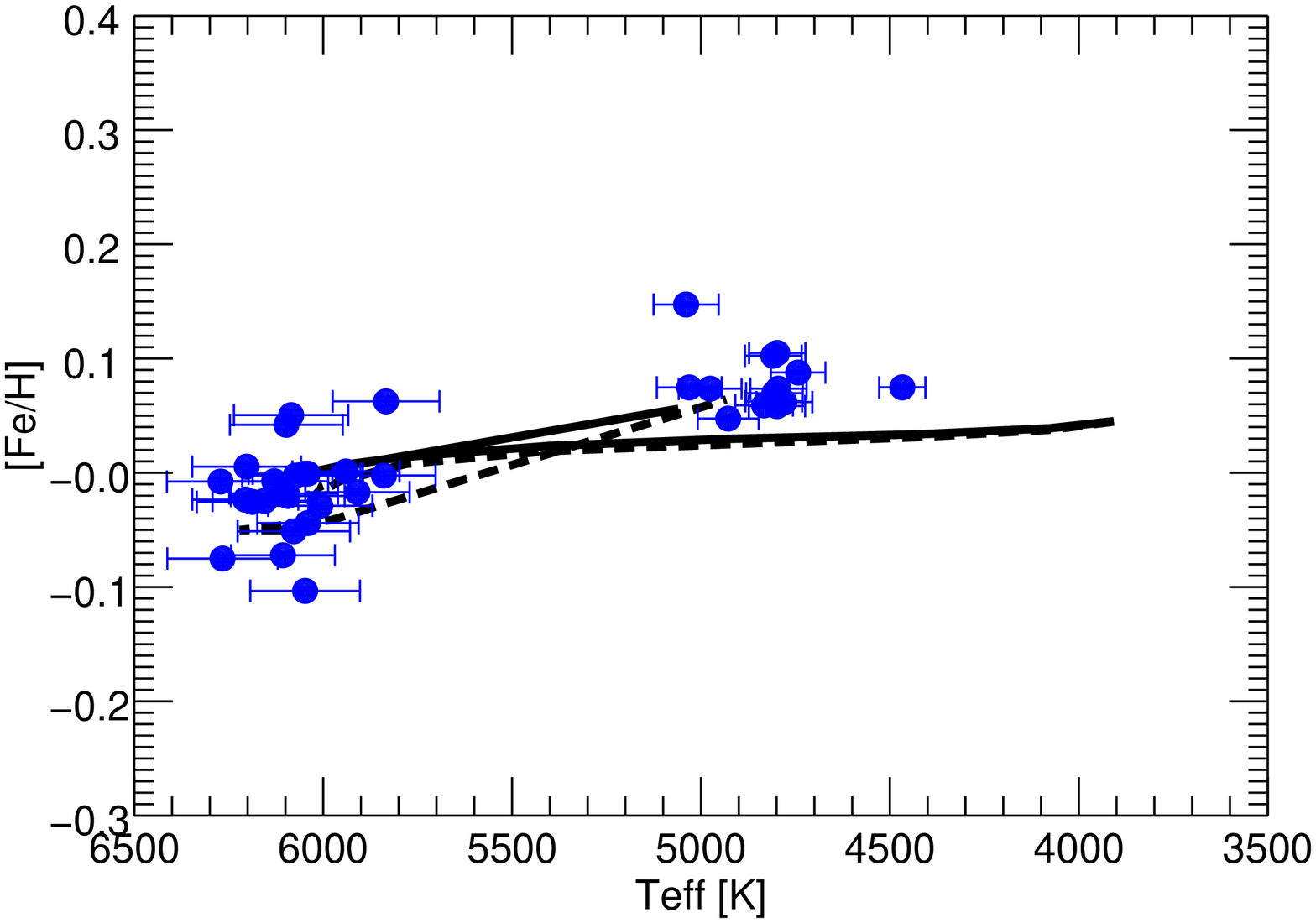}
	\includegraphics[width=0.69\columnwidth]{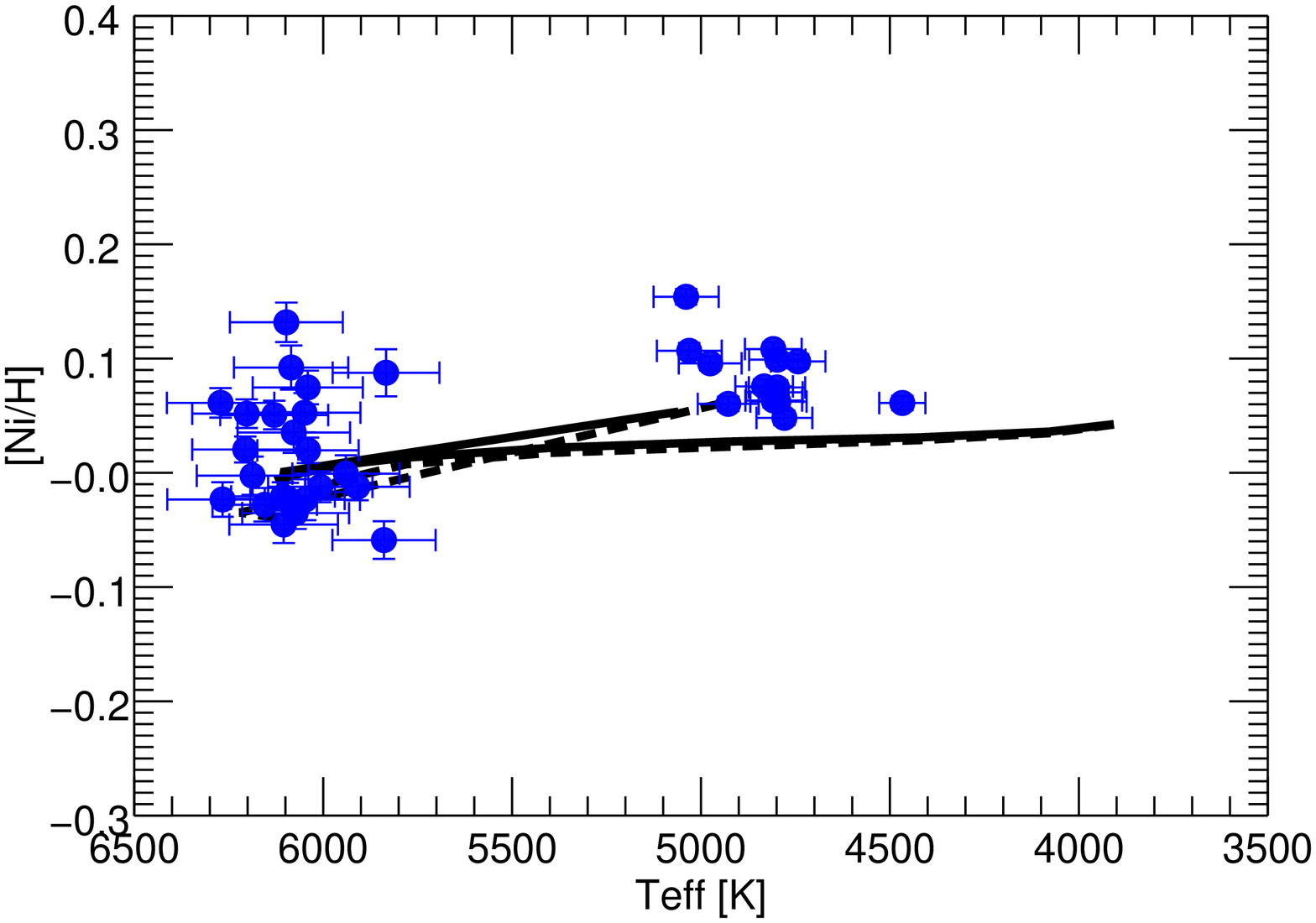}
	\caption{Uncalibrated abundances from APOGEE DR14 for M67 stars as a function of their (calibrated) effective temperature.}
	\label{fig:fig_ap_fe}
\end{figure*}

Figure~\ref{fig:fig_ap_iso} shows the CMD of the sample (not corrected for reddening) selected after a membership analysis based on radial velocity, proper motion, photometry and metallicity, and after all stars with the flags listed above not equal to zero were excluded. The stars are colour-coded based on their iron abundance. It is already clear from this plot that [Fe/H] on the TO is lower than on the RGB and on the RC. Since no calibrated $\log g$ is available for dwarf stars in APOGEE DR14, we plotted in Fig.~\ref{fig:fig_ap_fe} the abundances of several species as a function of the effective temperature together with the models from \citet{michaud2004}. Several of the abundances obtained by APOGEE DR14 show offsets between TO and RGB similar to those predicted by the models, although, as in GES iDR5, for some elements large spreads as well as offsets in comparison to the models are present. This is nevertheless an interesting result for our study, since it shows that abundances computed with a completely independent method from infrared rather than optical spectra yield similar trends to those obtained within GES iDR5.

During the refereeing process the work by \citet{souto2018} appeared, in which the authors analyse the APOGEE spectra of 8 M67 members in different evolutionary stages from the MS to the RGB. They derive abundances of 14 elements for these stars and come to a similar conclusion as our study, i.e. that diffusion effects seem to be present in the member stars of M67.

\section{Summary}
\label{sec:end}

We studied the surface chemical composition of member stars in the old open cluster M67  analysed by the Gaia-ESO Survey, searching for evidence of atomic diffusion processes. We compared the abundances of C, O, Mg, Al, Na, Si, Ca, Ti, Cr, Mn, Fe, and Ni with different stellar evolutionary models calculated by \citet{michaud2004} for the age and metallicity of M67 and found them to be roughly consistent with each other. Our results support the findings of \citet{onehag2014}, although the inclusion of three RGB stars compared to the sample of \citet{onehag2014} renders the results more conclusive, since in RGB stars the initial surface abundances are essentially restored, after the effects of atomic diffusion have been cancelled by convection. Furthermore, the predicted abundance variations between upper MS and RGB are larger than between upper MS and TO/early SGB and thus, at least in theory, easier to measure. Also \citet{casey2016} found a variation in metallicity [Fe/H] of $\sim0.05$ dex between TO and SGB/low-RGB stars of M67 from lower resolution spectra. Recently, \citet{souto2018} published similar results regarding a sample of 8 M67 stars that hint at the presence of atomic diffusion effects in M67. Although we find indications of ongoing diffusion processes, it is not possible to state whether the models with or without turbulent diffusion better represent the data, given the spread and the errors on the measured abundances on the upper MS and TO. 

We have also shown that the star S806 systematically presents higher surface abundances than its cluster companions. We suggest as a possible explanation that this star might be part of a binary system and have undergone accretion of expelled material from a putative evolved, more massive companion.

We have investigated the abundances in stars of the main sequence and RGB/RC of NGC 6633, a cluster young enough that we do not expect any signs of diffusion effects in the stars included in GES iDR5. As expected, for most elements the abundances of the two groups are consistent with each other within one sigma, except for few light elements of the Na group. For Na, the offset can be explained by mixing processes and NLTE effects, while at least for Si the measured abundance difference between dwarfs and giants might be enhanced due to analysis effects. The latter may also explain the large Si offset measured in M67 (Fig.~\ref{fig:fig_fe}).

We found that the latest APOGEE data release DR14 presents for the stars of M67 similar trends to those measured in GES iDR5. Nevertheless, the large scatter and offsets affecting some elements may be a combination of true abundance variations and uncertainties in the analysis that are difficult to disentangle at the present time. This calls for further refinements in the determination of abundances from observed spectra.

The detection of true variations in the surface abundances of stars due to stellar evolutionary effects has strong implications for the field of Galactic archaeology, since it puts a constraint on the precision achievable with chemical tagging of the order of $\sim0.1$ dex, similar to what was indicated in \citet{dotter2017}. As a consequence, future studies will need to re-think chemical tagging methods taking into account stellar evolution or they will be applicable only to stars in the same evolutionary stage.

\section*{Acknowledgements}

This work was supported by Sonderforschungsbereich SFB 881 "The Milky Way System" (subproject B5) of the German Research Foundation (DFG).  A.B. acknowledges support from the SFB 881 visitor program. C.B.M. thanks Elisabetta Caffau for fruitful discussions and Don VandenBerg for providing the isochrones shown in Fig~\ref{fig:fig_iso}. The authors would like to thank the anonymous referee for the constructive comments that helped improving the present work. 

M.T.C acknowledge the financial support from the Spanish Ministerio de Econom\'ia y Competitividad, through grant AYA2016-75931. A.D. and G.T. acknowledge support by the Research Council of Lithuania (MIP-082/2015). U.H. acknowledges support from the Swedish National Space Board (SNSB/Rymdstyrelsen). T.B. was supported by the project grant "The New Milky'' from the Knut and Alice Wallenberg foundation. R.S. acknowledges support from the Polish Ministry of Science and Higher Education. S.G.S and V.A. acknowledge the support by Funda\c{c}\~{a}o para a Ci\^{e}ncia e Tecnologia (FCT) through national funds and a research grant (project ref. 
UID/FIS/04434/2013, and PTDC/FIS-AST/7073/2014). S.G.S. also acknowledge the support from FCT through Investigador FCT contract of reference IF/00028/2014 and POPH/FSE (EC) by FEDER funding through the program "Programa Operacional de Factores de Competitividade -- COMPETE". V.A. acknowledges the support from FCT through Investigador FCT contract IF/00650/2015/CP1273/CT0001. E.J.A. acknowledges partial support from the Spanish Ministry for Economy and Competitiveness and FEDER funds through grant AYA2013-40611-P. A.R.C. is supported through an Australian Research Council Discovery Project under grant DP160100637. C.L. thanks the Swiss National Science Foundation for supporting this research through the Am- bizione grant number PZ00P2 168065. A.K. acknowledges support from the Swedish National Space Board (SNSB). X.F. acknowledges financial support from Premiale 2015 MITiC (PI B/ Garilli). S.L.M. acknowledges support from the Australian Research Council through grant DE140100598. A.B. acknowledges support from the Millennium Science Initiative (Chilean Ministry of Economy)

Based on data products from observations made with ESO Telescopes at the La Silla Paranal Observatory under programme ID 188.B-3002. These data products have been processed by the Cambridge Astronomy Survey Unit (CASU) at the Institute of Astronomy, University of Cambridge, and by the FLAMES/UVES reduction team at INAF/Osservatorio Astrofisico di Arcetri. These data have been obtained from the Gaia-ESO Survey Data Archive, prepared and hosted by the Wide Field Astronomy Unit, Institute for Astronomy, University of Edinburgh, which is funded by the UK Science and Technology Facilities Council.
This work was partly supported by the European Union FP7 programme through ERC grant number 320360 and by the Leverhulme Trust through grant RPG-2012-541. We acknowledge the support from INAF and Ministero dell' Istruzione, dell' Universit\`a' e della Ricerca (MIUR) in the form of the grant "Premiale VLT 2012",  and through PRIN-INAF 2014 
"The Gaia-ESO Survey". The results presented here benefit from discussions held during the Gaia-ESO workshops and conferences supported by the ESF (European Science Foundation) through the GREAT Research Network Programme.
This research has made use of the SIMBAD database, operated at CDS, Strasbourg, France and NASA's Astrophysics Data System.




\bibliographystyle{mnras}
\bibliography{references} 

\begin{thebibliography}{}
\makeatletter
\relax
\def\mn@urlcharsother{\let\do\@makeother \do\$\do\&\do\#\do\^\do\_\do\%\do\~}
\def\mn@doi{\begingroup\mn@urlcharsother \@ifnextchar [ {\mn@doi@}
  {\mn@doi@[]}}
\def\mn@doi@[#1]#2{\def\@tempa{#1}\ifx\@tempa\@empty \href
  {http://dx.doi.org/#2} {doi:#2}\else \href {http://dx.doi.org/#2} {#1}\fi
  \endgroup}
\def\mn@eprint#1#2{\mn@eprint@#1:#2::\@nil}
\def\mn@eprint@arXiv#1{\href {http://arxiv.org/abs/#1} {{\tt arXiv:#1}}}
\def\mn@eprint@dblp#1{\href {http://dblp.uni-trier.de/rec/bibtex/#1.xml}
  {dblp:#1}}
\def\mn@eprint@#1:#2:#3:#4\@nil{\def\@tempa {#1}\def\@tempb {#2}\def\@tempc
  {#3}\ifx \@tempc \@empty \let \@tempc \@tempb \let \@tempb \@tempa \fi \ifx
  \@tempb \@empty \def\@tempb {arXiv}\fi \@ifundefined
  {mn@eprint@\@tempb}{\@tempb:\@tempc}{\expandafter \expandafter \csname
  mn@eprint@\@tempb\endcsname \expandafter{\@tempc}}}

\bibitem[\protect\citeauthoryear{{Abolfathi} et~al.,}{{Abolfathi}
  et~al.}{2017}]{sdss14}
{Abolfathi} B.,  et~al., 2017, preprint, \href
  {http://adsabs.harvard.edu/abs/2017arXiv170709322A} {} (\mn@eprint {arXiv}
  {1707.09322})

\bibitem[\protect\citeauthoryear{{Altmann}, {Roeser}, {Demleitner}, {Bastian}
  \& {Schilbach}}{{Altmann} et~al.}{2017}]{altmann2017}
{Altmann} M.,  {Roeser} S.,  {Demleitner} M.,  {Bastian} U.,   {Schilbach} E.,
  2017, \mn@doi [\aap] {10.1051/0004-6361/201730393}, \href
  {http://adsabs.harvard.edu/abs/2017A%26A...600L...4A} {600, L4}

\bibitem[\protect\citeauthoryear{{Bellini} et~al.,}{{Bellini}
  et~al.}{2010}]{bellini2010}
{Bellini} A.,  et~al., 2010, \mn@doi [\aap] {10.1051/0004-6361/200913721},
  \href {http://adsabs.harvard.edu/abs/2010A%26A...513A..50B} {513, A50}

\bibitem[\protect\citeauthoryear{{Bertelli Motta}, {Salaris}, {Pasquali}  \&
  {Grebel}}{{Bertelli Motta} et~al.}{2017}]{bertelli2017}
{Bertelli Motta} C.,  {Salaris} M.,  {Pasquali} A.,   {Grebel} E.~K.,  2017,
  \mn@doi [\mnras] {10.1093/mnras/stw3252}, \href
  {http://adsabs.harvard.edu/abs/2017MNRAS.466.2161B} {466, 2161}

\bibitem[\protect\citeauthoryear{{Blanco-Cuaresma} et~al.,}{{Blanco-Cuaresma}
  et~al.}{2015}]{blanco2015}
{Blanco-Cuaresma} S.,  et~al., 2015, \mn@doi [\aap]
  {10.1051/0004-6361/201425232}, \href
  {http://adsabs.harvard.edu/abs/2015A%26A...577A..47B} {577, A47}

\bibitem[\protect\citeauthoryear{{Bragaglia}, {Gratton}, {Carretta}, {D'Orazi},
  {Sneden}  \& {Lucatello}}{{Bragaglia} et~al.}{2012}]{bragaglia2012}
{Bragaglia} A.,  {Gratton} R.~G.,  {Carretta} E.,  {D'Orazi} V.,  {Sneden} C.,
   {Lucatello} S.,  2012, \mn@doi [\aap] {10.1051/0004-6361/201220366}, \href
  {http://adsabs.harvard.edu/abs/2012A%26A...548A.122B} {548, A122}

\bibitem[\protect\citeauthoryear{{Bragaglia}, {Sneden}, {Carretta}, {Gratton},
  {Lucatello}, {Bernath}, {Brooke}  \& {Ram}}{{Bragaglia}
  et~al.}{2014}]{bragaglia2014}
{Bragaglia} A.,  {Sneden} C.,  {Carretta} E.,  {Gratton} R.~G.,  {Lucatello}
  S.,  {Bernath} P.~F.,  {Brooke} J.~S.~A.,   {Ram} R.~S.,  2014, \mn@doi
  [\apj] {10.1088/0004-637X/796/1/68}, \href
  {http://adsabs.harvard.edu/abs/2014ApJ...796...68B} {796, 68}

\bibitem[\protect\citeauthoryear{{Bressan}, {Marigo}, {Girardi}, {Salasnich},
  {Dal Cero}, {Rubele}  \& {Nanni}}{{Bressan} et~al.}{2012}]{bressan2012}
{Bressan} A.,  {Marigo} P.,  {Girardi} L.,  {Salasnich} B.,  {Dal Cero} C.,
  {Rubele} S.,   {Nanni} A.,  2012, \mn@doi [\mnras]
  {10.1111/j.1365-2966.2012.21948.x}, \href
  {http://adsabs.harvard.edu/abs/2012MNRAS.427..127B} {427, 127}

\bibitem[\protect\citeauthoryear{{Brucalassi} et~al.,}{{Brucalassi}
  et~al.}{2014}]{brucalassi2014}
{Brucalassi} A.,  et~al., 2014, \mn@doi [\aap] {10.1051/0004-6361/201322584},
  \href {http://adsabs.harvard.edu/abs/2014A%26A...561L...9B} {561, L9}

\bibitem[\protect\citeauthoryear{{Brucalassi} et~al.,}{{Brucalassi}
  et~al.}{2016}]{brucalassi2016}
{Brucalassi} A.,  et~al., 2016, \mn@doi [\aap] {10.1051/0004-6361/201527561},
  \href {http://adsabs.harvard.edu/abs/2016A%26A...592L...1B} {592, L1}

\bibitem[\protect\citeauthoryear{{Brucalassi} et~al.,}{{Brucalassi}
  et~al.}{2017}]{brucalassi2017}
{Brucalassi} A.,  et~al., 2017, \mn@doi [\aap] {10.1051/0004-6361/201527562},
  \href {http://adsabs.harvard.edu/abs/2017A%26A...603A..85B} {603, A85}

\bibitem[\protect\citeauthoryear{{Carraro}, {de Silva}, {Monaco}, {Milone}  \&
  {Mateluna}}{{Carraro} et~al.}{2014}]{carraro2014}
{Carraro} G.,  {de Silva} G.,  {Monaco} L.,  {Milone} A.~P.,   {Mateluna} R.,
  2014, \mn@doi [\aap] {10.1051/0004-6361/201423714}, \href
  {http://adsabs.harvard.edu/abs/2014A%26A...566A..39C} {566, A39}

\bibitem[\protect\citeauthoryear{{Carretta}, {Bragaglia}, {Gratton}, {D'Orazi}
  \& {Lucatello}}{{Carretta} et~al.}{2009}]{carretta2009}
{Carretta} E.,  {Bragaglia} A.,  {Gratton} R.,  {D'Orazi} V.,   {Lucatello} S.,
   2009, \mn@doi [\aap] {10.1051/0004-6361/200913003}, \href
  {http://adsabs.harvard.edu/abs/2009A%26A...508..695C} {508, 695}

\bibitem[\protect\citeauthoryear{{Casey}}{{Casey}}{2016}]{casey2016}
{Casey} A.~R.,  2016, \mn@doi [\apjs] {10.3847/0067-0049/223/1/8}, \href
  {http://adsabs.harvard.edu/abs/2016ApJS..223....8C} {223, 8}

\bibitem[\protect\citeauthoryear{{Dekker}, {D'Odorico}, {Kaufer}, {Delabre}  \&
  {Kotzlowski}}{{Dekker} et~al.}{2000}]{dekker2000}
{Dekker} H.,  {D'Odorico} S.,  {Kaufer} A.,  {Delabre} B.,   {Kotzlowski} H.,
  2000, in {Iye} M.,  {Moorwood} A.~F.,  eds,  \procspie Vol. 4008, Optical and
  IR Telescope Instrumentation and Detectors. pp 534--545,
  \mn@doi{10.1117/12.395512}

\bibitem[\protect\citeauthoryear{{Dotter}, {Conroy}, {Cargile}  \&
  {Asplund}}{{Dotter} et~al.}{2017}]{dotter2017}
{Dotter} A.,  {Conroy} C.,  {Cargile} P.,   {Asplund} M.,  2017, \mn@doi [\apj]
  {10.3847/1538-4357/aa6d10}, \href
  {http://adsabs.harvard.edu/abs/2017ApJ...840...99D} {840, 99}

\bibitem[\protect\citeauthoryear{{Freeman} \& {Bland-Hawthorn}}{{Freeman} \&
  {Bland-Hawthorn}}{2002}]{freeman2002}
{Freeman} K.,  {Bland-Hawthorn} J.,  2002, \mn@doi [\araa]
  {10.1146/annurev.astro.40.060401.093840}, \href
  {http://adsabs.harvard.edu/abs/2002ARA%26A..40..487F} {40, 487}

\bibitem[\protect\citeauthoryear{{Garc{\'{\i}}a P{\'e}rez}
  et~al.,}{{Garc{\'{\i}}a P{\'e}rez} et~al.}{2016}]{garciaperez2015}
{Garc{\'{\i}}a P{\'e}rez} A.~E.,  et~al., 2016, \mn@doi [\aj]
  {10.3847/0004-6256/151/6/144}, \href
  {http://adsabs.harvard.edu/abs/2016AJ....151..144G} {151, 144}

\bibitem[\protect\citeauthoryear{{Geller}, {Latham}  \& {Mathieu}}{{Geller}
  et~al.}{2015}]{geller2015}
{Geller} A.~M.,  {Latham} D.~W.,   {Mathieu} R.~D.,  2015, \mn@doi [\aj]
  {10.1088/0004-6256/150/3/97}, \href
  {http://adsabs.harvard.edu/abs/2015AJ....150...97G} {150, 97}

\bibitem[\protect\citeauthoryear{{Gilmore} et~al.,}{{Gilmore}
  et~al.}{2012}]{gilmore2012}
{Gilmore} G.,  et~al., 2012, The Messenger, \href
  {http://adsabs.harvard.edu/abs/2012Msngr.147...25G} {147, 25}

\bibitem[\protect\citeauthoryear{{Gratton}, {Sneden}  \& {Carretta}}{{Gratton}
  et~al.}{2004}]{gratton2004}
{Gratton} R.,  {Sneden} C.,   {Carretta} E.,  2004, \mn@doi [\araa]
  {10.1146/annurev.astro.42.053102.133945}, \href
  {http://adsabs.harvard.edu/abs/2004ARA%26A..42..385G} {42, 385}

\bibitem[\protect\citeauthoryear{{Gratton}, {Carretta}  \&
  {Bragaglia}}{{Gratton} et~al.}{2012}]{gratton2012}
{Gratton} R.~G.,  {Carretta} E.,   {Bragaglia} A.,  2012, \mn@doi [\aapr]
  {10.1007/s00159-012-0050-3}, \href
  {http://adsabs.harvard.edu/abs/2012A%26ARv..20...50G} {20, 50}

\bibitem[\protect\citeauthoryear{{Gruyters}, {Korn}, {Richard}, {Grundahl},
  {Collet}, {Mashonkina}, {Osorio}  \& {Barklem}}{{Gruyters}
  et~al.}{2013}]{gruyters2013}
{Gruyters} P.,  {Korn} A.~J.,  {Richard} O.,  {Grundahl} F.,  {Collet} R.,
  {Mashonkina} L.~I.,  {Osorio} Y.,   {Barklem} P.~S.,  2013, \mn@doi [\aap]
  {10.1051/0004-6361/201220821}, \href
  {http://adsabs.harvard.edu/abs/2013A%26A...555A..31G} {555, A31}

\bibitem[\protect\citeauthoryear{{Gruyters} et~al.,}{{Gruyters}
  et~al.}{2016}]{gruyters2016}
{Gruyters} P.,  et~al., 2016, \mn@doi [\aap] {10.1051/0004-6361/201527948},
  \href {http://adsabs.harvard.edu/abs/2016A%26A...589A..61G} {589, A61}

\bibitem[\protect\citeauthoryear{{Holtzman} et~al.,}{{Holtzman}
  et~al.}{2015}]{holtzman2015}
{Holtzman} J.~A.,  et~al., 2015, \mn@doi [\aj] {10.1088/0004-6256/150/5/148},
  \href {http://adsabs.harvard.edu/abs/2015AJ....150..148H} {150, 148}

\bibitem[\protect\citeauthoryear{{Jacobson} et~al.,}{{Jacobson}
  et~al.}{2016}]{jacobson2016}
{Jacobson} H.~R.,  et~al., 2016, \mn@doi [\aap] {10.1051/0004-6361/201527654},
  \href {http://adsabs.harvard.edu/abs/2016A%26A...591A..37J} {591, A37}

\bibitem[\protect\citeauthoryear{{Jofr{\'e}} \& {Weiss}}{{Jofr{\'e}} \&
  {Weiss}}{2011}]{jofre2011}
{Jofr{\'e}} P.,  {Weiss} A.,  2011, \mn@doi [\aap]
  {10.1051/0004-6361/201117131}, \href
  {http://adsabs.harvard.edu/abs/2011A%26A...533A..59J} {533, A59}

\bibitem[\protect\citeauthoryear{{Kayser}, {Hilker}, {Grebel}  \&
  {Willemsen}}{{Kayser} et~al.}{2008}]{kayser2008}
{Kayser} A.,  {Hilker} M.,  {Grebel} E.~K.,   {Willemsen} P.~G.,  2008, \mn@doi
  [\aap] {10.1051/0004-6361:200809446}, \href
  {http://adsabs.harvard.edu/abs/2008A%26A...486..437K} {486, 437}

\bibitem[\protect\citeauthoryear{{Kharchenko}, {Piskunov}, {Schilbach},
  {R{\"o}ser}  \& {Scholz}}{{Kharchenko} et~al.}{2013}]{kharchenko2013}
{Kharchenko} N.~V.,  {Piskunov} A.~E.,  {Schilbach} E.,  {R{\"o}ser} S.,
  {Scholz} R.-D.,  2013, \mn@doi [\aap] {10.1051/0004-6361/201322302}, \href
  {http://adsabs.harvard.edu/abs/2013A%26A...558A..53K} {558, A53}

\bibitem[\protect\citeauthoryear{{Korn}, {Grundahl}, {Richard}, {Mashonkina},
  {Barklem}, {Collet}, {Gustafsson}  \& {Piskunov}}{{Korn}
  et~al.}{2007}]{korn2007}
{Korn} A.~J.,  {Grundahl} F.,  {Richard} O.,  {Mashonkina} L.,  {Barklem}
  P.~S.,  {Collet} R.,  {Gustafsson} B.,   {Piskunov} N.,  2007, \mn@doi [\apj]
  {10.1086/523098}, \href {http://adsabs.harvard.edu/abs/2007ApJ...671..402K}
  {671, 402}

\bibitem[\protect\citeauthoryear{{Krishna Swamy}}{{Krishna
  Swamy}}{1966}]{krishna1966}
{Krishna Swamy} K.~S.,  1966, \mn@doi [\apj] {10.1086/148752}, \href
  {http://adsabs.harvard.edu/abs/1966ApJ...145..174K} {145, 174}

\bibitem[\protect\citeauthoryear{{Lada} \& {Lada}}{{Lada} \&
  {Lada}}{2003}]{lada2003}
{Lada} C.~J.,  {Lada} E.~A.,  2003, \mn@doi [\araa]
  {10.1146/annurev.astro.41.011802.094844}, \href
  {http://adsabs.harvard.edu/abs/2003ARA%26A..41...57L} {41, 57}

\bibitem[\protect\citeauthoryear{{Lind}, {Asplund}, {Barklem}  \&
  {Belyaev}}{{Lind} et~al.}{2011}]{lind2011}
{Lind} K.,  {Asplund} M.,  {Barklem} P.~S.,   {Belyaev} A.~K.,  2011, \mn@doi
  [\aap] {10.1051/0004-6361/201016095}, \href
  {http://adsabs.harvard.edu/abs/2011A%26A...528A.103L} {528, A103}

\bibitem[\protect\citeauthoryear{{Lind}, {Bergemann}  \& {Asplund}}{{Lind}
  et~al.}{2012}]{lind2012}
{Lind} K.,  {Bergemann} M.,   {Asplund} M.,  2012, \mn@doi [\mnras]
  {10.1111/j.1365-2966.2012.21686.x}, \href
  {http://adsabs.harvard.edu/abs/2012MNRAS.427...50L} {427, 50}

\bibitem[\protect\citeauthoryear{{Majewski} et~al.,}{{Majewski}
  et~al.}{2017}]{majewski2017}
{Majewski} S.~R.,  et~al., 2017, \mn@doi [\aj] {10.3847/1538-3881/aa784d},
  \href {http://adsabs.harvard.edu/abs/2017AJ....154...94M} {154, 94}

\bibitem[\protect\citeauthoryear{{Michaud}, {Richard}, {Richer}  \&
  {VandenBerg}}{{Michaud} et~al.}{2004}]{michaud2004}
{Michaud} G.,  {Richard} O.,  {Richer} J.,   {VandenBerg} D.~A.,  2004, \mn@doi
  [\apj] {10.1086/383001}, \href
  {http://adsabs.harvard.edu/abs/2004ApJ...606..452M} {606, 452}

\bibitem[\protect\citeauthoryear{{Michaud}, {Alecian}  \& {Richer}}{{Michaud}
  et~al.}{2015}]{michaud2015}
{Michaud} G.,  {Alecian} G.,   {Richer} J.,  2015, {Atomic Diffusion in Stars}.
Springer International Publishing, Switzerland,
  \mn@doi{10.1007/978-3-319-19854-5}

\bibitem[\protect\citeauthoryear{{Ness} et~al.,}{{Ness}
  et~al.}{2018}]{ness2018}
{Ness} M.,  et~al., 2018, \mn@doi [\apj] {10.3847/1538-4357/aa9d8e}, \href
  {http://adsabs.harvard.edu/abs/2018ApJ...853..198N} {853, 198}

\bibitem[\protect\citeauthoryear{{Nordlander} \& {Lind}}{{Nordlander} \&
  {Lind}}{2017}]{nordlander2017}
{Nordlander} T.,  {Lind} K.,  2017, \mn@doi [\aap]
  {10.1051/0004-6361/201730427}, \href
  {http://adsabs.harvard.edu/abs/2017A%26A...607A..75N} {607, A75}

\bibitem[\protect\citeauthoryear{{{\"O}nehag}, {Korn}, {Gustafsson}, {Stempels}
   \& {Vandenberg}}{{{\"O}nehag} et~al.}{2011}]{onehag2011}
{{\"O}nehag} A.,  {Korn} A.,  {Gustafsson} B.,  {Stempels} E.,   {Vandenberg}
  D.~A.,  2011, \mn@doi [\aap] {10.1051/0004-6361/201015138}, \href
  {http://adsabs.harvard.edu/abs/2011A%26A...528A..85O} {528, A85}

\bibitem[\protect\citeauthoryear{{{\"O}nehag}, {Gustafsson}  \&
  {Korn}}{{{\"O}nehag} et~al.}{2014}]{onehag2014}
{{\"O}nehag} A.,  {Gustafsson} B.,   {Korn} A.,  2014, \mn@doi [\aap]
  {10.1051/0004-6361/201322663}, \href
  {http://adsabs.harvard.edu/abs/2014A%26A...562A.102O} {562, A102}

\bibitem[\protect\citeauthoryear{{Pace}, {Pasquini}  \& {Fran{\c c}ois}}{{Pace}
  et~al.}{2008}]{pace2008}
{Pace} G.,  {Pasquini} L.,   {Fran{\c c}ois} P.,  2008, \mn@doi [\aap]
  {10.1051/0004-6361:200809969}, \href
  {http://adsabs.harvard.edu/abs/2008A%26A...489..403P} {489, 403}

\bibitem[\protect\citeauthoryear{{Pasquini} et~al.,}{{Pasquini}
  et~al.}{2002}]{pasquini2002}
{Pasquini} L.,  et~al., 2002, The Messenger, \href
  {http://adsabs.harvard.edu/abs/2002Msngr.110....1P} {110, 1}

\bibitem[\protect\citeauthoryear{{Pasquini} et~al.,}{{Pasquini}
  et~al.}{2012}]{pasquini2012}
{Pasquini} L.,  et~al., 2012, \mn@doi [\aap] {10.1051/0004-6361/201219169},
  \href {http://adsabs.harvard.edu/abs/2012A%26A...545A.139P} {545, A139}

\bibitem[\protect\citeauthoryear{{Randich}, {Sestito}, {Primas}, {Pallavicini}
  \& {Pasquini}}{{Randich} et~al.}{2006}]{randich2006}
{Randich} S.,  {Sestito} P.,  {Primas} F.,  {Pallavicini} R.,   {Pasquini} L.,
  2006, \mn@doi [\aap] {10.1051/0004-6361:20054291}, \href
  {http://adsabs.harvard.edu/abs/2006A%26A...450..557R} {450, 557}

\bibitem[\protect\citeauthoryear{{Randich}, {Gilmore}  \& {Gaia-ESO
  Consortium}}{{Randich} et~al.}{2013}]{randich2013}
{Randich} S.,  {Gilmore} G.,   {Gaia-ESO Consortium} 2013, The Messenger, \href
  {http://adsabs.harvard.edu/abs/2013Msngr.154...47R} {154, 47}

\bibitem[\protect\citeauthoryear{{Randich} et~al.,}{{Randich}
  et~al.}{2017}]{randich2017}
{Randich} S.,  et~al., 2017, preprint, \href
  {http://adsabs.harvard.edu/abs/2017arXiv171107699R} {} (\mn@eprint {arXiv}
  {1711.07699})

\bibitem[\protect\citeauthoryear{{Richard}, {Michaud}  \& {Richer}}{{Richard}
  et~al.}{2001}]{richard2001}
{Richard} O.,  {Michaud} G.,   {Richer} J.,  2001, \mn@doi [\apj]
  {10.1086/322264}, \href {http://adsabs.harvard.edu/abs/2001ApJ...558..377R}
  {558, 377}

\bibitem[\protect\citeauthoryear{{Richard}, {Michaud}, {Richer}, {Turcotte},
  {Turck-Chi{\`e}ze}  \& {VandenBerg}}{{Richard} et~al.}{2002}]{richard2002}
{Richard} O.,  {Michaud} G.,  {Richer} J.,  {Turcotte} S.,  {Turck-Chi{\`e}ze}
  S.,   {VandenBerg} D.~A.,  2002, \mn@doi [\apj] {10.1086/338952}, \href
  {http://adsabs.harvard.edu/abs/2002ApJ...568..979R} {568, 979}

\bibitem[\protect\citeauthoryear{{Sacco} et~al.,}{{Sacco}
  et~al.}{2014}]{sacco2014}
{Sacco} G.~G.,  et~al., 2014, \mn@doi [\aap] {10.1051/0004-6361/201423619},
  \href {http://adsabs.harvard.edu/abs/2014A%26A...565A.113S} {565, A113}

\bibitem[\protect\citeauthoryear{{Salaris} \& {Weiss}}{{Salaris} \&
  {Weiss}}{2001}]{salaris2001}
{Salaris} M.,  {Weiss} A.,  2001, in {von Hippel} T.,  {Simpson} C.,   {Manset}
  N.,  eds,  Astronomical Society of the Pacific Conference Series Vol. 245,
  Astrophysical Ages and Times Scales. p.~367

\bibitem[\protect\citeauthoryear{{Sanders}}{{Sanders}}{1977}]{sanders1977}
{Sanders} W.~L.,  1977, \aaps, \href
  {http://adsabs.harvard.edu/abs/1977A%26AS...27...89S} {27, 89}

\bibitem[\protect\citeauthoryear{{Sarajedini}, {Dotter}  \&
  {Kirkpatrick}}{{Sarajedini} et~al.}{2009}]{sarajedini2009}
{Sarajedini} A.,  {Dotter} A.,   {Kirkpatrick} A.,  2009, \mn@doi [\apj]
  {10.1088/0004-637X/698/2/1872}, \href
  {http://adsabs.harvard.edu/abs/2009ApJ...698.1872S} {698, 1872}

\bibitem[\protect\citeauthoryear{{Shetrone} \& {Sandquist}}{{Shetrone} \&
  {Sandquist}}{2000}]{shetrone2000}
{Shetrone} M.~D.,  {Sandquist} E.~L.,  2000, \mn@doi [\aj] {10.1086/301569},
  \href {http://adsabs.harvard.edu/abs/2000AJ....120.1913S} {120, 1913}

\bibitem[\protect\citeauthoryear{{Skrutskie} et~al.,}{{Skrutskie}
  et~al.}{2006}]{skrutskie2006}
{Skrutskie} M.~F.,  et~al., 2006, \mn@doi [\aj] {10.1086/498708}, \href
  {http://adsabs.harvard.edu/abs/2006AJ....131.1163S} {131, 1163}

\bibitem[\protect\citeauthoryear{{Smiljanic} et~al.,}{{Smiljanic}
  et~al.}{2014}]{smiljanic2014}
{Smiljanic} R.,  et~al., 2014, \mn@doi [\aap] {10.1051/0004-6361/201423937},
  \href {http://adsabs.harvard.edu/abs/2014A%26A...570A.122S} {570, A122}

\bibitem[\protect\citeauthoryear{{Smiljanic} et~al.,}{{Smiljanic}
  et~al.}{2016}]{smiljanic2016}
{Smiljanic} R.,  et~al., 2016, \mn@doi [\aap] {10.1051/0004-6361/201528014},
  \href {http://adsabs.harvard.edu/abs/2016A%26A...589A.115S} {589, A115}

\bibitem[\protect\citeauthoryear{{Souto} et~al.,}{{Souto}
  et~al.}{2018}]{souto2018}
{Souto} D.,  et~al., 2018, preprint, \href
  {http://adsabs.harvard.edu/abs/2018arXiv180304461S} {} (\mn@eprint {arXiv}
  {1803.04461})

\bibitem[\protect\citeauthoryear{{Tautvai{\v s}iene}, {Edvardsson}, {Tuominen}
  \& {Ilyin}}{{Tautvai{\v s}iene} et~al.}{2000}]{tautv2000}
{Tautvai{\v s}iene} G.,  {Edvardsson} B.,  {Tuominen} I.,   {Ilyin} I.,  2000,
  \aap, \href {http://adsabs.harvard.edu/abs/2000A%26A...360..499T} {360, 499}

\bibitem[\protect\citeauthoryear{{Turcotte}, {Richer}, {Michaud}, {Iglesias}
  \& {Rogers}}{{Turcotte} et~al.}{1998}]{turcotte1998}
{Turcotte} S.,  {Richer} J.,  {Michaud} G.,  {Iglesias} C.~A.,   {Rogers}
  F.~J.,  1998, \mn@doi [\apj] {10.1086/306055}, \href
  {http://adsabs.harvard.edu/abs/1998ApJ...504..539T} {504, 539}

\bibitem[\protect\citeauthoryear{{Yadav} et~al.,}{{Yadav}
  et~al.}{2008}]{yadav2008}
{Yadav} R.~K.~S.,  et~al., 2008, \mn@doi [\aap] {10.1051/0004-6361:20079245},
  \href {http://adsabs.harvard.edu/abs/2008A%26A...484..609Y} {484, 609}

\bibitem[\protect\citeauthoryear{{Yong}, {Carney}  \& {Teixera de
  Almeida}}{{Yong} et~al.}{2005}]{yong2005}
{Yong} D.,  {Carney} B.~W.,   {Teixera de Almeida} M.~L.,  2005, \mn@doi [\aj]
  {10.1086/430934}, \href {http://adsabs.harvard.edu/abs/2005AJ....130..597Y}
  {130, 597}

\bibitem[\protect\citeauthoryear{{Zhang}, {Shi}, {Pan}, {Allende Prieto}  \&
  {Liu}}{{Zhang} et~al.}{2016}]{zhang2016}
{Zhang} J.,  {Shi} J.,  {Pan} K.,  {Allende Prieto} C.,   {Liu} C.,  2016,
  \mn@doi [\apj] {10.3847/1538-4357/833/2/137}, \href
  {http://adsabs.harvard.edu/abs/2016ApJ...833..137Z} {833, 137}

\bibitem[\protect\citeauthoryear{{Zhang}, {Shi}, {Pan}, {Allende Prieto}  \&
  {Liu}}{{Zhang} et~al.}{2017}]{zhang2017}
{Zhang} J.,  {Shi} J.,  {Pan} K.,  {Allende Prieto} C.,   {Liu} C.,  2017,
  \mn@doi [\apj] {10.3847/1538-4357/835/1/90}, \href
  {http://adsabs.harvard.edu/abs/2017ApJ...835...90Z} {835, 90}

\makeatother
\end{thebibliography}

\section*{Affiliations}
	\footnotesize\textit{
	$^{1}$Astronomisches Rechen-Institut, Zentrum f\"ur Astronomie der Universit\"at Heidelberg, M\"onchhofstr. 12-14, 69120 Heidelberg, Germany\\
	$^{2}$D\'{e}partement de physique, Universit\'{e} de Montr\'{e}al, Montr\'{e}al, Qu\'{e}bec, H3C 3J7, Canada\\
	$^{3}$Astrophysics Research Institute, Liverpool John Moores University, 146 Brownlow Hill, Liverpool L3 5RF, UK\\
	$^{4}$INAF -- Osservatorio di Astrofisica e Scienza dello Spazio, via Gobetti 93/3, 40129 Bologna, Italy\\
	$^{5}$INAF -- Osservatorio Astrofisico di Arcetri, Largo E. Fermi, 5, I- 50125 Firenze, Italy\\
	$^{6}$Instituto de Astrof\'isica e Ci\^encias do Espa\c{c}o, Universidade do Porto, CAUP, Rua das Estrelas, 4150-762 Porto, Portugal\\
	$^{7}$Harvard-Smithsonian Center for Astrophysics, 60 Garden Street, Cambridge, MA 02138, USA\\
	$^{8}$Institute of Theoretical Physics and Astronomy, Vilnius University, Saul\.{e}tekio av. 3, LT-10257 Vilnius, Lithuania\\
	$^{9}$Dipartimento di Fisica \& Astronomia, Universit\`{a} degli Studi di Bologna,  via Gobetti 93/2, 40129 Bologna, Italy\\
	$^{10}$School of Physics, University of New South Wales, Sydney NSW 2052, Australia\\
	$^{11}$Institute of Astronomy, University of Cambridge, Madingley Road, Cambridge CB3 0HA, United Kingdom\\
	$^{12}$Instituto de Astrof\'{i}sica de Andaluc\'{i}a, CSIC, Glorieta de la Astronom\'{i}a, s/n, Granada, Spain\\
	$^{13}$Lund Observatory, Department of Astronomy and Theoretical Physics, Box 43, SE-221 00 Lund, Sweden\\
	$^{14}$INAF - Osservatorio Astronomico di Palermo, Piazza del Parlamento 1, 90134, Palermo, Italy\\
	$^{15}$Department of Physics, McWilliams Center for Cosmology, Carnegie Mellon University, 5000 Forbes Avenue, Pittsburgh, PA 15213, USA\\
	$^{16}$Department of Physics and Astronomy, Uppsala University, Box 516, SE-751 20 Uppsala, Sweden\\
	$^{17}$Dipartimento di Fisica e Astronomia, Sezione Astrofisica, Universit\`{a} di Catania, via S. Sofia 78, 95123, Catania, Italy\\
	$^{18}$Nicolaus Copernicus Astronomical Center, Polish Academy of Sciences, ul. Bartycka 18, 00-716, Warsaw, Poland\\
	$^{19}$Instituto de F\'{i}sica y Astronom\'{i}a, Universidad de Valpara\'{i}so, Chile\\
	$^{20}$N\'ucleo Milenio Formaci\'on Planetaria - NPF, Universidad de Valpara\'iso, Av. Gran Breta\~na 1111, Valpara\'iso, Chile\\
	$^{21}$Dipartimento di Fisica e Astronomia, Universit\`a di Padova, Vicolo dell'Osservatorio 3, 35122 Padova, Italy\\
	$^{22}$School of Physics and Astronomy, Monash University, Clayton 3800, Victoria, Australia\\
	$^{23}$Faculty of Information Technology, Monash University, Clayton 3800, Victoria, Australia\\
	$^{24}$Departamento de Did\'{a}ctica, Universidad de C\'{a}diz, 11519 Puerto Real, C\'{a}diz, Spain\\
	$^{25}$Observational Astrophysics, Department of Physics and Astronomy, Uppsala University, Box 516, 75120 Uppsala, Sweden\\
	$^{26}$N\'{u}cleo de Astronom\'{i}a, Universidad Diego Portales, Av. Ejercito 441, Santiago de Chile\\
	$^{27}$Laboratoire d'astrophysique, Ecole Polytechnique F\'ed\'erale de Lausanne (EPFL), Observatoire de Sauverny, CH-1290 Versoix, Switzerland\\
	$^{28}$Departamento de Ciencias Fisicas, Universidad Andres Bello, Fernandez Concha 700, Las Condes, Santiago, Chile\\
		}


\normalsize
\appendix

\section{}

In Table~\ref{tab:memb}, we present a summary of the main parameters for the 15 stars selected as members of M67 after our membership analysis. Besides GES-ID, RA, and Dec, the radial velocities, proper motions from the HSOY catalogue, effective temperatures and gravities derived within GES with the respective errors are listed. In Table~\ref{tab:memb_ab} the abundances of the different elements are summarised for each of the 15 stars.

\begin{figure*}
	\includegraphics[scale=0.94]{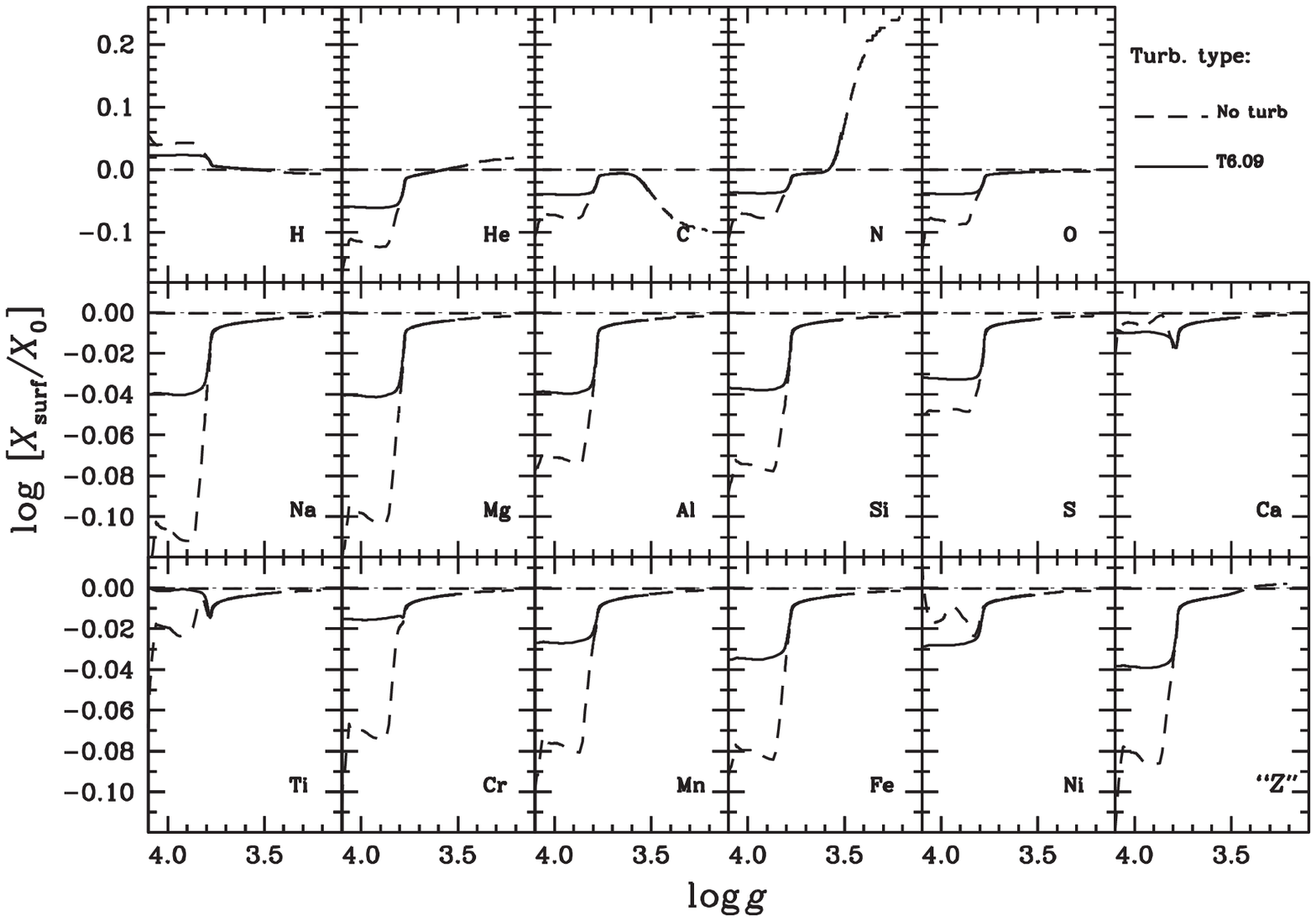}
	\caption{Comparison of the surface abundances of two different models with (solid black line) and without (dashed black line) turbulence for a star of $1.35M_{\odot}$ as a function of $\log g$. }
	\label{fig:fig_end}
\end{figure*}

Figure~\ref{fig:fig_end} shows the evolution of different surface abundances as a function of $\log g$ for two models with and without turbulence representing a star of mass $1.35M_{\odot}$. The model with turbulence did not reach the same age as the one without turbulence and hence is not as evolved as  the three giants in our sample. The plot aims to demonstrate that once the red giant branch is reached, both models present the same surface abundances. This is evident, since after  $\log g\sim3.6$ both models converge.

\begin{table*}
	\caption{List of the GES stars selected as members of M67 and their principal parameters: coordinates, proper motions from the HSOY catalogue, radial velocities, temperatures, and gravities as derived by the GES pipelines.}
\begin{tabular}{|l|l|l|l|l|l|l|l|l|l|l|l|l|}
	\hline
	\multicolumn{1}{|c|}{ID} &
	\multicolumn{1}{c|}{RA} &
	\multicolumn{1}{c|}{Dec} &
	\multicolumn{1}{c|}{Teff } &
	\multicolumn{1}{c|}{e\_Teff } &
	\multicolumn{1}{c|}{logg} &
	\multicolumn{1}{c|}{e\_logg} &
	\multicolumn{1}{c|}{RV} &
	\multicolumn{1}{c|}{e\_RV} &
	\multicolumn{1}{c|}{pmra} &
	\multicolumn{1}{c|}{pmde} &
	\multicolumn{1}{c|}{e\_pmra} &
	\multicolumn{1}{c|}{e\_pmde} \\
	\multicolumn{1}{|c|}{} &
	\multicolumn{1}{c|}{[hms]} &
	\multicolumn{1}{c|}{[dms]} &
	\multicolumn{1}{c|}{[K]} &
	\multicolumn{1}{c|}{[K]} &
	\multicolumn{1}{c|}{} &
	\multicolumn{1}{c|}{} &
	\multicolumn{1}{c|}{[km/s]} &
	\multicolumn{1}{c|}{[km/s]} &
	\multicolumn{1}{c|}{[mas/yr]} &
	\multicolumn{1}{c|}{[mas/yr]} &
	\multicolumn{1}{c|}{[mas/yr]} &
	\multicolumn{1}{c|}{[mas/yr]} \\
	\hline
	08514507+1147459 & 08:51:45.07 & +11:47:45.9 & 4798 & 146 & 3.031 & 0.226 & 33.609 & 0.364 & -11.08 & -1.55 & 0.85 & 0.86\\
	08513577+1153347 & 08:51:35.77 & +11:53:34.7 & 4913 & 147 & 3.296 & 0.228 & 34.472 & 0.364 & -9.70 & -3.43 & 0.65 & 0.68\\
	08510838+1147121 & 08:51:08.38 & +11:47:12.1 & 4915 & 147 & 3.381 & 0.227 & 33.887 & 0.364 & -10.11 & -3.20 & 0.65 & 0.65\\
	08510017+1154321 & 08:51:00.17 & +11:54:32.1 & 5423 & 56 & 3.824 & 0.112 & 34.591 & 0.364 & -9.53 & -3.66 & 0.66 & 0.66\\
	08510080+1148527 & 08:51:00.80 & +11:48:52.7 & 5733 & 57 & 4.422 & 0.112 & 34.690 & 0.364 & -8.68 & -1.98 & 0.85 & 0.85\\
	08511854+1149214 & 08:51:18.54 & +11:49:21.4 & 5873 & 57 & 3.732 & 0.111 & 35.739 & 0.364 & -10.02 & -3.15 & 0.72 & 0.77\\
	08510325+1145473 & 08:51:03.25 & +11:45:47.3 & 5887 & 57 & 3.762 & 0.112 & 35.455 & 0.364 & -9.58 & -3.10 & 0.65 & 0.65\\
	08514995+1149311 & 08:51:49.95 & +11:49:31.1 & 5915 & 56 & 3.726 & 0.112 & 33.675 & 0.364 & -10.07 & -3.23 & 0.72 & 0.76\\
	08513740+1150052 & 08:51:37.40 & +11:50:05.2 & 5946 & 56 & 3.835 & 0.112 & 32.802 & 0.364 & -10.03 & -3.79 & 0.65 & 0.65\\
	08510524+1149340 & 08:51:05.24 & +11:49:34.0 & 6001 & 56 & 4.174 & 0.112 & 35.750 & 0.364 & -9.76 & -5.37 & 0.95 & 1.07\\
	08514081+1149055 & 08:51:40.81 & +11:49:05.5 & 6022 & 57 & 4.253 & 0.112 & 34.439 & 0.364 & -10.42 & -3.29 & 0.71 & 0.71\\
	08505891+1148192 & 08:50:58.91 & +11:48:19.2 & 6036 & 57 & 4.229 & 0.112 & 35.408 & 0.364 & -9.26 & -2.73 & 0.77 & 0.71\\
	08505600+1153519 & 08:50:56.00 & +11:53:51.9 & 6064 & 56 & 4.186 & 0.111 & 35.828 & 0.364 & -8.77 & -4.52 & 0.71 & 0.75\\
	08514122+1154290 & 08:51:41.22 & +11:54:29.0 & 6065 & 56 & 3.854 & 0.111 & 33.997 & 0.365 & -9.27 & -3.53 & 0.65 & 0.65\\
	08512012+1146417 & 08:51:20.12 & +11:46:41.7 & 6069 & 57 & 3.865 & 0.113 & 34.409 & 0.364 & -8.78 & -3.66 & 0.65 & 0.65\\
	\hline\end{tabular}

	\label{tab:memb}
\end{table*}

\begin{table*}
	\caption{List of the GES stars selected as members of M67 together with the respective abundances derived within GES and investigated in this work.}
\begin{tabular}{|l|l|l|l|l|l|l|l|l|l|l|l|l|}
	\hline
	\multicolumn{1}{|c|}{ID} &
	\multicolumn{1}{c|}{C} &
	\multicolumn{1}{c|}{e\_C} &
	\multicolumn{1}{c|}{O} &
	\multicolumn{1}{c|}{e\_O} &
	\multicolumn{1}{c|}{Na} &
	\multicolumn{1}{c|}{e\_Na} &
	\multicolumn{1}{c|}{Mg} &
	\multicolumn{1}{c|}{e\_Mg} &
	\multicolumn{1}{c|}{Al} &
	\multicolumn{1}{c|}{e\_Al} &
	\multicolumn{1}{c|}{Si} &
	\multicolumn{1}{c|}{e\_Si} \\
\hline
08514507+1147459 & 8.22 & 0.05 & 8.72 & 0.09 & 6.31 & 0.04 & 7.51 & 0.06 & 6.42 & 0.04 & 7.57 & 0.06\\
08513577+1153347 & 8.24 & 0.02 & 8.70 & 0.10 & 6.20 & 0.07 & 7.50 & 0.05 & 6.39 & 0.04 & 7.54 & 0.05\\
08510838+1147121 & 8.26 & 0.02 & 8.69 & 0.10 & 6.20 & 0.07 & 7.54 & 0.05 & 6.41 & 0.03 & 7.55 & 0.07\\
08510017+1154321 & 8.23 & 0.03 & 8.65 & 0.02 & 6.27 & 0.06 & 7.59 & 0.07 & 6.45 & 0.06 & 7.58 & 0.05\\
08510080+1148527 & 8.23 & 0.02 & 8.60 & 0.10 & 6.14 & 0.06 & 7.53 & 0.10 & 6.33 & 0.04 & 7.43 & 0.08\\
08511854+1149214 & 8.19 & 0.09 & 8.55 & 0.04 & 6.18 & 0.11 & 7.45 & 0.12 & 6.22 & 0.10 & 7.43 & 0.08\\
08510325+1145473 & 8.18 & 0.06 & 8.61 & 0.02 & 6.20 & 0.11 & 7.50 & 0.05 & 6.28 & 0.04 & 7.47 & 0.06\\
08514995+1149311 & 8.17 & 0.05 & 8.50 & 0.09 & 6.16 & 0.14 & 7.43 & 0.07 & 6.19 & 0.04 & 7.39 & 0.08\\
08513740+1150052 & 8.17 & 0.08 & 8.39 & 0.04 & 6.14 & 0.10 & 7.48 & 0.06 & 6.22 & 0.04 & 7.43 & 0.08\\
08510524+1149340 & 8.18 & 0.05 & 8.70 & 0.09 & 6.10 & 0.09 & 7.43 & 0.08 & 6.17 & 0.04 & 7.38 & 0.08\\
08514081+1149055 & 8.20 & 0.06 & 8.57 & 0.01 & 6.15 & 0.08 & 7.47 & 0.10 & 6.19 & 0.04 & 7.41 & 0.08\\
08505891+1148192 & 8.18 & 0.05 & 8.50 & 0.09 & 6.13 & 0.08 & 7.48 & 0.10 & 6.19 & 0.04 & 7.39 & 0.08\\
08505600+1153519 & 8.19 & 0.07 & 8.60 & 0.09 & 6.11 & 0.10 & 7.48 & 0.06 & 6.19 & 0.05 & 7.43 & 0.10\\
08514122+1154290 & 8.16 & 0.07 & 8.61 & 0.01 & 6.17 & 0.13 & 7.47 & 0.06 & 6.18 & 0.03 & 7.41 & 0.08\\
08512012+1146417 & 8.15 & 0.07 & 8.53 & 0.05 & 6.12 & 0.10 & 7.42 & 0.09 & 6.17 & 0.03 & 7.39 & 0.06\\
\hline\end{tabular}

\begin{tabular}{|l|l|l|l|l|l|l|l|l|l|l|l|l|}
	\hline
	\multicolumn{1}{|c|}{ID} &
	\multicolumn{1}{c|}{Ca} &
	\multicolumn{1}{c|}{e\_Ca} &
	\multicolumn{1}{c|}{Ti} &
	\multicolumn{1}{c|}{e\_Ti} &
	\multicolumn{1}{c|}{Cr} &
	\multicolumn{1}{c|}{e\_Cr} &
	\multicolumn{1}{c|}{Mn} &
	\multicolumn{1}{c|}{e\_Mn} &
	\multicolumn{1}{c|}{Fe} &
	\multicolumn{1}{c|}{e\_Fe} &
	\multicolumn{1}{c|}{Ni} &
	\multicolumn{1}{c|}{e\_Ni} \\
	\hline
	08514507+1147459 & 6.46 & 0.08 & 4.91 & 0.10 & 5.62 & 0.08 & 5.48 & 0.09 & 7.43 & 0.06 & 6.30 & 0.09\\
	08513577+1153347 & 6.44 & 0.10 & 4.88 & 0.09 & 5.61 & 0.09 & 5.45 & 0.11 & 7.42 & 0.06 & 6.28 & 0.08\\
	08510838+1147121 & 6.44 & 0.11 & 4.90 & 0.09 & 5.61 & 0.09 & 5.45 & 0.13 & 7.43 & 0.06 & 6.29 & 0.09\\
	08510017+1154321 & 6.54 & 0.06 & 5.00 & 0.07 & 5.70 & 0.07 & 5.50 & 0.07 & 7.54 & 0.06 & 6.31 & 0.06\\
	08510080+1148527 & 6.42 & 0.06 & 4.90 & 0.09 & 5.60 & 0.08 & 5.39 & 0.06 & 7.43 & 0.06 & 6.20 & 0.06\\
	08511854+1149214 & 6.40 & 0.10 & 4.78 & 0.14 & 5.51 & 0.11 & 5.26 & 0.08 & 7.40 & 0.08 & 6.09 & 0.09\\
	08510325+1145473 & 6.47 & 0.09 & 4.86 & 0.12 & 5.56 & 0.10 & 5.32 & 0.08 & 7.44 & 0.08 & 6.15 & 0.09\\
	08514995+1149311 & 6.41 & 0.11 & 4.77 & 0.11 & 5.50 & 0.11 & 5.25 & 0.11 & 7.37 & 0.07 & 6.06 & 0.09\\
	08513740+1150052 & 6.45 & 0.11 & 4.82 & 0.12 & 5.51 & 0.11 & 5.29 & 0.11 & 7.41 & 0.08 & 6.09 & 0.11\\
	08510524+1149340 & 6.36 & 0.07 & 4.76 & 0.10 & 5.50 & 0.10 & 5.24 & 0.06 & 7.34 & 0.07 & 6.07 & 0.08\\
	08514081+1149055 & 6.37 & 0.07 & 4.80 & 0.11 & 5.53 & 0.10 & 5.29 & 0.05 & 7.39 & 0.07 & 6.11 & 0.09\\
	08505891+1148192 & 6.39 & 0.08 & 4.81 & 0.10 & 5.53 & 0.10 & 5.30 & 0.05 & 7.38 & 0.07 & 6.10 & 0.09\\
	08505600+1153519 & 6.37 & 0.09 & 4.79 & 0.12 & 5.52 & 0.11 & 5.27 & 0.11 & 7.39 & 0.08 & 6.09 & 0.10\\
	08514122+1154290 & 6.35 & 0.14 & 4.76 & 0.12 & 5.48 & 0.12 & 5.30 & 0.17 & 7.36 & 0.08 & 6.05 & 0.10\\
	08512012+1146417 & 6.39 & 0.13 & 4.75 & 0.12 & 5.48 & 0.10 & 5.25 & 0.16 & 7.35 & 0.08 & 6.04 & 0.10\\
	\hline\end{tabular}

	\label{tab:memb_ab}
\end{table*}

\bsp	
\label{lastpage}
\end{document}